\documentclass[12pt]{article}

\usepackage[fleqn]{amsmath}
\usepackage{amsfonts}
\usepackage{amssymb}
\usepackage{epsfig}
\usepackage{geometry}
\usepackage{ctable}

\newcommand{\Langle}{{\left\{\!\left[\right.\right.}}
\newcommand{\Rangle}{{\left.\left.\right]\!\right\}}}

\newcommand{\bcdot}{{\,\boldsymbol{\cdot}\,}}

\newcommand{\nullV}{{\boldsymbol{0}}}
\newcommand{\ID}{\boldsymbol{\mathcal{I}}}
\newcommand{\ONE}{\boldsymbol{1}}

\newcommand{\cP}{{\mathcal{P}}}
\newcommand{\cI}{{\mathcal{I}}}

\newcommand{\ul}{\underline}
\newcommand{\ol}{\overline}

\newcommand{\crprd}{\boldsymbol{\times}}
\newcommand{\eps}{\epsilon}

\newcommand{\alphaS}{\alpha_{\mbox{\tiny{\textrm{S}}}}}

\newcommand{\otl}{{\otimes_\ell}}

\newcommand{\aV}{{\boldsymbol{a}}}
\newcommand{\AV}{{\boldsymbol{A}}}

\newcommand{\dV}{{\boldsymbol{d}}}
\newcommand{\eV}{{\boldsymbol{e}}}
\newcommand{\jV}{{\boldsymbol{j}}}
\newcommand{\JV}{{\boldsymbol{J}}}
\newcommand{\JVv}{\vec{\boldsymbol{J}}}
\newcommand{\vVv}{\vec{\boldsymbol{v}}}
\newcommand{\kV}{{\boldsymbol{k}}}
\newcommand{\nV}{{\boldsymbol{n}}}
\newcommand{\pV}{{\boldsymbol{p}}}
\newcommand{\PV}{{\boldsymbol{P}}}
\newcommand{\qV}{{\boldsymbol{q}}}
\newcommand{\QV}{{\boldsymbol{Q}}}
\newcommand{\qVv}{\vec{\boldsymbol{q}}}

\newcommand{\sV}{{\boldsymbol{s}}}
\newcommand{\vV}{{\boldsymbol{v}}}

\newcommand{\PsiV}{{\boldsymbol{\Psi}}}
\newcommand{\PsiVL}{{\boldsymbol{\Psi}}^L}
\newcommand{\PsiVa}{{\boldsymbol{\Psi}}^{L-1}}

\newcommand{\EV}{{\boldsymbol{E}}}
\newcommand{\BV}{{\boldsymbol{B}}}

\newcommand{\Cset}{{\mathbb{C}}}
\newcommand{\Nset}{{\mathbb{N}}}
\newcommand{\Rset}{{\mathbb{R}}}
\newcommand{\Sset}{{\mathbb{S}}}

\newcommand{\eEL}{{e}} 
\newcommand{\mEL}{{m}_{\rm{e}}}       
\newcommand{\drm}{\mathrm{d}}

\newcommand{\abs}[1]{\big\vert #1 \big\vert}

\newcommand{\Efrak}{{\mathfrak E}} 
\newcommand{\Ffrak}{{\mathfrak F}} 

\begin{document}
\title{Semi-relativistic $N$-body quantum mechanics\\ of electrons and photons, with fixed nuclei} 

\author{\sc{Michael K.-H. Kiessling}\\ 
    {\small{Department of Mathematics}}\\
    {\small{Rutgers, The State University of New Jersey}}\\
     {\small{110 Frelinghuysen Rd., Piscataway, NJ 08854}}}

\date{May 05, 2021\vspace{-1truecm}} 

\maketitle

\begin{abstract}\noindent
  It is argued that by the end of the 1920s a quantum-mechanical model could have been in place, that not only 
produces the atomic and molecular energy levels of the many-body Pauli equation with Coulomb interactions and external 
classical electro- and magneto-static fields without putting these interactions in by hand, but that also accurately describes 
the interaction of charged particles with electromagnetic radiation, in particular the transitions between atomic or molecular 
energy levels associated with emission or absorption of radiation.
 This model suggests a re-interpretation of Maxwell's electromagnetic field equations on spacetime as 
quantum-mechanical expected values of wave equations on time $\times$ configuration space for photons and electrons.
 The creation / annihilation formalism for photons emerges without invoking second-quantizing the classical Maxwell equations,
and without involving the concept of creation / annihilation, thus suggesting an alternative physical interpretation of this formalism.
 Furthermore, the model suggests that Lorentz covariance of macroscopic physics models emerges through a law of large numbers 
from a fundamental microscopic model that is not itself Lorentz covariant.
\end{abstract}

\vfill

\hrule
\smallskip
\copyright {\footnotesize{(2021) The author. Reproduction, in its entirety, for non-commercial purposes is permitted.}}

\newpage

\section{Introduction}\vspace{-.4truecm}

 Dirac reportedly once said about the so-called \emph{standard model of} ``\emph{everyday matter}'' that it
covers all of the chemistry and most of the physics of systems from the size of atoms and molecules all the way up 
on the number-of-particles scale to objects the size of the moon (and beyond, a little bit).
 The scare quotes around \emph{everyday matter} are meant to remind us that
atoms and molecules are of course not part of our everyday experience. 
 Yet it is one and the same model which gives equally accurate results for atoms, molecules, etc., as well as for
objects the size of the moon.
 This standard model has also allowed mathematical physicists to prove quantum field-theoretical results which are out 
of reach in QED; cf. \cite{JuergA}, \cite{SpohnBOOKb}.

 Incidentally, the name \emph{standard model} in physics usually simply means that the model sets \emph{the 
standard of accuracy and efficiency in computing} quantitative answers to questions concerning the subject
matter, in essential agreement with empirical data.
 Any such standard model typically consists of a patchwork of partial theories, strung together, plus 
heuristic rules of procedure.
 It does \emph{not} mean that it sets the standard for a conceptually unified fundamental theory, although (ideally) it should.

 In particular, the practically successful rules of the standard model of everyday matter
are a curious mix of non-relativistic classical and quantum mechanics, and of relativistic quantum field theory:
the spin-$\frac12$ electrons are treated with the non-relativistic Pauli equation;
the nuclei are treated in the Born--Oppenheimer approximation, which means 
their positions are classical parameters in the Pauli equation for the electrons;
the photons are treated with the quantized electromagnetic Maxwell fields, minimally coupled to the Pauli spinors;
in addition, there are externally generated (applied) classical electromagnetic fields, also minimally coupled to the Pauli
spinors.
 Predictions extracted from the model are based on the usual measurement axioms formulated by von Neumann \cite{vN}.

 The non-relativistic quantum-mechanical Schr\"odinger--Pauli equation for electrons and nuclei (the latter of which 
could be either effectively bosons or fermions) is expected to be a reasonably accurate approximation to deal with the
 matter part of the system. 
 The energy densities of everyday matter are way below the matter-antimatter 
pair creation threshold of the involved matter particles so that quantum field-theoretical effects should be negligible
for the matter part of the model.

 The Born--Oppenheimer approximation for the nuclei is merely a convenient yet unnecessary further simplification;
the nuclear position degrees of freedom can easily be included at the level of nonrelativistic quantum mechanics.
 
 However, it is widely believed that the emission / absorption of photons 
by atoms can only be described quantum field-theoretically, cf. \cite{BohmBOOKa}, \cite{WeinbergBOOKqft}.
 This unproven yet widely held belief should be greeted with a healthy dose of skepticism.

 In this paper we inquire into a purely quantum-mechanical alternative to this standard model of atoms, molecules, etc., 
all the way up to objects as large as the moon (and a little bit further).
 For convenience we keep the Born--Oppenheimer approximation; yet again, this can be relaxed.
 The quantum-mechanical model proposed in this paper can be easier controlled with rigorous mathematical 
techniques, perturbatively as well as non-perturbatively, than the standard model of everyday matter.
 It produces exactly the same atomic and molecular (etc.) energy spectra as the 
many-body Schr\"odinger, respectively Pauli equation with Coulomb interactions and external electro- and magneto-static fields, 
 without putting those interactions into the Schr\"odinger / Pauli equation by hand, 
yet it also describes the emission of photons with the right frequencies.
 Furthermore, its physical predictions do not require von Neumann's measurement axioms of orthodox quantum theory.

 Another point we emphasize is that photons appear naturally in our approach, without resorting to
second-quantizing the classical Maxwell field equations.
 The photon emerges by analyzing Schr\"odinger's 1926 findings from the perspective
of Born's 1926 re-interpretation of Schr\"odinger's wave function, and by pursuing this lead to its logical conclusion.
 Of course, 
to have a viable many particle theory requires the input of principles  
discovered after 1926, in particular Pauli's 1927 theory that electrons are spin $\frac12$ fermions requiring permutation-antisymmetric 
spinor wave functions evolved by his generalization of Schr\"odinger's equation, and that photons are spin $1$ bosons, requiring 
permutation-symmetric wave functions.
 
 Thus, and initially ignoring electron spin, 
we begin by recalling Schr\"odinger's 1926 ``$\Psi$ as matter-wave'' theory, first for a non-relativistic hydrogen atom 
coupled with the actual classical electromagnetic fields, emphasizing its  initial successes and also its ultimate failure,
then for a non-relativistic spinless multi-electron atom.
 Next we recall Born's 1926 ``$\Psi$ as a probability amplitude'' re-interpretation of Schr\"odinger's formulas for the
charge and current densities and how he justified it with his ``$\Psi$ as a guiding field'' interpretation, which
Born wrote was inspired by Einstein's earlier speculations that photons are particles which are guided by the electromagnetic Maxwell fields.
 Einstein's ideas will also play a role in our model.
  Einstein's speculations also inspired de Broglie, who already in 1923 postulated a first-order guiding
equation for massive particles involving a guiding phase wave $\Phi$, though without having a wave equation for $\Phi$; in 1926 
he then found his $\Phi$ in Schr\"odinger's $\Psi$ through the polar presentation $\Psi = |\Psi|e^{i\Phi}$ and presented his
theory at the 5th Solvay Conference in 1927 \cite{Solvay}.
 De Broglie's deterministic guiding equation was later re-discovered by Bohm, who developed the theory further.
 While Born himself did not propose a guiding equation, he wrote that he was convinced that it had to be a 
non-deterministic equation. 
 Such a stochastic guiding law was eventually supplied by Nelson and further developed by Guerra et al.
 We use the deterministic de Broglie--Bohm law; a stochastic (Born--)Nelson--Guerra law may do just as well.

 Then, by analyzing Schr\"odinger's calculations of the electromagnetic radiation produced by solutions of his $\Psi$ equation 
from the perspective of Born's interpretation of $\Psi$ we deduce the existence of generalized electromagnetic fields
which depend not only on space and time, but also on the generic position variables of the electrons.
 The field equations for the generalized electromagnetic fields can be put together easily.
 After Fourier transform, they can be solved by the method of characteristics, and the de Broglie--Bohm-type
guiding equation is an integral part of these characteristic equations!
 Next, well-known results from classical electromagnetic field theory then suggest how to couple the generalized electromagnetic fields
back into Schr\"odinger's equation. 
 The model not only produces the same Schr\"odinger spectra as used in the (accordingly simplified) standard model of everyday matter,
electron spin is easily accomodated by working with the Pauli equation instead of Schr\"odinger's spinless equation, in which case
the spectra agree with those of the standard model.
 Moreover, it also describes emission of an electromagnetic radiation field with the right frequencies. 
 Those radiation fields are spread out and cannot explain the ``clicks'' of some localized photon detector.
 Yet the mathematical equations suggest a re-interpretation of the generalized electromagnetic fields
as actually living on many particle configuration space, with the photon position 
being part of the configuration space variables, and we propose a guiding equation for a photon.
 This change of physical perspective now does have the potential of explaining localized detector ``clicks.''
 Furthermore, it also suggests a re-interpretation of the empirial electron charge and current densities in terms
of ``photon creation operators'' in this model.
 This leads at once to a model of an atom coupled with many photons.
 Its generalization to a system of many nuclei, electrons, and photons is straightforward.

 Photon annihilation operators are also hinted at, in our non-relativistic Schr\"o\-dinger--Pauli equation, 
but our semi-relativistic model will have to be developed further to involve the relativistic Dirac operator
before it can take a putatively final form in which it can possibly compete with the standard model of everyday matter. 
 The many aspects which our tentative model gets exactly right already give us reason to be optimistic 
that such a purely quantum-mechanical model is feasible.

 We close with a brief discussion of perhaps the most intriguing finding of our work,
how the putatively final model could account for Lorentz covariance \emph{in the mean}.
 By a law of large numbers, which holds for all everyday phenomena,
a many body system will essentially behave like the mean, so that at the many-body level the special theory of relativity emerges 
as an apparent law of nature.
 Yet at a few-body level significant deviations may appear, as demonstrated in Bell-type experiments. \vspace{-20pt}

\newpage

\section{Schr\"odinger's matter-wave theory of radiating\\ atoms, molecules, etc.}\label{sec:Erwin}

 Schr\"odinger's notion of ``$\Psi$ as a matter wave,''\footnote{This folklore is too simplistic and should not
  be taken literally. In the course of the discussion in this section it will be made precise 
what Schr\"odinger meant by matter waves. See also \cite{ValiaETal}.}
when coupled with the (classical) electromagnetic fields in a self-consistent manner, ultimately
yields the Schr\"odinger--Maxwell system, a (neo-)classical field theory 
that is not a physically acceptable system of equations of a
few-electron atom or molecule, coupled with electromagnetic radiation --- hopes expressed to the contrary \cite{KomechLECT} notwithstanding.
 Fortunately Schr\"odinger worked at first with a truncated system which produced the well-known striking results that became part of 
textbook QM. 
 Since Schr\"odinger's results are important stepping stones on our way to a coherent quantum mechanics of electrons, nuclei, and photons,
we briefly recall them below. 
 While we do pay attention to the developments in 1926, we emphasize that our presentation is not meant as a strictly 
historical account; rather, we hope to convey a plausible train of thought. 
 For a historian's account, see \cite{Renn}.

\subsection{Hydrogen}

 We first work with the non-relativistic Schr\"odinger equation for a hydrogen atom, deferring spin and 
the Pauli equation to a later section.
 Like Schr\"odinger, we treat it first in isolation from the electromagnetic radiation fields, 
switching on the coupling subsequently in a perturbative manner, then address the non-perturbative 
Schr\"odinger--Maxwell system.

 Schr\"odinger's equation for the ``matter-wave'' function $\Psi(t,\sV)\in\Cset$ of an electron of mass $\mEL$ and charge $-e$
in the Coulomb field of a point proton of charge $e$, fixed at the origin, is found in every quantum-mechanics textbook; it reads \cite{ErwinWMd}
\begin{equation}
i \hbar \partial_t\Psi(t,\sV) = \tfrac{1}{2\mEL}\big(-i\hbar\nabla_\sV\big)^2\Psi(t,\sV) - \tfrac{e^2}{|\sV|}\Psi(t,\sV).
 \label{eq:ERWINeqnMatterWaveBOhydrogen}
\end{equation}
 Here, $\partial_t =\frac{\partial}{\partial t}$, and $\nabla_\sV$ is the gradient operator w.r.t. the space
vector $\sV\in\Rset^3$, while $\hbar$ is Planck's constant divided by $2\pi$.
 In \cite{ErwinWMa} Schr\"odinger discussed only the stationary version of (\ref{eq:ERWINeqnMatterWaveBOhydrogen}), whose
solutions $\psi_{n,\ell,m}(\sV)$ map into solutions of (\ref{eq:ERWINeqnMatterWaveBOhydrogen}) via
$\Psi_{n,\ell,m}(t,\sV) = e^{-i E_n t/\hbar}\psi_{n,\ell,m}(\sV)$, with 
$\psi_{n,\ell,m}(\sV) = R_{n,\ell}(r)Y_\ell^{m}(\vartheta,\varphi)$ (see the Appendix), for
$n\in\Nset$ and $\ell\in\{0,...,n-1\}$ and $m\in\{-\ell,...,0,...,\ell\}$, and where
$E_n = E^{\mbox{\textrm{\tiny{Bohr}}}}_n$ are the familiar Bohr energies of hydrogen (in Born--Oppenheimer approximation),
\begin{equation}
E^{\mbox{\textrm{\tiny{Bohr}}}}_n 
= - \tfrac12 \tfrac{e^4\mEL}{\hbar^2n^2};
\label{eq:EperBOHR}
\end{equation}
see \cite{BohrHatomA}.
 The Bohr spectrum meant that Schr\"odinger was onto something.
 But, of course, it had been obtained previously also by de Broglie; 
more importantly, it had been obtained by Pauli \cite{PauliH} who solved Heisenberg's matrix formulation of the hydrogen problem.
 However, thanks to the linearity of Schr\"odinger's equation (\ref{eq:ERWINeqnMatterWaveBOhydrogen}) the general bound state solution 
of (\ref{eq:ERWINeqnMatterWaveBOhydrogen}) is readily obtained as
\begin{equation}
\Psi(t,\sV) = \sum_{n\in\Nset} e^{-i E_n t/\hbar}\sum_{\ell=0}^{n-1}\sum_{m=-\ell}^\ell c^{}_{n,\ell,m}\psi^{}_{n,\ell,m}(\sV),
\label{eq:PSIboundGENERAL}
\end{equation}
and with this Schr\"odinger was in the position to obtain further significant results.

\noindent
\textbf{Remark}:
\emph{Even though the dynamical equation \eqref{eq:ERWINeqnMatterWaveBOhydrogen} appears only in the fourth communication of
 Schr\"odinger's series ``Quantisierung als Eigenwertproblem,'' see Eq.($4^{\prime\prime}$) in} \cite{ErwinWMe}, \emph{an equivalent version of
formula \eqref{eq:PSIboundGENERAL} does appear in} \cite{ErwinWMc} \emph{as  Eq.(35); the ensuing Eq.(36) there makes it plain, 
though, that Schr\"odinger at that time must
 have been considering the relativistic (subsequently so-called) Klein--Gordon equation, which also appears as Eq.(36) in} \cite{ErwinWMe}.

 As is well-known, equation \eqref{eq:ERWINeqnMatterWaveBOhydrogen} implies the conservation of the $L^2$ norm of $\Psi$.
 With $\Im\ $ meaning \emph{imaginary part} and ${}^*$ meaning \emph{complex conjugate},
Schr\"odinger showed in \cite{ErwinWMd} that  $\varrho(t,\sV)\! := \Psi^*(t,\sV) \Psi(t,\sV)$ and 
$\JV(t,\sV) :=\frac{\hbar}{\mEL} \Im \left(\Psi^* \nabla_\sV \Psi\right)(t,\sV)$
satisfy 
\begin{alignat}{1}
 \partial_t{\varrho(t,\sV)}  + \nabla_\sV\cdot\JV(t,\sV) = \label{eq:probCONSERVATIONs} 0,
\end{alignat}
and this continuity equation implies that $\int_{\Rset^3}\varrho(t,\sV)\drm^3{s}$ is conserved if it is finite initially.
 Inserting the general bound state solution (\ref{eq:PSIboundGENERAL}) into 
the bilinear formulas for $\varrho$ and $\JV$ revealed that they are sums of terms which oscillate harmonically with the 
Bohr angular frequencies $\omega_{n,n'} = \frac1\hbar (E_{n'}-E_n)$ for hydrogen.

 This finding must have suggested to Schr\"odinger that the electron charge density at the space point $\sV$ at time $t$ is 
$\rho_{\mathrm{el}}(t,\sV) = -e\Psi^*(t,\sV) \Psi(t,\sV)$, and that
the electron's electric current  vector density $\jV_{\mathrm{el}}(t,\sV)
= -e \frac{\hbar}{\mEL} \Im \left(\Psi^*(t,\sV) \nabla_\sV \Psi(t,\sV)\right)$.
 By (\ref{eq:probCONSERVATIONs}) this identification satisfies the electron charge conservation, viz.
\begin{alignat}{1}
 \partial_t{\rho_{\mathrm{el}}(t,\sV)}  + \nabla_\sV\cdot\jV_{\mathrm{el}}(t,\sV) = \label{eq:chargeCONSERVATION} 0,
\end{alignat}
which we should have. 
 Since the charge density of an electron, $\rho_{\mathrm{el}}$, has to integrate to $-e$, this 
requires the normalization $\int_{\Rset^3}\varrho(t,\sV)\drm^3{s}= 1$.
 And since $\rho_{\mathrm{el}}(t,\sV)$ and $\jV_{\mathrm{el}}(t,\sV)$ are sums of terms which oscillate harmonically with the 
Bohr angular frequencies $\omega_{n,n'} = \frac1\hbar (E_{n'}-E_n)$ for hydrogen, using these expressions for
the charge and current densities as source terms in the inhomogeneous Maxwell--Lorentz equations for the electromagnetic fields of the electron,
\begin{alignat}{1}
 - \partial_t{\EV_{\mathrm{el}}(t,\sV)} + c\nabla\times\BV_{\mathrm{el}}(t,\sV)  &= \label{eq:MdotE} 4\pi \jV_{\mathrm{el}}(t,\sV),\\
   \nabla\cdot\EV_{\mathrm{el}}(t,\sV)  &= \label{eq:MdivE}  4\pi \rho_{\mathrm{el}}(t,\sV)\, ,
\end{alignat}
coupled with the homogeneous Maxwell equations
\begin{alignat}{1}
 \partial_t{\BV_{\mathrm{el}}(t,\sV)} + c \nabla\times\EV_{\mathrm{el}}(t,\sV)  &= \label{eq:MdotB}   \nullV\, ,  \\ 
    \nabla\cdot \BV_{\mathrm{el}}(t,\sV) &= \label{eq:MdivB} 0\, , 
\end{alignat}
the electric field $\EV_{\mathrm{el}}(t,\sV)$ and the magnetic induction field $\BV_{\mathrm{el}}(t,\sV)$
which solve this system of Maxwell--Lorentz equations are also sums of fields which oscillate with the same Bohr hydrogen frequencies,
plus an arbitrary vacuum field solution.
 This is a striking result that Bohr --- not in possession of a dynamical theory --- could only obtain from his hydrogen energies
$E^{\mbox{\textrm{\tiny{Bohr}}}}_n$ by invoking Planck's postulate $h\nu = \triangle E$ for the emitted / absorbed electromagnetic
radiation through matter.

 Not all is well, though.
 The problem is that this calculation says that the hydrogen atom is oscillating forever with the 
superposition of its eigenmodes, and likewise the electromagnetic radiation is a superposition of incoming and outgoing waves
forever. 
 This is not surprising, though, for the feedback from the Maxwell--Lorentz field equations for $\EV_{\mathrm{el}},\BV_{\mathrm{el}}$ into Schr\"odinger's 
matter-wave equation for $\Psi$ is absent. 

 Schr\"odinger in \cite{ErwinWMe} 
 used minimal coupling to inject $\EV_{\mathrm{el}},\BV_{\mathrm{el}}$ into the matter-wave equation for $\Psi$.
 Thus he introduced the  potentials $(\phi_{\mathrm{el}}(t,\sV),\AV_{\mathrm{el}}(t,\sV))$ of the electromagnetic fields 
$\BV_{\mathrm{el}}(t,\sV)$ and $\EV_{\mathrm{el}}(t,\sV)$, 
which are solutions to the inhomogeneous, linear partial differential equations
\begin{alignat}{1}
 \textstyle -\frac1c\partial_t{\AV_{\mathrm{el}}(t,\sV)}  - \nabla_\sV\phi_{\mathrm{el}}(t,\sV) & = \EV_{\mathrm{el}}(t,\sV), \label{Aevolve}\\
 \textstyle \nabla_\sV\crprd {\AV_{\mathrm{el}}(t,\sV)}  & =  \BV_{\mathrm{el}}(t,\sV).\label{Aconstraint}
\end{alignat}
 Note that these two equations comprise an evolution equation for $\AV_{\mathrm{el}}$, given $\EV_{\mathrm{el}}$ and $\phi_{\mathrm{el}}$, 
plus a constraint equation for $\AV_{\mathrm{el}}$, given $\BV_{\mathrm{el}}$. 
 Another equation is needed, for $\phi_{\mathrm{el}}$.
 A compelling choice from the perspective of relativity is the \emph{Lorenz gauge}
\begin{alignat}{1}\label{LorLorGAUGE}
 \textstyle \frac1c\partial_t{\phi_{\mathrm{el}}(t,\sV)}  + \nabla\cdot\AV_{\mathrm{el}}(t,\sV) & = 0,
\end{alignat}
which is an evolution equation for $\phi_{\mathrm{el}}$.
 Also the Coulomb gauge condition $\nabla_\sV\cdot \AV_{\mathrm{el}} = 0$ is popular, although it is not Lorentz covariant.
 Aside from demanding that all fields decay to zero when $|\sV|\to\infty$ together with their derivatives, we need
initial data for $\AV_{\mathrm{el}}$ and for $\phi_{\mathrm{el}}$, but let's not digress.

 The minimal-coupling substitution for energy $E\mapsto E +e\phi_{\mathrm{el}}$ and momentum $\pV\mapsto\pV +\frac1c e\AV_{\mathrm{el}}$,
known from classical mechanics of the motion of a \emph{test electron}, a point particle with charge $-e$ in \emph{given} 
electromagnetic fields, changes (\ref{eq:ERWINeqnMatterWaveBOhydrogen}) into
\begin{equation}
\left(i \hbar \partial_t + e \phi_{\mathrm{el}}(t,\sV)\right)\Psi(t,\sV)
= \tfrac{1}{2\mEL}\left(-i\hbar \nabla_\sV + \textstyle\frac{e}{c}\AV_{\mathrm{el}}(t,\sV)\right)^2\Psi(t,\sV) 
- \tfrac{e^2}{|\sV|}\Psi(t,\sV).
 \label{eq:ERWINeqnAphi}
\end{equation}
 By inserting electromagnetic potential fields
with simple periodic time dependence $\sin(\omega_{n,n'}t)$ Schr\"odinger was able to estimate  that 
indeed the solution of (\ref{eq:ERWINeqnAphi}) will show a resonance and consist predominantly of a superposition of
the eigenmodes for $E_n$ and $E_{n'}$. 
 In 1927 Dirac \cite{GoldenRule} computed perturbatively that the solution of (\ref{eq:ERWINeqnAphi}) will transit
from an initially $n$-th eigenstate to the $n'$-th eigenstate of (\ref{eq:ERWINeqnMatterWaveBOhydrogen}), or
to the continuum. 
  His formula for the transition probability was later sanctioned ``Golden Rule'' by Fermi.
 Note though that in this calculation the external electromagnetic potential fields oscillate forever, \emph{by assumption}. 

 A definitive assessment of Schr\"odinger's matter-wave can
only be obtained by non-perturbatively studying the self-consistent model, nowadays known as 
the Schr\"odinger--Maxwell system of equations, consisting of Schr\"odinger's equation
(\ref{eq:ERWINeqnAphi}) and the electromagnetic potential equations (\ref{Aevolve}), (\ref{Aconstraint}), (\ref{LorLorGAUGE}), 
plus Maxwell's equations (\ref{eq:MdotE})--(\ref{eq:MdivB}), now with 
$\jV_{\mathrm{el}}(t,\sV) = -e \Im \bigl(\Psi^*[ \frac{\hbar}{\mEL} \nabla + i\frac{e}{\mEL c}\AV_{\mathrm{el}}]\Psi\bigr)(t,\sV)$ 
as electron current vector density pertinent to minimal coupling between $\Psi$ and $(\phi_{\mathrm{el}},\AV_{\mathrm{el}})$. 
 The Schr\"odinger--Maxwell system is expected to yield emission of an electromagnetic wave at the expense of the electron-proton 
system's energy, and in the process resulting in $\Psi$ to settle down in a ground state.
 This expectation is based on the hyperbolicity of the Maxwell field equations in concert with the 
conservation of the energy functional 
\begin{eqnarray} \label{eq:SMenergyFctl}
\hspace{-3pt}\Efrak(\Psi,\EV_{\mathrm{el}},\BV_{\mathrm{el}}) 
\!\!& := &\!\!
 \frac{1}{2\mEL}
\displaystyle\int_{\Rset^3}\!\! \abs{\left(-i\hbar \nabla_\sV + \textstyle\frac{e}{c}\AV_{\mathrm{el}}(t,\sV)\right)\Psi(t,\sV)}^2 d^3\!s  
\\ 
\notag
&& \displaystyle
-e^2\int_{\Rset^3}\!\! \frac{|{\Psi}|^2(t,\sV) }{|\sV|}
d^3\!s  
+ e^2 \frac12\int_{\Rset^3}\! \int_{\Rset^3} \!\!\!\! \frac{|\Psi|^2(t,\sV)|\Psi|^2(t,\sV')}{\abs{\sV-\sV'}}d^3\!s d^3\!s'
\\
\notag
&& \displaystyle
+  \frac{1}{8\pi}\int_{\Rset^3}\!\left(\abs{\EV_{\mathrm{el}}^{\mbox{\tiny{rad}}}(t,\sV)}^2 + 
\abs{\BV_{\mathrm{el}}^{\mbox{\tiny{rad}}}(t,\sV)}^2 \right)d^3\!s,
\end{eqnarray}
where ${}^{\mbox{\tiny{rad}}}$ refers to the fields without the Coulomb field of the nucleus.
 This functional can easily be shown to be bounded below.
 Emission of electromagnetic radiation by a localized oscillating $\rho_{\mathrm{el}}(t,\sV)$ and $\jV_{\mathrm{el}}(t,\sV)$
will increase the electromagnetic field energy integral at the expense of the $\Psi$-energy, which is bounded below, and so
the emission process should cease eventually.
 However, while plausible, to the best of our knowledge this has not yet been established rigorously.
 For a mathematical review of this model (though in the Coulomb gauge), see \cite{KomechLECT}.

 Even if the just described scenario can be established rigorously, which would be a fine result,
this model of hydrogen coupled with the classical electromagnetic Maxwell fields 
does not seem to produce quantitatively acceptable results.

 For instance, the energy ground state in this model corresponds to minimizing $\Efrak(\Psi,\EV_{\mathrm{el}},\BV_{\mathrm{el}})$
for $\AV_{\mathrm{el}}\equiv\nullV$ and vanishing electromagnetic radiation fields, and $\Psi(t,\sV) = e^{-i Et/\hbar}\psi(\sV)$. 
 We set $\Efrak(e^{-iEt/\hbar}\psi,\nullV,\nullV)=:\Ffrak(\psi)$, thus
\begin{eqnarray} \label{eq:psiFctl}
\hspace{-3pt}\Ffrak(\psi) := \!\!
\frac{\hbar^2}{2\mEL}
\int_{\Rset^3}\!\! \abs{\nabla \psi}^2(\sV) d^3\!s  
-e^2\!\!\!\int_{\Rset^3}\!\! \frac{|\psi|^2(\sV) }{\abs{\sV}}d^3\!s  
+
e^2\frac{1}{2}\int_{\Rset^3}\! \int_{\Rset^3} \!\!\!\!
\frac{|\psi|^2(\sV)|\psi|^2(\sV')}{\abs{\sV-\sV'}}d^3\!s d^3\!s'\!.\
\end{eqnarray}
 As shown in  \cite{BenguriaBrezisLieb}, \cite{BenguriaLieb}, 
the functional $\Ffrak$ has a necessarily spherically symmetric minimizer $\psi_1$ on the Sobolev space $H^1(\Rset^3)$,
satisfying $\int_{\Rset^3} |\psi|^2(\sV) d^3\!s =1$; spherical symmetry follows from uniqueness by 
convexity. 
 The same spherical-symmetry-through-uniqueness-by-convexity argument for any putative minizer was subsequently 
rediscovered and emphasized in \cite{KK}.
 All these authors noticed that, while $\psi\mapsto \Ffrak(\psi)$ is neither convex nor concave, the map 
$|\psi|^2\mapsto \Ffrak(\psi)$ is convex, hence any minimizer must be unique, and uniqueness implies its spherical symmetry.

 By the virial theorem for systems with Coulomb interactions one has 
$\Ffrak(\psi_1)=-\tfrac{\hbar^2}{2\mEL} \int_{\Rset^3}\!\! |{\nabla \psi}|^2(\sV) d^3\!s < 0$, as expected.

 The minimizer $\psi_1$ satisfies the pertinent Euler--Lagrange equation \cite{BenguriaLieb}, which 
is a special case of (\ref{eq:ERWINeqnAphi}), with
$\Psi(t,\sV)=e^{-iE_gt/\hbar}\psi(\sV)$ and $\phi_{\mathrm{el}}$ the electrostatic Coulomb potential of $\rho_{\mathrm{el}}$, viz.
\begin{equation}
 -\frac{\hbar^2}{2\mEL}\Delta_\sV\psi(\sV) - e^2\frac{1}{|\sV|}\psi(\sV) +
e^2 {\displaystyle\int_{\Rset^3}} \frac{1}{|\sV-\sV'|}|\psi|^2(\sV')d^3s'\psi(\sV) = E_g \psi(\sV);
 \label{eq:ERWINeqnEphiEXPLICIT}
\end{equation}
the eigenvalue $E_g$ is also the Lagrange multiplier for the  constraint $\|\psi\|_{L^2}=1$.
 However, in this \emph{nonlinear eigenvalue problem} the eigenvalue $E_g$ does not
coincide with the minimum of $\Ffrak(\psi)$, which is easily seen as follows.

Setting $\psi=\psi_1$ in \eqref{eq:ERWINeqnEphiEXPLICIT}, then 
multiplying \eqref{eq:ERWINeqnEphiEXPLICIT} by $\psi_1$ and integrating over $\Rset^3$, and recalling
the normalization of $\psi_1$, for the ground state energy $E_g$ one obtains 
 \begin{eqnarray} \label{eq:energy}
\hspace{-10pt} 
E_g =\frac{\hbar^2}{2\mEL} \int_{\Rset^3}\!\! \abs{\nabla \psi_1}^2(\sV) d^3\!s  
-
e^2\int_{\Rset^3}\!\! \frac{|{\psi_1}|^2(\sV)}{\abs{\sV}}d^3\!s  
+
e^2 \! \int_{\Rset^3}\!\! \int_{\Rset^3} \!\!\!\!
\frac{|\psi_1|^2(\sV)|\psi_1|^2(\sV')}{\abs{\sV-\sV'}}d^3\!s d^3\!s',\
\end{eqnarray}
i.e.
 \begin{eqnarray} \label{eq:energyB}
\hspace{-10pt} 
E_g =
\Ffrak(\psi_1) 
+
e^2 \frac12 \int_{\Rset^3}\!\! \int_{\Rset^3} \!\!\!\!
\frac{|\psi_1|^2(\sV)|\psi_1|^2(\sV')}{\abs{\sV-\sV'}}d^3\!s d^3\!s';\
\end{eqnarray}
cf. Eq.(6) in \cite{BazleySeydel}.
 And so we have
 \begin{eqnarray} \label{eq:EgBIGGERthanFCTLmin}
E_g > \Ffrak(\psi_1).
\end{eqnarray}
\newpage

By inequality \eqref{eq:EgBIGGERthanFCTLmin} the energetic significance of $E_g$ is obscure in this theory; 
recall that $\Ffrak(\psi_1)$ is the ground state energy of the conserved energy functional $\Efrak$, while $E_g$ is
the lowest eigenvalue of the nonlinear eigenvalue problem.

 Moreover, for all non-vanishing $\psi$, we obviously also have
\begin{eqnarray} \label{eq:psiFctlBIGGERthan}
\hspace{-3pt}\Ffrak(\psi) > 
\frac{\hbar^2}{2\mEL}
\int_{\Rset^3}\!\! \abs{\nabla \psi}^2(\sV) d^3\!s  
-\int_{\Rset^3}\!\! \frac{e^2}{\abs{\sV}}\abs{ \psi}^2(\sV) d^3\!s , 
\end{eqnarray}
which is the usual energy functional for the textbook QM Schr\"odinger equation of the hydrogen eigenvalue problem
in Born--Oppenheimer approximation, and so 
 \begin{eqnarray} \label{eq:FCTLminBIGGERthanEminBOHR}
\Ffrak(\psi_1)
 > E^{\mbox{\textrm{\tiny{Bohr}}}}_1.
\end{eqnarray}
 Since the conservation law for the energy functional makes it plain that in this theory the
ionization energy has to be identified with $|\Ffrak(\psi_1)|$, and since by \eqref{eq:FCTLminBIGGERthanEminBOHR}
this is smaller than $|E^{\mbox{\tiny{Bohr}}}_1|$, while $|E^{\mbox{\tiny{Bohr}}}_1|$ agrees quite well with the empirical ionization energy,
it follows that this theory gives an incorrect ionization energy for hydrogen.

 Of course, the inequality $\Ffrak(\psi_1) > E^{\mbox{\tiny{Bohr}}}_1$ is only qualitative; in principle it leaves
 the possibility that the quantitative discrepancy could be unnoticeable. 
 To rigorously eliminate the possibility of a quantitatively insignificant shift, one needs an explicit lower bound on $\Ffrak(\psi_1)$
which is strictly bigger than $E^{\mbox{\tiny{Bohr}}}_1$ by a significant amount. 
 This should be possible to accomplish with the technique of \cite{BazleySeydel}, which relies on a clever choice of
a trial function; our first attempt has not produced the desired bound, and so we defer producing one to some later time. 
 However, if we do not insist on a rigorous proof and are willing to accept numerical results, a verdict exists!

 Indeed, the self-consistent Schr\"odinger matter-wave ground state problem for hydrogen (\ref{eq:ERWINeqnEphiEXPLICIT}) is 
mathematically identical with the Hartree \cite{Hartree} and with the
Hartree--Fock (HF) approximation to the traditional Schr\"odinger ground state of the hydrogen 
anion $H^-$, a.k.a. hydride in the chemical literature \cite{CBSM}, with the Schr\"odinger matter-wave ground state energy $E_g$ 
equal to one-half of the HF ground state energy of hydride ($E_{H^-}^{\mbox{\tiny{HF}}}$, say), so $E_g = \frac12 E_{H^-}^{\mbox{\tiny{HF}}}$. 
 The HF ground state energy of hydride has been computed numerically to an astonishing precision (see references 60-66 in \cite{CBSM};
see also \cite{Rau}), which (retaining just a few decimal places precision) translates
into $E_g \approx 0.488 E^{\mbox{\textrm{\tiny{Bohr}}}}_1$ for the ground state energy of (\ref{eq:ERWINeqnEphiEXPLICIT}).
 This is so far off of the empirical data that I don't see how the model could possibly account for the electromagnetic
radiation energy released when an electron and a proton recombine into a hydrogen atom.

 The upshot of the above discussion, cautiously expressed, is: 
\begin{quote}
\emph{Schr\"odinger's ``matter-wave'' interpretation of his wave function leads to a non-linear theory of hydrogen
that does not seem to be physically viable.}
\end{quote}

\noindent
\textbf{Remark}:
\emph{We have noted that the ground state problem for hydrogen in Schr\"odinger's matter-wave model is mathematically
identical (up to rescaling of the eigenvalue by a factor 2) with the Hartree(--Fock) approximation to 
the conventional (textbook) QM Schr\"odinger equation (see (\ref{eq:ERWINeqnZ}), (\ref{eq:HAMzATOM}) below) 
for the ground state problem of hydride.
 Incidentally, since 
$E_{H^-}^{\mbox{\tiny{HF}}}\approx 0.976 E^{\mbox{\textrm{\tiny{Bohr}}}}_1$ the Hartree--Fock approximation to the 
conventional Schr\"odinger ground state of hydride fails to predict the existence of a bound state for this two-electron
problem, which has exactly one bound state} \cite{Hill}.
 \emph{Hydride is the lightest two-electron ion in the isoelectronic family of helium, and the 
Hartree(--Fock) functional and Hartree(--Fock) equation for the Hartree(--Fock) approximation to the ground state of this
family is obtained by replacing $-e^2$ by $-Ze^2$ in the attractive potential of the nucleus in the
hydrogen problem \eqref{eq:psiFctl} and \eqref{eq:ERWINeqnEphiEXPLICIT}.
 A unique positive solution of the Hartree equation for helium satisfying  $\|\psi\|_{L^2}=1$ was first shown to exist in} \cite{Reeken}, 
\emph{and that it coincides with the minimizer of the Hartree functional was shown in} \cite{BazleySeydel}. 
 \emph{The Hartree--Fock model is a widely used mathematical approximation to the textbook Schr\"odinger equation for many-electron
problems, successfully so when $Z>1$, but
 that should not be misconstrued as supporting Schr\"odinger's matter-wave ontology, his own hopes to the contrary,
expressed in 1927} (see p.472 of \cite{Solvay}), \emph{notwithstanding.} 
\medskip

 Schr\"odinger himself eventually realized that a matter-wave ontology was not viable, but before he came to this 
conclusion he generalized his wave equation \eqref{eq:ERWINeqnMatterWaveBOhydrogen} to the many body problem and 
made further important discoveries. 
 In the next subsection we briefly summarize Schr\"odinger's attempt at a matter-wave interpretation of multi-electron atoms, 
again in Born--Oppenheimer approximation, for it helps to appreciate the subsequent change of perspective offered by Born.
\vspace{-.2truecm}

\subsection{Helium, Lithium, Beryllium, etc. as per Schr\"odinger}\vspace{-.1truecm}

 We begin with an English translation of Schr\"odinger's own words:
\begin{quote}
``We have repeatedly called attention to the fact that the $\Psi$-function itself cannot and may not be interpreted directly in terms of 
three-dimensional space --- however much the one-electron problem tends to mislead us on this point --- because it is in
general a function in configuration space, not real space.'' (Quoted from \cite{WaveMechanics}, p.120/1.)
\end{quote}

\noindent
 In 1926, when Schr\"odinger arrived at his equation for 
an $N$ particle system of many nuclei and electrons, which may form any ordinary piece of everyday matter, he
obtained a $t$-dependend $\Psi$ function on $N$-particle {configuration space}.
 So, at time $t$, $\Psi$ is a function of a high-dimensional vector variable $\qVv=(\qV_1,...,\qV_N)\in \Rset^{3N}$ 
formed by the generic position variables of the $N$ point particles. 
 Restricting ourselves to an $N$-electron atom or ion with a nucleus of charge $Ze$ fixed at the origin, with $Z\in\Nset$ 
(though $Z\leq 92$ in nature),  Schr\"odinger's $N$-body generalization of (\ref{eq:ERWINeqnMatterWaveBOhydrogen}) reads
\begin{equation}
i \hbar \partial_t\Psi(t,\qVv) = H \Psi(t,\qVv)
 \label{eq:ERWINeqnZ}
\end{equation}
\vspace{-.5truecm}

\noindent
with \vspace{-.3truecm}
\begin{equation}
H = \sum_{k=1}^N \frac{1}{2\mEL} \big(- i\hbar \nabla_{\qV_k}\big)^2 
 - \sum_{k=1}^N \frac{Ze^2}{|\qV_k|} +
	\sum\sum_{\hskip-.7truecm  1 \leq j < k \leq N} \frac{e^2}{|\qV_j-\qV_k|}.
 \label{eq:HAMzATOM}
\end{equation}
 It is known that $H$ has infinitely many discrete eigenvalues $E_1<E_2<\cdots< 0$ below the essential
spectrum $\sigma_{\mbox{\tiny{ess}}}$ (which itself has a strictly negative minimum), each one at most finitely 
degenerate, whose eigenfunctions represent bound states \cite{ReedSimonBOOKiv}.
 Let $\dV(n)$, for $n\in\Nset$ denote a set of labels 
that label the eigenstates with same eigenvalue $E_n$, 
i.e. write $H\psi^{}_{n,\dV(n)}(\qVv) = E_n\psi^{}_{n,\dV(n)}(\qVv)$.
 Then the general bound state solution is given by
\begin{equation}
\Psi(t,\qVv) = \sum_{n\in\Nset} e^{-i E_n t/\hbar}\sum_{\dV(n)} c^{}_{n,\dV}\psi^{}_{n,\dV(n)}(\qVv).
\label{eq:PSIboundGENERALz}
\end{equation}

 The same year, Sommerfeld's (former) students Uns\"old and Heisenberg made
quantitatively promising calculations for the $N=2$ helium problem ($Z=2$), using first-order perturbation techniques
with hydrogenic wave functions.
 Hylleraas a few years later obtained more accurate spectral results for helium by a variational method that included
the distance between the electron as a variable for the wave function, and also the continuous part of the hydrogenic 
spectrum; see the charming article \cite{Hylleraas} for Hylleraas' own reminiscences. 
 
 Soon after, many other empirically known facts about matter were quite accurately extracted from Schr\"odinger's
many body equation (\ref{eq:ERWINeqnZ}) with Hamiltonian $H$ given by (\ref{eq:HAMzATOM}), most notably the periodic table
of the chemical elements by using Slater's determinantal wave functions built with hydrogenic eigenfunctions plus an extra bit
(explained with spin, subsequently); see \cite{GeroETal} for a more recent rigorous contribution.
 There was no doubt that (\ref{eq:ERWINeqnZ}) with Hamiltonian given by (\ref{eq:HAMzATOM}) is an important
equation for computing answers to questions about large atoms / ions, the atomic spectra among them.

 While the hydrogen $\Psi(t,\qV)$ could be confused with a matter-wave $\Psi(t,\sV)$ on physical space and time, 
the fact that the $N$-body $\Psi(t,\qV_1,...,\qV_N)$ lives on $N$-particle configuration space $\Rset^{3N}$ (at time $\Rset$) 
should have made it plain for Schr\"odinger that it was absurd to think that
$\Psi$ was associated with a fundamental matter-wave ontology on physical space (and time). 
 Yet for a while he continued to maintain that there are no particles
at the fundamental subatomic level and instead proposed that his $\Psi$ does supply a matter-wave ontology, 
not directly so but indirectly, as follows.

 Schr\"odinger showed in \cite{ErwinWMd} that the function
$\varrho(t,\qVv) := \Psi^*(t,\qVv) \Psi(t,\qVv)$ on $\Rset^{3N}$ and the $3N$-dimensional vector 
function $\JVv(t,\qVv):=\frac{\hbar}{\mEL} \Im \left(\Psi^*(t,\qVv) \nabla_{\qVv} \Psi(t,\qVv)\right)$ on $\Rset^{3N}$, 
jointly satisfy the continuity equation
\begin{alignat}{1}
 \partial_t{\varrho(t,\qVv)}  + \nabla\bcdot\JVv(t,\qVv) = \label{eq:probCONSERVATIONqN} 0;
\end{alignat}
here, $\nabla\bcdot\ $ is a $3N$-dimensional divergence operation; i.e. it acts on $\Rset^{3N}$-dimensional vectors.
 Equation (\ref{eq:probCONSERVATIONqN}) has the important implication that
$\int_{\Rset^{3N}}|\Psi|^2(t,\qV_1,...,\qV_N) \drm^{3N}q$ is constant in time if it is finite at $t=0$.

 Schr\"odinger \cite{ErwinWMd} now proposed that $\Psi$ yields a
matter-wave ontology in physical space (and time) through a many-electron charge density
\begin{alignat}{1}
 \rho_{\mathrm{el}} (t,\sV) := - e \sum_n \int_{\Rset^{3(N-1)}}\varrho(t,\qV_1,...,\sV,...,\qV_N) \drm^{3(N-1)}q,
\label{eq:chargeRHO}
\end{alignat}
where $\sV$ at r.h.s.(\ref{eq:chargeRHO}) is in the $n$-th position slot. 
 For (\ref{eq:chargeRHO}) to yield the total charge $\int_{\Rset^3} \rho_{\mathrm{el}} (t,\sV)\drm^3s = -Ne$,
Schr\"odinger stipulated that $\int_{\Rset^{3N}}|\Psi|^2(t,\qV_1,...,\qV_N) \drm^{3N}q~=~1$. 
 Similarly, in \cite{ErwinWMd} he defined the many-electron current vector density as
\begin{alignat}{1}
 \jV_{\mathrm{el}} (t,\sV) : = -e \sum_n \int_{\Rset^{3(N-1)}}\jV_n(t,\qV_1,...,\sV,...,\qV_N) \drm^{3(N-1)}q,
\label{eq:chargeJ}
\end{alignat}
with 
\begin{alignat}{1}\label{eq:jOFsDEF}
\jV_n(t,\qV_1,...,\sV,...,\qV_N) :=
\tfrac{\hbar}{\mEL} \Im\! \left(\Psi^*(t,\qV_1,...,\sV,...\qV_N) \nabla_{\sV} \Psi(t,\qV_1,...,\sV,...\qV_N)\right) 
\end{alignat}
and $\sV$ in the $n$-th position slot. 
  He noted that $\rho_{\mathrm{el}}$ and $\jV_{\mathrm{el}}$ jointly satisfy the continuity equation (\ref{eq:chargeCONSERVATION}).
 When evaluated with the general bound state solution (\ref{eq:PSIboundGENERALz}), again one finds harmonically
oscillatory terms with Bohr-type frequencies $\propto (E_n-E_{n'})$.
 Inserted as source terms for the inhomogeneous Maxwell equations (\ref{eq:MdotE}), (\ref{eq:MdivE}), coupled with 
the homogeneous equations (\ref{eq:MdotB}), (\ref{eq:MdivB}), one finds fields which oscillate with these Bohr-type 
frequencies, extending Schr\"odinger's striking hydrogen result to many-electron atoms.

 Yet again, thinking of $\EV_{\mathrm{el}}$ and $\BV_{\mathrm{el}}$ as actual fields produced by all the electrons, their
minimally coupled feedback into the Schr\"odinger equation leads to a contradiction with the just mentioned
striking results obtained from (\ref{eq:ERWINeqnZ}) with Hamiltonian $H$ given by (\ref{eq:HAMzATOM}).

 Schr\"odinger himself must have had conflicting thoughts in his mind already in 1926, for in \cite{ErwinWMe} he writes that 
``$\Psi\overline{\Psi}$ is a kind of \emph{weight-function} in the system's configuration space. The \emph{wave-mechanical}
configuration of the system is a \emph{superposition} of many, strictly speaking of \emph{all}, point-mechanical configurations
kinematically possible.''
  This rings more akin to a many-worlds type theory; for more on this, see \cite{ValiaETal}.
 As already mentioned, Schr\"odinger would eventually abandon his matter-wave ontology, and also his ``many-worlds type'' thoughts.
 Yet Schr\"odinger would continue to reject a particle ontology of matter. 

\section{
Born -- de Broglie -- Bohm-inspired 
approach to \\ a quantum mechanics of radiating atoms (etc.)}

 We begin with an English translation of de Broglie's own words:
\begin{quote}
``We cannot recall here the successes obtained by this method (papers by Messrs Schr\"odinger, ..., etc.), but we must insist on 
the difficulties of a conceptual type that it raises. Indeed let us consider, for simplicity, a system of $N$ material points each 
possessing three degrees of freedom. The configuration space is in an essential way formed by means of the coordinates of the points, 
and yet Mr. Schr\"odinger assumes that in atomic systems material points no longer have a clearly defined position. It seems a little 
paradoxical to construct a configuration space with points that do not exist.''
 (Quoted on p.379 of the arXiv version in \cite{Solvay}.)
\end{quote}

 De Broglie and Born had no problems with a particle ontology, and thought of $\Psi$ not ontologically
but nomologically. 
 Max Born seems to have been the first to interpret $\Psi$  as a guiding field for the particles, followed suit by de Broglie whose
insights were later rediscovered and advanced by Bohm;
note though that the existence of a guiding field for electrons was first postulated by de Broglie two or three years
earlier, inspired by Einstein's ideas \cite{EinsteinPHOTON} that photons are guided in their motion by ``ghost fields.''
 Born also re-interpreted $|\Psi|^2$ not as a weight-function in Schr\"odinger's sense but as  a probability density. 
 This probability interpretation became part of the standard textbook narrative which, unfortunately, usually 
leaves out the guiding field interpretation, and without it the practical success of using $|\Psi|^2$ as a probability density
becomes mysterious, rather incomprehensible. 
 We will retain the guiding field interpretation.

\subsection{Particles and guiding fields}

\subsubsection{Born's probability rule}

 If one has $N$ electrons with generic positions $\qV_n\in \Rset^3$, then Schr\"odinger's 
many-electron ``charge density function'' (\ref{eq:chargeRHO}) can be rewritten as 
\begin{alignat}{1}
 \rho_{\mathrm{el}} (t,\sV) =\int_{\Rset^{3N}} \Big({\textstyle\sum\limits_n} - e \delta_{\qV_n}(\sV) \Big) \varrho(t,\qVv) \drm^{3N}q,
\label{eq:chargeRHOexpect}
\end{alignat}
and since $\varrho \geq 0$ integrates to 1 (as Schr\"odinger had to stipulate), this 
looks like the expected value of the \emph{generic empirical charge density} $\sum_n - e \delta_{\qV_n}(\sV)$
of the electrons (a singular measure, to be petty), computed w.r.t. a probability measure $\varrho(t,\qVv) \drm^{3N}\!q$.
 So Born in \cite{BornsPSISQUAREpapersB} proposed that $|\Psi|^2(t,\qVv)$, normalized to integrate to 1 as Schr\"odinger had stipulated,
is a \emph{probability density} for the first particle being at $\qV_1$, the second one at $\qV_2$, and so on.\footnote{This soon 
  was rendered incomprehensible by positivistic interpretations of Heisenberg's uncertainty relations. 
  Rather than probability density for a particle being at $\qV$, it was insisted that $|\Psi|^2$ is the probability 
  density of finding a particle at $\qV$ in a measurement (which is still comprehensible), but that it would not make sense 
 to say a particle is at $\qV$ without a measurement (which is not).} 


 To see that also Schr\"odinger's 
\begin{alignat}{1}
\hspace{-1truecm}
 \jV_{\mathrm{el}} (t,\sV) =
{\textstyle\sum\limits_n}\! \int_{\Rset^{3(N-1)}}\!\!\!\!\! -e
\tfrac{\hbar}{\mEL} \Im\! \left(\Psi^*(t,\qV_1,...,\sV,...\qV_N) \nabla_{\sV} \Psi(t,\qV_1,...,\sV,...\qV_N)\right) \drm^{3(N-1)}q
\!\!\!\!\!\!
\label{eq:chargeJexpect}
\end{alignat}
is an expected value w.r.t. $|\Psi|^2$, we recall that the polar representation 
$\Psi = |\Psi|e^{i\Phi}$ yields $\Im\!\left( \Psi^* \nabla \Psi\right) = |\Psi|^2\nabla\Phi$.
 Thus for each $n$ we have
\begin{alignat}{1}\label{eq:JisRHOgradPHI}
\hspace{-.6truecm}
\Im\!\left( \Psi^*(t,\qVv) \nabla_{\qV_n}\Psi(t,\qVv)\right) = |\Psi|^2(t,\qVv)\nabla_{\qV_n}\Phi(t,\qVv),
\end{alignat}
and so, with $\cI_n: T_{\qV_n}\Rset^3\to T_\sV\Rset^3$ denoting the natural injection map from the $n$-th component of 
tangent space of configuration space at $\qVv=(\qV_1,...,\qV_N)$ into the tangent space of physical space at $\sV$, 
(\ref{eq:chargeJexpect}) becomes 
\begin{alignat}{1}
 \jV_{\mathrm{el}} (t,\sV)\! =\! \int_{\Rset^{3N}}\!\! \Big({\textstyle\sum\limits_n} - e
\tfrac{\hbar}{\mEL} \cI_n\big(\nabla_{\qV_n} \Phi(t,\qVv)\big)\delta_{\qV_n}(\sV)\Big) |\Psi|^2(t,\qVv) \drm^{3N}q,
\label{eq:chargeJexpectMANIFEST}
\end{alignat}
which is the expected value w.r.t. $|\Psi|^2$ of the electrons' \emph{generic electrical current vector density} 
$\sum_n -e \frac{\hbar }{\mEL}\cI_n\big(\nabla_{\qV_n} \Phi(t,\qVv)\big)\delta_{\qV_n}(\sV)$. 
 We remark that $\Phi$, while part of the polar representation of $\Psi$, \emph{is generally not a function of} $|\Psi|^2$. 

 Note that (\ref{eq:JisRHOgradPHI}) and (\ref{eq:probCONSERVATIONqN}) 
imply that $\frac{\hbar}{\mEL}\nabla_{\qV_n} \Phi(t,\qVv)$ must be interpreted as the $n$-th 
component of a \emph{generic velocity field} on configuration space $\Rset^{3N}\!$.

 Now, probabilities usually reflect our ignorance of something, otherwise we would speak about those things with certainty.
 Our ignorance may simply be due to technical limitations for accessing in-principle-available information, or
it may be a matter of principle, if nature herself throws rocks our way of accessing the information. 
 Either way, why should our ignorance satisfy such an equation like Schr\"odinger's which has the amazing feature of producing
an energy spectrum whose level differences agree to great precision with the empirical frequencies of spectral lines 
registered by chemists and atomic physicists? 
 Max Born, in \cite{BornsPSISQUAREpapersB}, offered the following way out of this dilemma.
\subsubsection{$\Psi$ as a guiding field (Born and de Broglie on the ``same'' page)}
 Like de Broglie a couple years earlier in his doctoral work, Born picked up on Einstein's speculation that photons are
particles which are guided by the electromagnetic field, which becomes a guiding field in Einstein's interpretation.
 Born now proposed that $\Psi$ is a guiding field for the matter particles of atomic physics: the electrons and the nuclei.
 He emphasized that all that was needed was to assume that $\Psi$ guides the particles more likely to where $|\Psi|^2$ is large 
and less likely to where it's small. 
 Born's explanation makes it plain that when $\Psi$ somehow guides the particles in this manner, then
$|\Psi|^2$ only \emph{appears to be} a probability density \emph{for all practical purposes} --- it is \emph{not fundamentally}
a probability density. 

 At the end of \cite{BornsPSISQUAREpapersB} Born wrote that he thought it was unlikely that a detailed description of the 
actual dynamics of positions and momenta (the ``phases'') of the particles was possible, but that Frenkel had pointed out 
to him that it might be possible. 
 In any event, he emphasized that he was convinced that it had to be a non-deterministic law.

 Born's ideas about a non-deterministic guiding role played by $\Psi$ were eventually implemented by Nelson \cite{NelsonBOOKa,NelsonBOOKb}
and further developped by Guerra and his collaborators \cite{GuerraBOOK} and became known as ``Stochastic Mechanics.'' 
 The generic velocity field whose $n$-th component is $\frac{\hbar}{\mEL}\nabla_{\qV_n} \Phi(t,\qVv)$ 
appears as the so-called ``current velocity field'' in Nelson's stochastic mechanics; in addition there is an ``osmotic velocity field.'' 

 In the meantime, at the 1927 Solvay Conference de Broglie (see \cite{deBroglieSOLVAY})
proposed that Schr\"odinger's equation supplied the very equation for the guiding
field whose existence he had postulated in 1923, but could not nail down. 
 Already in his 1924 thesis de Broglie had suggested as guiding equation for the \emph{actual electron positions} (here 
in non-relativistic approximation) precisely the guiding equation of the Hamilton--Jacobi reformulation of Newton's mechanics, viz.
\begin{alignat}{1}\label{dBguidingEQ}
\forall\ n:\ \frac{\drm \qV_n(t)}{\drm t} = \frac{\hbar}{\mEL}\cI_n\Big(\Bigl.\nabla_{\qV_n}\Phi(t,\qV_1,...,\qV_N)\Big)\Bigr|_{\qVv=\qVv(t)},
\end{alignat}
where often one writes $S$ instead of $\hbar \Phi$. 
 De Broglie's point was that the Hamilton--Jacobi partial differential equation for $S$ 
had to be an approximation to the fundamental equation for $\Phi$, but he did not know how it should be formulated. 
 After Schr\"odinger published his papers on ``wave mechanics,'' de Broglie figured that Schr\"odinger's 
continuity equation (\ref{eq:probCONSERVATIONqN}) in concert with the identity (\ref{eq:JisRHOgradPHI}) suggest that 
$\frac{\hbar}{\mEL}\nabla_{\qV_n}\Phi(t,\qVv)$ is the $n$-th three-vector component of a velocity field on configuration
space. 
 Thus de Broglie put one-plus-one together and proposed that if one evaluates this generic velocity field 
at the \emph{actual configuration} $\qVv(t)=(\qV_1(t),...,\qV_N(t))$, one gets the actual velocity of the $n$-th particle, given by
the system of equations (\ref{dBguidingEQ}).
 For $\Phi(t,\qV_1,...,\qV_N) = \sum_n \kV_n\cdot\qV_n$ one recovers the de Broglie relation $\pV_j = \hbar \kV_j$
for the $j$-th particle, where $\pV_j$ has been defined per Newton's formula $\pV_j(t) = m_j\dot\qV(t)$.
 For a discussion of de Broglie's theory, and its reception in 1927, see \cite{Solvay}.

 Facing criticism from Pauli and many others, and encouragement only by a few, most notably by Einstein and Brillouin, 
and furthermore hitting road blocks on his way to finding a deeper justification for his theory through his pursuit of ``the double solution,''  
de Broglie abandoned his proposal 
--- until his theory was rediscovered 25 years after the 1927 Solvay Conference, 
by Bohm \cite{BohmsHIDDENvarPAPERS,BohmsREPLYtoCRITICSa,BohmsREPLYtoCRITICSb}.
 As relayed by Bricmont \cite{BricmontBOOK}, Bohm's work was received by his peers with an irrational hostility, 
which discouraged also Bohm from working on this guiding wave theory for many years. 
 For a long time John Stewart Bell \cite{BellBOOK} seems to have been one of only a few physicists who
promoted the virtues of the de Broglie--Bohm theory.
 Nowadays there are several excellent expositions \cite{DuerrEtalA,DuerrTeufelBOOK,BricmontBOOK};
see also \cite{BoHi,Holland,Solvay}.

 In the following we will for simplicity work with the de Broglie--Bohm guiding law, but a perfectly analogous treatment
would seem possible with a Born--Nelson--Guerra-type stochastic guiding principle.
 It may be helpful to think of the de Broglie--Bohm theory as a deterministic limit of the Born--Nelson--Guerra theory. 
 As far as the empirical output is concerned, to the best of our knowledge the theories are equivalent (so far).
 We emphasize already that the guiding equation will play an integral part in solving the generalized
electromagnetic field equations.
\newpage

\subsection{Electromagnetic fields with given generic sources}\label{sec:sharpFIELDS}

 When Born proposed that $|\Psi|^2(t,\qVv)$ should be considered as a joint probability density for the $N$ particle
positions at time $t$ he chose to demonstrate the utility of his proposal by studying the scattering of particles off of 
each other \cite{BornsPSISQUAREpapersA,BornsPSISQUAREpapersB,BornsPSISQUAREpapersC,BornsPSISQUAREpapersD}; cf. \cite{ReedSimonBOOKiii}.
 He could also have revisited Schr\"odinger's ``$\Psi$ as matter wave''-inspired calculations for hydrogen and for
many-electron atoms coupled to the electromagnetic Maxwell fields (cf. \cite{WaveMechanics}) and deduce
from his ``$\Psi$ as probability amplitude and guiding wave field'' perspective many of the insights which we deduce below,
but apparently he did not. 
 In fact, the present author is not aware of any publication which details the following considerations and deductions.

 In this vein, we revisit the four Maxwell field equations (\ref{eq:MdotE})--(\ref{eq:MdivB})  with Schr\"odinger's 
expression (\ref{eq:chargeRHO}) at r.h.s.(\ref{eq:MdivE}) and his (\ref{eq:chargeJ}), (\ref{eq:jOFsDEF}) at r.h.s. (\ref{eq:MdotE}),
computed from the general bound state solution (\ref{eq:PSIboundGENERALz}) of (\ref{eq:ERWINeqnZ}),
with Hamiltonian $H$ given by (\ref{eq:HAMzATOM}).
 Thus for now we continue our discussion of many-electron atoms in the Born--Oppenheimer approximation;
the generalization to many-nuclei \&\ many-electron systems (i.e. molecules, solids, ...) in the Born--Oppenheimer approximation
is straightforward and will be briefly addressed in section \ref{manyatoms}.

 In Born's widely accepted probability interpretation of $\Psi$ the 
expressions (\ref{eq:chargeJ}) at r.h.s. (\ref{eq:MdotE}) and (\ref{eq:chargeRHO}) at r.h.s.(\ref{eq:MdivE})
must be seen as \emph{expected} values of the empirical current and charge densities, not as 
actual ``matter-wave'' values which Schr\"odinger had proposed.
 To emphasize this in our notation, we introduce the abbreviations 
\begin{alignat}{1}
 \rho_{\mathrm{el}}^{\mbox{\tiny{emp}}} (\sV;\qVv) 
&:= {\textstyle\sum\limits_n} - e \delta_{\qV_n}(\sV) ,
\label{eq:chargeRHOemp}\\
 \jV_{\mathrm{el}}^{\mbox{\tiny{emp}}} (t,\sV;\qVv) 
& := {\textstyle\sum\limits_n} - e \delta_{\qV_n}(\sV) 
\tfrac{\hbar}{\mEL} \cI_n\big(\nabla_{\qV_n} \Phi(t,\qVv)\big)
\label{eq:chargeJemp}
\end{alignat}
for the generic empirical charge and current vector densities of the electrons,
and we abbreviate Born's expected values $\int_{\Rset^3} G(t,\sV;\qVv)|\Psi(t,\qVv)|^2\drm^{3N}q=: \langle G\rangle(t,\sV)$. 
 Thus we have $\rho_{\mathrm{el}}(t,s) = \langle  \rho_{\mathrm{el}}^{\mbox{\tiny{emp}}}\rangle (t,\sV)$ at r.h.s.(\ref{eq:MdivE})
and $\jV_{\mathrm{el}}(t,s) = \langle  \jV_{\mathrm{el}}^{\mbox{\tiny{emp}}} \rangle (t,\sV)$ at r.h.s.(\ref{eq:MdotE}).

 But then the solutions to the field equations (\ref{eq:MdotE})--(\ref{eq:MdivB}) with Born's expression 
(\ref{eq:chargeJemp}) at r.h.s.(\ref{eq:MdotE}) and his (\ref{eq:chargeRHOemp}) at r.h.s.(\ref{eq:MdivE}) 
are not, as Schr\"odinger originally thought, the actual fields of nature which are generated by some 
actual electric charge and current densities. 
 Rather, Maxwell's equations with the expected values $\langle  \rho_{\mathrm{el}}^{\mbox{\tiny{emp}}}\rangle (t,\sV)$ at r.h.s.(\ref{eq:MdivE})
and $\langle  \jV_{\mathrm{el}}^{\mbox{\tiny{emp}}} \rangle (t,\sV)$ at r.h.s.(\ref{eq:MdotE}) 
as source terms should be seen as the \emph{expected} values of some equations for electric and magnetic fields which depend not only on space 
$\sV$ and time $t$ but also on the generic position variables $\qV_1,...,\qV_N$ of the electrons.
 Those fields will be written with a superscript ${}^\sharp$, viz.: $\EV^\sharp(t,\sV;\qVv)$ and $\BV^\sharp(t,\sV;\qVv)$. 

 In 2004 the present author already introduced related $\sharp$-field equations in an attempt 
to formulate a divergence problem-free 
classical relativistic theory of point charges coupled with the nonlinear Maxwell--Born--Infeld field equations \cite{KieJSPa}.
 Their nonlinearity has been (and still is) a major road block for progress, though.
 Here we adapt this precursor work to formulate the $\sharp$-field equations pertinent to the 
linear Maxwell--Lorentz field equations for regularized point sources and will ultimately
succeed in formulating a semi-relativistic quantum theory of particle motion.
 To avoid infinite ``self-field'' energies (etc.), instead of point charges assume that the $\qV_k$ are the centers of tiny uniformly 
charged balls of radius $a$.
 Likewise let the nucleus be such a uniformly charged ball, centered at the origin. 
 We aim at a semi-relativistic theory, so this regularization is acceptable.
 $\!\!$We next state the $\sharp$-field equations for generic empirical sources with a velocity field $\vV_n$ for the $n$-th source. 
 We do not assume that the $n$-th velocity field component $\vV_n(t,\qVv)$ is given by $\frac{\hbar}{m_n}\nabla_{\qV_n}\Phi(t,\qVv)$, 
where $\Phi$ is the phase of a Schr\"odinger wave function; it will be defined subsequently.

 The $\sharp$-field equations with ball instead of point charges as sources 
comprise two {inhomogeneous equations} 
\begin{alignat}{1}\hspace{-.3truecm}
 - \partial_t{\EV^\sharp} - \bigl({\textstyle\sum\limits_k}\vV_k\!\cdot\!\nabla_{\qV_k}\bigr)\EV^\sharp
+ c\nabla_\sV\times\BV^\sharp   &= \label{eq:aMdotEsharp}   
 4\pi e {\textstyle\sum\limits_n} - \cI_n \vV_n \delta^{(a)}_{\qV_n}(\sV),\hspace{-.3truecm}  \\
\hspace{-.5truecm}
\nabla_\sV\cdot\EV^\sharp &=\label{eq:aMdivEsharp} 4\pi e\bigl(Z \delta^{(a)}_{\nullV}(\sV) + {\textstyle\sum\limits_n} -\delta^{(a)}_{\qV_n}(\sV) 
\bigr),
\end{alignat}
and two {homogeneous equations}
\begin{alignat}{1}
 \partial_t{\BV^\sharp} +\bigl({\textstyle\sum\limits_k}\vV_k\!\cdot\!\nabla_{\qV_k}\bigr)\BV^\sharp
  + c \nabla_\sV\times\EV^\sharp  &= \label{eq:MdotBsharp} \nullV  \, ,  \\ 
    \nabla_\sV\cdot \BV^\sharp  &= \label{eq:MdivBsharp} 0\, ,
\end{alignat}
where, for brevity, we have suppressed the arguments from $\EV^\sharp(t,\sV;\qVv)$ and $\BV^\sharp(t,\sV;\qVv)$,
and we wrote $\vV_n$ for $\vV_n(t,\qVv)$; moreover, 
$\delta_{\qV_n}^{(a)}(\sV)$ is the indicator function of a ball of radius $a$ centered at $\qV_n$, normalized by $\frac43\pi a^3$.
 We note that \eqref{eq:aMdivEsharp} and \eqref{eq:MdivBsharp} are simply constraint equations on the initial data. 
 For \eqref{eq:MdivBsharp} this is immediately seen by applying $\nabla_\sV\cdot$ to \eqref{eq:MdotBsharp}, which reveals that 
$\nabla_\sV\cdot \BV^\sharp$ does not change with time. 
 Similarly, for \eqref{eq:aMdivEsharp} this is seen by applying $\nabla_\sV\cdot$ to \eqref{eq:aMdotEsharp} and noting that 
$\nabla_\sV \delta^{(a)}_{\qV_n}(\sV) = - \nabla_{\qV_n} \delta^{(a)}_{\qV_n}(\sV)$.


 We now show that by averaging them w.r.t. $|\Psi|^2 = \varrho$ they turn into  
\begin{alignat}{1}
\quad - \partial_t{\langle\EV^\sharp\rangle(t,\sV)} + c\nabla_\sV\times\langle\BV^\sharp\rangle(t,\sV)   &= \label{eq:MdotEmean}   
 4\pi\; \langle \jV_{\mathrm{el},a}^{\mbox{\tiny{emp}}} \rangle(t,\sV),  \\
    \nabla_\sV\cdot \langle\EV^\sharp\rangle(t,\sV)  &= \label{eq:MdivEmean}  
4\pi \left(\langle  \rho_{\mathrm{el},a}^{\mbox{\tiny{emp}}} \rangle(t,\sV)+ Ze\delta^{(a)}_{\nullV}(\sV)\right),\\
\quad \partial_t\langle\BV^\sharp\rangle(t,\sV) + c \nabla_\sV\times\langle\EV^\sharp\rangle(t,\sV)   &= \label{eq:MdotBmean}   
\nullV \, ,  \\ 
    \nabla_\sV\cdot \langle\BV^\sharp\rangle(t,\sV)  &= \label{eq:MdivBmean} 
0\, ,
\end{alignat} 
where
\begin{alignat}{1}
 \rho_{\mathrm{el},a}^{\mbox{\tiny{emp}}} (\sV;\qVv) 
& := {\textstyle\sum\limits_n} - e \delta^{(a)}_{\qV_n}(\sV) ,
\label{eq:achargeRHOemp}\\
 \jV_{\mathrm{el},a}^{\mbox{\tiny{emp}}} (t,\sV;\qVv) 
& := {\textstyle\sum\limits_n} - e \delta^{(a)}_{\qV_n}(\sV)  \cI_n\vV_n(t,\qVv),
\label{eq:achargeJemp}
\end{alignat}
are the ball-regularized generic empirical charge and current densities. 

 Thus the $\varrho$-averaged $\sharp$-field equations for the generic empirical sources are precisely 
Schr\"odinger's version of the  Maxwell field equations (\ref{eq:MdotE})--(\ref{eq:MdivB}) for the electromagnetic field,
with Schr\"odinger's expression (\ref{eq:chargeJ}) at r.h.s. (\ref{eq:MdotE}) and his (\ref{eq:chargeRHO}) at r.h.s.(\ref{eq:MdivE}),
except that here we have also included the charge density of the nucleus at r.h.s.(\ref{eq:MdivEmean}), and 
we have replaced the Dirac point sources by Abraham-type tiny ball sources.

 To demonstrate that the averaged $\sharp$-field equations are identical with (\ref{eq:MdotEmean})--(\ref{eq:MdivBmean}), 
all we will need in the following is that (a) $\vV_n$ is the $n$-th $\Rset^3$ component of a velocity field $\vVv$ 
in the tangent space of $\Rset^{3N}$, and that (b) $\vVv$ is defined by $\JVv = \varrho \vVv$, where $\varrho$ and 
$\JVv$ jointly satisfy the continuity equation (\ref{eq:probCONSERVATIONqN}). 
 We now multiply the $\sharp$-field equations with $|\Psi(t,\qVv)|^2$ and integrate over $\drm^{3N}q$.
 Integrations by parts, and use of the 
continuity equation (\ref{eq:probCONSERVATIONqN})  together with the  definition $\JVv = \varrho \vVv$ yields
\begin{alignat}{1}
\hspace{-1truecm}
\langle\partial_t\BV^\sharp\rangle(t,\sV)   &= \label{eq:aveMdotB}  
\partial_t \langle\BV^\sharp\rangle(t,\sV)  - \int_{\Rset^{3N}}\big(\partial_t\varrho(t,\qVv)\big)\BV^\sharp(t,\sV;\qVv)\drm^{3N}q \\ 
&=\partial_t\langle\BV^\sharp\rangle(t,\sV)+
\int_{\Rset^{3N}}\big({\textstyle\sum\limits_n} \nabla_{\qV_n}\cdot[\varrho(t,\qVv)\vV_n(t,\qVv)]\big)\BV^\sharp(t,\sV;\qVv)\drm^{3N}q \\ 
&=\partial_t\langle\BV^\sharp\rangle(t,\sV) -
\int_{\Rset^{3N}}\bigl(\varrho(t,\qVv){\textstyle\sum\limits_n} \vV_n(t,\qVv)\cdot\nabla_{\qV_n}\bigr)\BV^\sharp(t,\sV;\qVv)\drm^{3N}q \\
 &=\partial_t\langle\BV^\sharp\rangle(t,\sV) -
\bigl\langle \bigl({\textstyle\sum\limits_n} \vV_n\cdot\nabla_{\qV_n}\bigr)\BV^\sharp\bigr\rangle(t,\sV),
\end{alignat} 
and similarly for $\langle\partial_t\EV^\sharp\rangle(t,\sV)$.
 And so,
\begin{alignat}{1}
\langle\partial_t\BV^\sharp\rangle(t,\sV) + \bigl\langle \bigl({\textstyle\sum\limits_n} \vV_n\cdot\nabla_{\qV_n}\bigr)\BV^\sharp\bigr\rangle(t,\sV)
& = \label{eq:aveMdotBfinal}  
\partial_t\langle\BV^\sharp\rangle(t,\sV) ,\\
\langle\partial_t\EV^\sharp\rangle(t,\sV) + \bigl\langle \bigl(
{\textstyle\sum\limits_n} \vV_n\cdot\nabla_{\qV_n}\bigr)\EV^\sharp\bigr\rangle(t,\sV)
& = \label{eq:aveMdotEfinal}  
\partial_t\langle\EV^\sharp\rangle(t,\sV) .
\end{alignat} 
 On the other hand, since $\sV$ derivatives and $\qVv$ integration commute, we have
\begin{alignat}{1}
\langle\nabla_\sV\times\BV^\sharp\rangle(t,\sV)   &= \label{eq:curBmean}  \nabla_\sV\times\langle\BV^\sharp\rangle(t,\sV) ,\\
\langle\nabla_\sV\times\EV^\sharp\rangle(t,\sV)   &= \label{eq:curEmean}  \nabla_\sV\times\langle\EV^\sharp\rangle(t,\sV). 
\end{alignat} 
 Lastly, the $\varrho$ average of r.h.s.(\ref{eq:aMdotEsharp}) 
is by definition equal to $ 4\pi\; \langle \jV_{\mathrm{el},a}^{\mbox{\tiny{emp}}} \rangle(t,\sV)$,
that of r.h.s.(\ref{eq:aMdivEsharp}) equal to 
$4\pi \big(Ze \delta^{(a)}_{\nullV}(\sV)+ \langle \rho_{\mathrm{el},a}^{\mbox{\tiny{emp}}} \rangle(t,\sV) \big)$.
 Our demonstration is complete.
\newpage

 \emph{Thus, what Schr\"odinger thought are the electromagnetic fields of the electrons, here appear as the expected values of the 
$\sharp$-fields for generic empirical sources, to which we added the electrostatic Coulomb field}
$-Ze \nabla_{\!\sV}\, \frac{1}{|\sV|}$ ($|\sV|>a$) \emph{of the nucleus.}
\smallskip

 Schr\"odinger's findings obtained with what he thought is a ``matter-wave'' theory allows us to instead conclude
that the general bound state solution (\ref{eq:PSIboundGENERALz}) of (\ref{eq:ERWINeqnZ}),
with Hamiltonian $H$ given by (\ref{eq:HAMzATOM}), produces expected charge and current densities which are given by a
sum of terms that oscillate with the Bohr-type frequencies $\propto (E_n-E_{n'})$, and that this implies the same for the
expected values of the $\sharp$-fields. 
 This is not a mere change of name for the same mathematical expressions.
 Rather, the change from Schr\"odinger's ``$\Psi$ as matter-wave'' perspective to Born's ``$\Psi$ as probability amplitude / guiding 
wave'' perspective implies: \emph{Not} the $\varrho$-averaged $\sharp$-fields but the $\sharp$-fields should be coupled back into 
Schr\"odinger's wave equation.
\smallskip

 \emph{This is a decisive change of perspective!}
\smallskip

 Before we come to discuss the back-coupling of the $\sharp$-fields into Schr\"odinger's equation, we here add another observation 
about the $\sharp$-field equations that is the analog of what has been observed already in \cite{KieJSPa,KieJSPb}.
 Namely, if instead of taking the $\varrho$ average of the $\sharp$-field equations we 
evaluate the $\sharp$-fields at the actual position of the electrons, i.e. substituting $\qV_n(t)$ for $\qV_n$ in the 
generic position slots, we obtain electromagnetic fields $\EV^\sharp(t,\sV;\qVv(t)) =:\EV(t,\sV)$ and $\BV^\sharp(t,\sV;\qVv(t))=: \BV(t,\sV)$
which satisfy the classical Maxwell--Lorentz field equations
\begin{alignat}{1}
- \partial_t{\EV(t,\sV)} + c \nabla_\sV\times\BV(t,\sV)  &= \label{eq:MdotEactual}   
 4\pi  {\textstyle\sum\limits_n} -e \dot\qV_n(t) \delta^{(a)}_{\qV_n(t)}(\sV),\\
    \nabla_\sV\cdot \EV(t,\sV)  &= \label{eq:MdivEactual}  4\pi \bigl(Ze\delta^{(a)}_{\nullV}(\sV) +
  {\textstyle\sum\limits_n} -e \delta^{(a)}_{\qV_n(t)}(\sV) \bigr)\, ,  \\
  \partial_t{\BV(t,\sV)} + c \nabla_\sV\times\EV(t,\sV)  &= \label{eq:MdotBactual} \nullV \, ,  \\ 
    \nabla_\sV\cdot \BV(t,\sV)  &= \label{eq:MdivBactual} 0.
\end{alignat}
 These field equations with rigidly transported tiny charged ball sources are Max Abraham's version of
the Maxwell--Lorentz equations for the actual electromagnetic fields $\BV(t,\sV)$ and $\EV(t,\sV)$ in spacetime, although
$t\mapsto \qV_n(t)$ will not obey Abraham's classical equations of motions. 
 For an assessment of Abraham's theory, see \cite{KiePLA}, the appendix in \cite{AppKieAOP}, and the monographs \cite{SpohnBOOKb},
 \cite{Yaghjian}.

 A final remark before we move on to compute the feedback from the $\sharp$-fields into Schr\"odinger's equation: 
 All computations in this subsection are valid also in the limit $a\to 0$ when the ball-type sources become proper
point sources, and \eqref{eq:MdotEactual}--\eqref{eq:MdivBactual} become the Maxwell--Lorentz field equations for moving
point charge sources.
 The regularization with $a>0$ is needed for what comes next.

\newpage
\subsection{Coupling the $\sharp$-fields back into the Schr\"odinger equation}\vspace{-.2truecm}

 We next show that the Coulomb interaction term in (\ref{eq:HAMzATOM}) can naturally be obtained from the \emph{electrostatic}
solutions to these $\sharp$-field equations. 
 This is encouraging, and suggests the necessary modifications in the Schr\"odinger equation (\ref{eq:ERWINeqnZ}) to also couple 
dynamical $\sharp$-fields back into the dynamics of $\Psi$. 
 The $\sharp$-field equations themselves do not need to be modified, except that the velocity field $\vVv$, while still defined by
$\JVv = \varrho \vVv$, is not generally $\propto\nabla\Phi$ because $\JVv$ will be given by a modified expression.

\subsubsection{The electrostatic $\sharp$-field energy}

 We begin with the electrostatic special case.
 Although very special, it will produce the Schr\"odinger eigenvalue spectra of large
atoms (in Born--Oppenheimer approximation) with purely Coulombic interactions of the electrons among each other and with the nucleus, 
when an electrostatic external field is acting or not.

 We suppress $t$ as argument in the $\sharp$-fields, in this case, 
and now recall a well-known result from the classical theory of electrostatics. 
 Assume that pairwise $|\qV_j-\qV_k| > 2a$, and that all $|\qV_k|>2a$, so that no two charge balls overlap.
 Then the electrostatic field energy of such a generic $N+1$ charge configuration, with the field being the
sum of the Coulomb fields of all charged balls, is given by (cf. \cite{JacksonBOOKb})
\begin{equation}
\frac{1}{8\pi} \int_{\Rset^3} \big|\EV^\sharp(\sV;\qVv)\big|^2 \drm^3s
 = 
E_{\mbox{\tiny{self}}} 
- \sum_{n=1}^N \frac{Z e^2}{|\qV_n|} + 
	\sum\sum_{\hskip-.7truecm  1 \leq  j < k \leq N} \frac{e^2}{|\qV_j-\qV_k|},
 \label{eq:HAMfromFIELDenergyZ}
\end{equation}
and except for the configuration-independent ``self-field'' energy $E_{\mbox{\tiny{self}}}=\frac35\frac{e^2 }{a} \big( z^2 +  N \big)$
this is precisely the interaction term in Schr\"odinger's (\ref{eq:ERWINeqnZ}) with Hamiltonian $H$ given by (\ref{eq:HAMzATOM}).
 The configuration-independent ``self-field'' energy term is a constant which shifts the whole spectrum by this constant, but it will 
cancel in differences of energy eigenvalues of $H$ and therefore not be seen in the spectral frequencies $\omega_j-\omega_k$.

 When any $|\qV_n|\leq 2a$ or $|\qV_j-\qV_k|<2a$, the interactions are mollified and the field energy stays bounded
so long as $a>0$.

 Next, since the charged nucleus is treated in Born--Oppenheimer approximation, it plays the role of an
``external'' source --- i.e. not part of the system of $N$ electrons whose Schr\"odinger wave function $\Psi$ we are concerned with. 
 Therefore (\ref{eq:HAMfromFIELDenergyZ}) at once suggests a generalization when in addition to the $N$ electron atom with its nucleus 
also another ``external'' electrostatic source is introduced whose charge distribution is regular and compactly supported
(e.g. the field produced by a charged capacitor in the laboratory, used to study the \emph{Stark effect}) --- a calculation 
is supplied further below.
 The electrostatic $\sharp$-field $\EV^\sharp(\sV;\qVv)$ is then the sum
of such an external electrostatic field $\EV^{\mbox{\tiny{ext}}}(\sV) = -\nabla_\sV \phi^{\mbox{\tiny{ext}}}(\sV)$, which
includes the field of the nucleus, and all the Coulomb fields of the $N$ electrons.
 If no two balls overlap and no ball overlaps with the support of this extra external charge distribution, one now finds 
(cf. (\ref{eq:FIELDenergySTATICext})--(\ref{eq:FIELDenergySTATICextII}))
\begin{alignat}{1}
\frac{1}{8\pi} \int_{\Rset^3} \big|\EV^\sharp(\sV;\qVv)\big|^2 \drm^3s
 =  
 C \label{eq:HAMfromFIELDenergyZext}  
- \sum_{n=1}^N e \phi^{\mbox{\tiny{ext}}}(\qV_n) 
  + \sum\sum_{\hskip-.7truecm  1 \leq  j < k \leq N} \frac{e^2}{|\qV_j-\qV_k|}, 
\end{alignat}
where $C = N \tfrac35\tfrac{e^2 }{a}  + \tfrac{1}{8\pi} \int_{\Rset^3} \big|\EV^{\mbox{\tiny{ext}}}(\sV)\big|^2 \drm^3s$ (note that
the external field energy integral includes the self energy $\frac35 z^2e^2/a$ of the nucleus), and where the external potential at
the $n$-th position $\phi^{\mbox{\tiny{ext}}}(\qV_n) = \phi^{\mbox{\tiny{ext}}}_{\mbox{\tiny{lab}}}(\qV_n) + \frac{Z e}{|\qV_n|}$ 
includes the Coulomb potential 
of the nucleus.
 We conclude:

 \emph{Except for an irrelevant additive constant (the $C$ at r.h.s.(\ref{eq:HAMfromFIELDenergyZext}))
which shift the whole spectrum (which cancels in the eigenvalue differences that yield the spectral frequencies of the atom),
the r.h.s.(\ref{eq:HAMfromFIELDenergyZext}) is recognized as the usual interaction term in Schr\"odinger's equation for
the Born--Oppenheimer approximation of a many-electron atom exposed to an additional applied external electrostatic potential field.}

\subsubsection{The general electromagnetic feedback $\sharp$-field $\to \Psi$}

 The formulas obtained in the previous subsection for the electrostatic special case suggest
that at least part of the back-coupling of the electromagnetic fields into the Schr\"odinger equation is obtained
from the energy of the $\sharp$-fields sourced by generic point charge densities and current densities, given by
\begin{equation}
E^\sharp(t,\qVv):=\frac{1}{8\pi}\int_{\Rset^3}\left(\big|\EV^\sharp(t,\sV;\qVv)\big|^2 + \big|\BV^\sharp(t,\sV;\qVv)\big|^2 \right)\drm^3s,
 \label{eq:FIELDenergyZ}
\end{equation}
which takes the place of a ``minimally coupled external electric potential.'' 

 This now suggests that the field momentum of the electromagnetic $\sharp$-fields, 
\begin{equation}
\PV^\sharp(t,\qVv):=\frac{1}{4\pi c}\int_{\Rset^3} \EV^\sharp(t,\sV;\qVv)\times\BV^\sharp(t,\sV;\qVv)\drm^3s,
 \label{eq:FIELDmomentumZ}
\end{equation}
injected into the $n$-th component of $T_{\qVv}\Rset^{3N}$, may 
take the place of a ``minimally coupled external magnetic vector potential,'' but this would overcount the contributions
to the $n$-th canonical momentum.
 The linearity of the $\sharp$-field equations comes to the rescue.
 We decompose $\EV^\sharp(t,\sV;\qVv) =  \EV^{\mbox{\tiny{ext}}}(t,\sV) + \sum_{n=1}^N\EV^\sharp_n(t,\sV;\qVv)$ and
$\BV^\sharp(t,\sV;\qVv) =  \BV^{\mbox{\tiny{ext}}}(t,\sV) + \sum_{n=1}^N\BV^\sharp_n(t,\sV;\qVv)$.
 Here,
$\EV^{\mbox{\tiny{ext}}}(t,\sV)$ and $\BV^{\mbox{\tiny{ext}}}(t,\sV)$ are classical electromagnetic Maxwell fields sourced
by the charge density $Ze\delta_{\nullV}^{(a)}$ of the nucleus and possibly other compactly supported 
external sources $\rho^{\mbox{\tiny{ext}}}_{\mbox{\tiny{lab}}}(t,\sV)$ and $\jV^{\mbox{\tiny{ext}}}_{\mbox{\tiny{lab}}}(t,\sV)$ 
located far away from the atom, satisfying the continuity equation for external charge conservation.
 Hence,  these external fields satisfy the Maxwell--Lorentz field equations
 \begin{alignat}{1}
- \partial_t{\EV^{\mbox{\tiny{ext}}}(t,\sV)} + c \nabla_\sV\times\BV^{\mbox{\tiny{ext}}}(t,\sV)  &= \label{eq:MdotEexternal}   
 4\pi  \jV^{\mbox{\tiny{ext}}}_{\mbox{\tiny{lab}}}(t,\sV), \\
    \nabla_\sV\cdot \EV^{\mbox{\tiny{ext}}}(t,\sV)  &= \label{eq:MdivEexternal}  4\pi \bigl(Ze\delta^{(a)}_{\nullV}(\sV) +
   \rho^{\mbox{\tiny{ext}}}_{\mbox{\tiny{lab}}}(t,\sV) \bigr)\, ,  \\
  \partial_t{\BV^{\mbox{\tiny{ext}}}(t,\sV)} + c \nabla_\sV\times\EV^{\mbox{\tiny{ext}}}(t,\sV)  &= \label{eq:MdotBexternal} \nullV \, ,  \\ 
    \nabla_\sV\cdot \BV^{\mbox{\tiny{ext}}}(t,\sV)  &= \label{eq:MdivBexternal} 0.
\end{alignat}
 Again suppressing, for brevity, the arguments from $\EV^\sharp(t,\sV;\qVv)$ and $\BV^\sharp(t,\sV;\qVv)$,
and from the $n$-th velocity field component $\vV_n(t,\qVv)$, to be defined below, 
the $n$-th $\sharp$-fields satisfy the two {inhomogeneous equations} 
\begin{alignat}{1}\hspace{-.3truecm}
 - \partial_t{\EV^\sharp_n} 
- \bigl({\textstyle\sum\limits_k}\vV_k\!\cdot\!\nabla_{\qV_k}\bigr)\EV^\sharp_n
+ c\nabla_\sV\times\BV^\sharp_n   &= \label{eq:MdotEsharpn}   
  4\pi   \cI_n \big(-e\vV_n \delta^{(a)}_{\qV_n}(\sV)\big),\hspace{-.3truecm}  \\
\hspace{-.5truecm}
 \nabla_\sV\cdot \EV^\sharp_n &= \label{eq:MdivEsharpn}  4\pi\big(- e \delta^{(a)}_{\qV_n}(\sV)\big),
\end{alignat}
and the two {homogeneous equations}, 
\begin{alignat}{1}
 \partial_t{\BV^\sharp_n} +\bigl({\textstyle\sum\limits_k}\vV_k\!\cdot\!\nabla_{\qV_k}\bigr)\BV^\sharp_n
  + c \nabla_\sV\times\EV^\sharp_n  &= \label{eq:MdotBsharpn} \nullV  \, ,  \\ 
    \nabla_\sV\cdot \BV^\sharp_n  &= \label{eq:MdivBsharpn} 0\, .
\end{alignat}

 We now define (tentatively)
\begin{alignat}{1}
 \label{eq:FIELDmomentumZn}
\PV^\sharp_n(t,\qVv):= &\cI_n^{-1}\frac{1}{4\pi c}\!\int_{\Rset^3} 
 \EV^\sharp_n \times\BV^\sharp
(t,\sV;\qVv)\drm^3s,
\end{alignat}
and propose 
\begin{equation}
\boxed{\left(i \hbar \partial_t - E^\sharp (t,\qVv)\right)\Psi(t,\qVv) 
= {\textstyle\sum\limits_{n=1}^N}
\tfrac{1}{2\mEL}\left(-i \hbar \nabla_{\qV_n} - \PV^\sharp_n(t,\qVv) \right)^2 \Psi(t,\qVv)}
 \label{eq:ERWINeqnNEW}
\end{equation}
as Schr\"odinger wave equation for $\Psi(t,\qVv)$, coupled to the $\sharp$-fields.
 Note that $E^\sharp$ and the $\PV^\sharp_n$ occupy the slots of the external electromagnetic potentials used 
in the convential minimal coupling procedure.

 We now multiply (\ref{eq:ERWINeqnNEW}) with $\Psi^*$, and
the complex conjugate of (\ref{eq:ERWINeqnNEW}) with $\Psi$, then subtract the second from the first equation, 
and after some standard manipulations, arrive at the continuity equation (\ref{eq:probCONSERVATIONqN}), 
again with $\varrho(t,\qVv) := \Psi^*(t,\qVv) \Psi(t,\qVv)$, but now with $\JVv$ having $n$-th component
\begin{equation}
\JV_n(t,\qVv) :=\Im \left(\Psi^*(t,\qVv)\frac{1}{\mEL} \bigl(\hbar \nabla_{\qV_n} - i\PV^\sharp_n(t,\qVv)\bigr) \Psi(t,\qVv)\right).
\end{equation}
 As a consequence, the $L^2(\Rset^{3N})$ norm of $\Psi(t,\qVv)$ is preserved in time.
Moreover, using the polar decomposition $\Psi = |\Psi|e^{i\Phi}$, we have the familiar
 $\Im\left( \Psi^*(t,\qVv) \nabla_{\qVv}\Psi(t,\qVv)\right) = |\Psi|^2(t,\qVv)\nabla_{\qVv}\Phi(t,\qVv)$, 
and therefore $\JVv(t,\qVv) = \varrho(t,\qVv)\vVv(t,\qVv)$ with $\vVv$ given by
\begin{alignat}{1}\label{dBvelonMINcoup}
\forall\; n:\ \vV_n(t,\qVv) = \frac{1}{\mEL}\left(\hbar\nabla_{\qV_n} \Phi(t,\qVv) - \PV^\sharp_n(t,\qVv)\right),
\end{alignat}
which is to be used in (\ref{eq:aMdotEsharp}) and (\ref{eq:MdotEsharpn}).

 The actual positions of the electrons, $\qV_n(t)$, are now postulated to evolve in time according to the
pertinent de Broglie--Bohm-type guiding equation
\begin{alignat}{1}
\forall\;n:\quad \frac{\drm \qV_n(t)}{\drm t}
 =  \frac{1}{\mEL}\cI_n\Bigl.\Bigl(\hbar\nabla_{\qV_n}\Phi(t,\qVv)-\PV^\sharp_n(t,\qVv)\Bigr)\Bigr|_{\qVv=\qVv(t)}.
\end{alignat}
 
\subsection{The Schr\"odinger--Maxwell\, ${}^\sharp$ bound states and spectrum}

 We saw that the $\Psi$-dependent Hamiltonian $H$ of the Schr\"odinger--Maxwell system does not produce the correct hydrogen spectrum. 
 Worse, it does not produce the correct Schr\"odinger spectrum of any many-electron atom with Coulomb interactions, 
although in the semi-classical large $N$ limit some aspects of it may be recovered. 

 By contrast, as we will now show, after subtraction of an additive constant the Schr\"odinger spectra 
of many-electron atoms with Coulomb interactions are \emph{arbitrarily precisely} reproduced by the Hamiltonian 
\begin{equation}
H =  {\textstyle\sum\limits_{n=1}^N}
\frac{1}{2\mEL}\left(-i \hbar \nabla_{\qV_n} - \PV^\sharp_n(t,\qVv) \right)^2  + E^\sharp (t,\qVv)
 \label{eq:HerwinSHARP}
\end{equation}
of the Schr\"odinger--Maxwell\,$^\sharp$ system when the $\sharp$-fields are static and of finite energy. 

\subsubsection{Ground state energy in the absence of non-nuclear external fields}

 We begin with a definition.
\smallskip

\noindent
\textbf{Definition}:
\emph{Let the energy functional $W(\Psi,\EV^\sharp,\BV^\sharp)$ be given by}
\begin{eqnarray}
W =\int_{\Rset^{3N}}\!\!
\Bigl({\textstyle\sum\limits_{n=1}^N}\tfrac{1}{2\mEL}\left|\left(-i \hbar \nabla_{\!\qV_n} - \PV^\sharp_n(t,\qVv) \right)\Psi(t,\qVv)\right|^2
\!+\! E^\sharp(t,\qVv)|\Psi(t,\qVv)|^2\! \Bigr)\drm^{3N}q,
 \label{eq:energyFUNC}
\end{eqnarray}
\emph{where $E^\sharp$ is given in (\ref{eq:FIELDenergyZ}) and $\PV^\sharp_n$ in (\ref{eq:FIELDmomentumZn}), and with
$\Psi\in H^1(\drm^{3N}q)$, satisfying $\|\Psi\|_{L^2}=1$, and 
with $\EV^\sharp\in L^2(\drm^3s)$ and $\BV^\sharp\in L^2(\drm^3s)$ satisfying the constraint equations
(\ref{eq:aMdivEsharp}), (\ref{eq:MdivBsharp}), with $E^\sharp(t,\qVv)\in L^1(|\Psi|^2\drm^{3N}q)$ and 
$\vV_n(t,\qVv)\in L^2(|\Psi|^2\drm^{3N}q)$.
 Then a solution $(\Psi(t,\qVv), \EV^\sharp(t,\sV;\qVv),\BV^\sharp(t,\sV;\qVv))$ of the Schr\"odinger--Maxwell\,${}^\sharp$
system is called a \underline{\emph{ground state}} if $W$ evaluated with this solution takes the 
smallest possible value among all solutions of the same system of equations.}
 \newpage

 We are now ready to state our first theorem.

\noindent
\textbf{Theorem}:
\emph{In the absence of non-nuclear external fields, the ground state is of the form
$\Big( e^{-iE_gt/\hbar}\psi_g(\qVv), -\nabla_\sV \phi^\sharp(\sV;\qVv),\nullV\Big)$, where 
\begin{alignat}{1}
\phi^\sharp(\sV;\qVv) 
= \label{eq:phisharpSTATIC}  
 e \int_{\Rset^3} \left(Z\delta^{(a)}_{\nullV}(\sV') - 
{\textstyle\sum\limits_{n=1}^N} \delta^{(a)}_{\qV_n}(\sV')\right)\frac{1}{|\sV-\sV'|} \drm^3s',
\end{alignat}
and where $E_g \; (= E_1 >0)$ is the lowest eigenvalue of the Hamiltonian 
\begin{alignat}{1}\label{HAMnoEXTERNALfields}
 H : = 
{\textstyle\sum\limits_{n=1}^N}-\tfrac{\hbar^2}{2\mEL}\Delta_{\qV_n}
\!+\!
\tfrac{1}{8\pi}\int_{\Rset^3} \big|\nabla_\sV\phi^\sharp(\sV;\qVv)\big|^2 \drm^3s,
\end{alignat}
and $\psi_g(\qVv) $ is the associated real eigenfunction. }

\noindent
\emph{Proof}:
 We use the polar representation
$\Psi(t,\qVv) = |\Psi(t,\qVv)|e^{i\Phi(t,\qVv)}$ and find, for a solution of the 
Schr\"odinger--Maxwell\,${}^\sharp$ system,
\begin{eqnarray}
W =
\int_{\Rset^{3N}}\Bigl({\textstyle\sum\limits_{n=1}^N}\Bigl[\tfrac{\hbar^2}{8\mEL}|\nabla_{\qV_n}\ln \varrho(t,\qVv)|^2 + 
 \tfrac{1}{2}\mEL|\vV_n(t,\qVv)|^2 \Bigr]
 + E^\sharp(t,\qVv)\Bigr)\varrho(t,\qVv) \drm^{3N}q,
 \label{eq:energyFUNCrewrite}
\end{eqnarray}
with $\varrho(t,\qVv) = |\Psi(t,\qVv)|^2$.
 Equation \eqref{eq:energyFUNCrewrite} makes it plain that $W$ as defined earlier is well-defined. 
 Next, by \eqref{eq:energyFUNCrewrite} it is manifest that the value of $W$ can be lowered for the 
same $\varrho$ and same $\sharp$-fields by setting $\vV_n(t,\qVv)\equiv \nullV\ \forall\;n$. 
 This means $\hbar\nabla_{\qV_n}\Phi(t,\qVv) = \PV^\sharp_n(t,\qVv)$, by (\ref{dBvelonMINcoup}).
 Compatible with $\vV_n(t,\qVv)\equiv\nullV\ \forall\;n$ we can now further lower the  ${}^\sharp$-field energy 
$\frac{1}{8\pi}\int_{\Rset^3}\bigl(\big|\EV^\sharp(t,\sV;\qVv)\big|^2 + \big|\BV^\sharp(t,\sV;\qVv)\big|^2 \bigr)\drm^3s$
by setting $\BV^\sharp(t,\sV;\qVv)\equiv \nullV$.
 This now means $\PV^\sharp_n(t,\qVv)\equiv\nullV$, by (\ref{eq:FIELDmomentumZn}),
and therefore $\nabla_{\qV_n}\Phi(t,\qVv) \equiv \nullV\ \forall\;n$, meaning $\Phi(t,\qVv)=\Phi(t)$.
 Moreover, with $\vV_n(t,\qVv)\equiv\nullV\ \forall\; n$ we also have $\partial_t \varrho(t,\qVv)=0$, and so
$\varrho(t,\qVv)\equiv\varrho(\qVv)$.
 Thus, $\Psi(t,\qVv) = e^{i\varphi(t)}\psi(\qVv)$.

 Next, $\EV^\sharp(t,\sV;\qVv)$ cannot be set identically zero because of its contraint equation. 
 Yet, with $\vV_n(t,\qVv)\equiv\nullV\ \forall\; n$ and $\BV^\sharp(t,\sV;\qVv)\equiv \nullV$, we have 
$\EV^\sharp(t,\sV;\qVv)\equiv \EV^\sharp(\sV;\qVv)$, a static field, and thus  $E^\sharp(t,\qVv)=E^\sharp(\qVv)$.
 But for a static solution of the $\sharp$-field equations, $\nabla_\sV \crprd \EV^\sharp(\sV;\qVv)=\nullV$, 
which means $\EV^\sharp(\sV;\qVv)=-\nabla_\sV \phi^\sharp(\sV;\qVv)$. 
 Insertion of this representation into the divergence equation (\ref{eq:aMdivEsharp}) yields the Poisson equation 
\begin{alignat}{1}
   -\Delta_\sV\phi^\sharp(\sV;\qVv)  &= \label{eq:LAPphisharpSTATIC}  4\pi  e \Big(Z\delta^{(a)}_{\nullV}(\sV) -
{\textstyle\sum\limits_{n=1}^N} \delta^{(a)}_{\qV_n}(\sV)\Big).
\end{alignat}
 This Poisson equation for $\phi^\sharp(\sV;\qVv)\in\dot{H}^1(\drm^3s)$ is solved by (\ref{eq:phisharpSTATIC}).

 Therefore, with $\phi^\sharp(\sV;\qVv)$ given in (\ref{eq:phisharpSTATIC}), 
the ground state wave function $\Psi_g(t,\qVv) = e^{i\varphi_g(t)}\psi_g(\qVv)$ minimizes the reduced energy functional 
(abusing notation)
\begin{eqnarray}
{W}(\psi) =\int_{\Rset^{3N}}\!
\Bigl({\textstyle\sum\limits_{n=1}^N}\tfrac{\hbar^2}{2\mEL}\left|\nabla_{\!\qV_n}\psi(\qVv)\right|^2
\!+\! \tfrac{1}{8\pi}\int_{\Rset^3} \big|\nabla_\sV\phi^\sharp(\sV;\qVv)\big|^2 \drm^3s\; |\psi(\qVv)|^2\! \Bigr)\drm^{3N}q
 \label{eq:energyFUNCreduced}
\end{eqnarray}
among all $\psi\in \dot{H}^1(\drm^{3N}q)$ for which $\|\psi\|_{L^2}=1$. 
 The Euler--Lagrange equation for this minimization problem is $H\psi_g = E_g\psi_g$, with $H$ given in (\ref{HAMnoEXTERNALfields}),
with $E_g$ the lowest eigenvalue and $\psi_g$ the associated real eigenfunction (unique up to sign).
 Since here we have not insisted on an antisymmetric electron wave function, the ground state wave function $\psi_g$ has no nodes, 
hence we can assume it to be positive. 
 But then $\Psi_g(t,\qVv) = e^{i\varphi_g(t)}\psi_g(\qVv) = e^{i\varphi_g(t)}|\psi_g(\qVv)|$, so $\varphi_g(t) = \Phi_g(t)$. 
 
 Lastly, for $\Psi_g(t,\qVv) = e^{i\Phi_g(t)}\psi_g(\qVv)$ to solve the corresponding Schr\"odinger equation $i\hbar\partial_t\Psi_g(t,\qVv)
= H \Psi_g(t,\qVv)$, it follows that $\Phi_g(t) = - E_g t/\hbar$.  \hfill $\square$


 If no two balls overlap, the  ${}^\sharp$-field energy $ \tfrac{1}{8\pi}\int_{\Rset^3} \big|\nabla_\sV\phi^\sharp(\sV;\qVv)\big|^2 \drm^3s$
is given by (\ref{eq:HAMfromFIELDenergyZ}).
 We have arrived at the usual variational principle for the ground state $\Psi_g(t,\qVv)$ of a many-electron atom in 
Born--Oppenheimer approximation, except for the irrelevant additive electrostratic self-energy of the regularized 
electron and proton charge distributions (which only shifts the whole spectrum by an additive constant), and except for
the regularization of the point charges by tiny balls.
 The additive constant self-energy we can subtract from the Hamiltonian (\ref{HAMnoEXTERNALfields}). 
 In this Hamiltonian without self-energy contribution we may now let $a\searrow 0$, for 
 the difference between the correct regularized interaction term and the Coulomb interaction for true point
charges makes a practically negligible difference in its eigenvalues (shown rigorously in 
the textbooks of Thirring), so we may as well work with r.h.s.(\ref{eq:HAMfromFIELDenergyZ}) for all $\qVv\in\Rset^{3N}$. 

\subsubsection{Excited bound state energies without non-nuclear external fields}

 The eigenvalue equations for the excited states result also from the Ansatz 
$\Psi(t,\qVv)= e^{-i Et/\hbar}\psi(\qVv)$ with static $\sharp$-fields. 
 This  does not reveal their variational characterization.
 In fact, the spectrum can be defined by a Courant min-max principle, thusly.

 The spectrum can be build up successively by seeking the minimum of $W$ under
the constraint that the minimizing $\Psi$ be $L^2$ orthogonal to all previously found so-constrained minimizers.
 The same reasoning as used in the previous subsection reduces the $W$ functional to the usual Schr\"odinger variational
problem with ($a$-regularized) Coulomb interaction between all particles (electrons and nucleus), except for the same
additive constant as in the ground state energy problem.
 Thus the orthogonality condition is the same as used in the usual atomic $N$-body problem.

\subsubsection{\hspace{-5.pt}
Atomic spectra when static external fields $\phi^{\mbox{\tiny{ext}}}_{\mbox{\tiny{lab}}}$ and $\AV^{\mbox{\tiny{ext}}}_{\mbox{\tiny{lab}}}$ 
are present}

 In addition to the  fixed nucleus, which is an ``external source'' of a Coulomb field for the many electron system, other
``external fields'' may be generated in a laboratory, e.g. by a charged capacitor, or by Helmholtz coils with a stationary
electrical current flowing through them.
 Far away from these additional field-generating external sources the external fields may be assumed to decay to zero sufficiently rapidly 
to make their field energy finite.
 Away from their (regular macroscopic) external sources these are harmonic fields. 

 We introduce the notation $\phi^{\mbox{\tiny{ext}}}(\sV)
=  \phi^{\mbox{\tiny{ext}}}_{\mbox{\tiny{lab}}}(\sV) +  Ze \int_{\Rset^3} \delta^{(a)}_{\nullV}(\sV') \tfrac{1}{|\sV-\sV'|} \drm^3s'$
to explicitly exhibit the electrostatic Coulomb field of the nucleus; here the suffix ${}_{\mbox{\tiny{lab}}}$ emphasizes the Coulomb
field generated in the laboratory.
 As long as we do not introduce a magnetic moment of the nucleus, the external magnetic potential is entirely due to the laboratory equipment.
 In the presence of such an additional external, static electromagnetic field with potentials $\phi^{\mbox{\tiny{ext}}}_{\mbox{\tiny{lab}}}(\sV)$ and 
$\AV^{\mbox{\tiny{ext}}}_{\mbox{\tiny{lab}}}(\sV)$, satisfying $\nabla_\sV\cdot\AV^{\mbox{\tiny{ext}}}_{\mbox{\tiny{lab}}}(\sV)=0$, 
the eigenvalue equations are again obtained from the
 Ansatz of stationary $\Psi(t,\qVv)= e^{-i Et/\hbar}\psi(\qVv)$ and static $\sharp$-fields.
 Thus we have
\begin{alignat}{1}
&\phi^\sharp(\sV;\qVv) = \label{eq:phisharpSTATICext}    \phi^{\mbox{\tiny{ext}}}_{\mbox{\tiny{lab}}}(\sV)
+  e \int_{\Rset^3} \Bigl(Z\delta^{(a)}_{\nullV}(\sV') - {\textstyle\sum\limits_n}\delta^{(a)}_{\qV_n}(\sV')\Bigr)\tfrac{1}{|\sV-\sV'|} \drm^3s',
\\
&\AV^\sharp(\sV;\qVv) = \AV^{\mbox{\tiny{ext}}}_{\mbox{\tiny{lab}}}(\sV) .
\end{alignat}

 These potentials produce the electric and magnetic fields 
\begin{alignat}{1}
& \EV^\sharp(\sV;\qVv)  = \label{eq:EsharpSTATICext}  -\nabla_\sV \phi^{\mbox{\tiny{ext}}}_{\mbox{\tiny{lab}}}(\sV) +
e\int_{\Rset^3} \Bigl(Z\delta^{(a)}_{\nullV}(\sV')-{\textstyle\sum\limits_n}\delta^{(a)}_{\qV}(\sV')\Bigr)\tfrac{\sV-\sV'}{|\sV-\sV'|^3} \drm^3s',\\
\label{eq:BsharpSTATICext}
& \BV^\sharp(\sV;\qV) = \nabla_\sV\times \AV^{\mbox{\tiny{ext}}}_{\mbox{\tiny{lab}}}(\sV) .
\end{alignat}

 The ${}^\sharp$-field energy (for $|\qV_n|>2a$ and $|\qV_j-\qV_k|>2a$) changes to 
\begin{alignat}{1}
&\frac{1}{8\pi}\int_{\Rset^3}\left(\big|\EV^\sharp(\sV;\qVv)\big|^2 + \big|\BV^\sharp(\sV;\qVv)\big|^2 \right)\drm^3s =
 \label{eq:FIELDenergySTATICext} \\
& \frac{1}{8\pi}\int_{\Rset^3}\left[\big|\EV^{\mbox{\tiny{ext}}}_{\mbox{\tiny{lab}}}(\sV)\big|^2 
+\big|\BV^{\mbox{\tiny{ext}}}_{\mbox{\tiny{lab}}}(\sV)\big|^2\right] \drm^3s 
+ N \frac35\frac{e^2 }{a} + \sum\sum_{\hskip-.7truecm  1 \leq  j < k \leq N} \frac{e^2}{|\qV_j-\qV_k|} \notag \\ 
& -  {\sum\limits_n} \frac{Ze^2}{|\qV_n|} + \frac{e}{4\pi}\int_{\Rset^3}\!\! \nabla_\sV\int_{\Rset^3} 
\frac{Z\delta^{(a)}_{\nullV}(\sV') - \sum_n \delta^{(a)}_{\qV_n}(\sV')}{|\sV-\sV'|} \drm^3s' \cdot 
\nabla_\sV \phi^{\mbox{\tiny{ext}}}_{\mbox{\tiny{lab}}}(\sV) \drm^3s 
 = \\
&  C + \sum\!\sum_{\hskip-.7truecm  1 \leq  j < k \leq N} \frac{e^2}{|\qV_j-\qV_k|} - {\sum\limits_n} \Bigl(\frac{Ze^2}{|\qV_n|}
- \frac{\eEL}{4\pi}\!\int_{\Rset^3}\!\!\! \Delta_\sV\!\!
\int_{\Rset^3}\!\! \frac{\delta^{(a)}_{\qV_n}(\sV')}{|\sV-\sV'|} \drm^3s'\phi^{\mbox{\tiny{ext}}}_{\mbox{\tiny{lab}}}(\sV) \drm^3s\Bigr) = 
 \label{eq:FIELDenergySTATICextII} \\
&  C + \sum\sum_{\hskip-.7truecm  1 \leq  j < k \leq N} \frac{e^2}{|\qV_j-\qV_k|}
- e\sum_{n=1}^N \Big(\frac{Z e}{|\qV_n|} + \phi^{\mbox{\tiny{ext}}}_{\mbox{\tiny{lab}}}(\qV_n)\Big), 
\notag
\end{alignat}
where $C: = N\frac35\frac{\eEL^2 }{a}+
\frac{1}{8\pi}\int_{\Rset^3}\bigl[\big|\EV^{\mbox{\tiny{ext}}}_{\mbox{\tiny{lab}}}(\sV)\big|^2+
\big|\BV^{\mbox{\tiny{ext}}}_{\mbox{\tiny{lab}}}(\sV)\big|^2\bigr] \drm^3s$ 
is a constant that can be absorbed into the energy eigenvalues.
 We used that $\ -\Delta_\sV \frac{1}{|\sV-\sV'|} = 4\pi \delta_{\sV'}(\sV)$, and that
$\int_{\Rset^3} \delta^{(a)}_{\qV}(\sV)\phi^{\mbox{\tiny{ext}}}_{\mbox{\tiny{lab}}}(\sV) \drm^3s=\phi^{\mbox{\tiny{ext}}}_{\mbox{\tiny{lab}}}(\qV)$,
for $\phi^{\mbox{\tiny{ext}}}_{\mbox{\tiny{lab}}}(\sV)$ is harmonic on the support of the ``charged ball'' particles.
 R.h.s.(\ref{eq:FIELDenergySTATICextII}) vindicates what we anticipated in (\ref{eq:HAMfromFIELDenergyZext}). 

 Also, now there is a non-vanishing $t$-independent field momentum $\PV^\sharp_n$.
 Since $\BV^\sharp_n\equiv\nullV$ in our setting, the only contribution comes from the external field, viz.
\begin{alignat}{1}
&\hspace{-1truecm} 
\int_{\Rset^3} \EV^\sharp_n(\sV;\qVv)\times\BV^\sharp(\sV;\qVv)\drm^3s =\\
&\hspace{-.1truecm} - 
\int_{\Rset^3} \nabla_\sV\phi^\sharp_n(\sV;\qVv)\times\nabla_\sV\times \AV^{\mbox{\tiny{ext}}}_{\mbox{\tiny{lab}}}(\sV)\drm^3s =\\
&\hspace{-.1truecm}  - 
\int_{\Rset^3}\left[ \nabla_\sV\times\left(\phi^\sharp_n(\sV;\qVv) \nabla_\sV\times \AV^{\mbox{\tiny{ext}}}_{\mbox{\tiny{lab}}}(\sV)\right)-
 \phi^\sharp_n(\sV;\qVv)\nabla_\sV\times\nabla_\sV\times \AV^{\mbox{\tiny{ext}}}_{\mbox{\tiny{lab}}}(\sV)\right]\drm^3s = \\
&\hspace{-.1truecm}  - 
\int_{\Rset^3} \phi^\sharp_n(\sV;\qVv)\Delta_\sV \AV^{\mbox{\tiny{ext}}}_{\mbox{\tiny{lab}}}(\sV)\drm^3s =\\
&\hspace{-.1truecm}  - 
\int_{\Rset^3}  \AV^{\mbox{\tiny{ext}}}_{\mbox{\tiny{lab}}}(\sV) \Delta_\sV \phi^\sharp_n(\sV;\qVv)\drm^3s = 
-\eEL 
4\pi\AV^{\mbox{\tiny{ext}}}_{\mbox{\tiny{lab}}}(\qV_n).
 \label{eq:FIELDmomentumEXT}
\end{alignat}
 When substituting our results (\ref{eq:FIELDmomentumEXT}) and (\ref{eq:FIELDenergySTATICext}) into (\ref{eq:ERWINeqnNEW}), together with 
$\Psi(t,\qVv) = e^{-i Et/\hbar}\psi(\qVv)$, where $\psi(\qVv)\in\Rset$, we obtain (for $|\qV_n|>2a$ and $|\qV_j-\qV_k|>2a$)
\begin{eqnarray}\!\!\!\!
\Bigl[{\textstyle\sum\limits_{n=1}^N}\Big(
\!\tfrac{1}{2\mEL}\!\left[\tfrac{\hbar}{i}\nabla_{\qV_n}^{}\!\! +
\tfrac{\eEL}{c}\cI_n^{-1}\!\AV^{\mbox{\tiny{ext}}}_{\mbox{\tiny{lab}}}(\qV_n)\right]^2\!\!
- e\phi^{\mbox{\tiny{ext}}}_{\mbox{\tiny{lab}}}(\qV_n) - \tfrac{Ze^2}{|\qV_n|}\! \Big)
 \!+\! {\textstyle\sum\!\!\!\sum\limits_{\hskip-.6truecm  1 \leq  j < k \leq N}} \tfrac{e^2}{|\qV_j-\qV_k|}
\!\Bigr]\psi(\qVv) 
= \label{eq:ErwinEQstationaryRELext}
\widetilde{E} \psi(\qVv)
\end{eqnarray}
where 
$\widetilde{E} = E - N\frac35\frac{\eEL^2 }{a} -
\frac{1}{8\pi}\int_{\Rset^3}\big[\big|\EV^{\mbox{\tiny{ext}}}_{\mbox{\tiny{lab}}}(\sV)\big|^2 
+\big|\BV^{\mbox{\tiny{ext}}}_{\mbox{\tiny{lab}}}(\sV)\big|^2\big]\drm^3s$, 
while for $|\qV_n|\leq 2a$ the interaction term between the $n$-th electron and the nucleus
converges monotone down to $- 2\frac35\frac{z\eEL^2 }{a}$ when $|\qV_n|\downarrow 0$; an analogous result with $+$ and $Z$ replaced by $1$
holds for any electron-electron interaction term when $|\qV_j-\qV_k|\leq 2a$.

 As mentioned, for tiny $a$ the difference between the regularized 
interaction term and the true Coulomb interaction for two point 
charges makes a practically negligible difference in the eigenvalues, so we may 
as well work with (\ref{eq:ErwinEQstationaryRELext}) for all $\qVv\in\Rset^{3N}$. 

 Since all eigenfunctions of (\ref{eq:ErwinEQstationaryRELext}) can be chosen real, they 
produce a trivial generic velocity field, consistent with an electro-magneto-static 
$\sharp$-field which gave us (\ref{eq:ErwinEQstationaryRELext}).

 We summarize our observation as follows: 

\noindent
\emph{The assumption of a stationary solution $\Psi(t,\qVv) = e^{-i Et/\hbar}\psi(\qVv)$
with $\psi(\qVv)\in\Rset$ is compatible with purely electrostatic $\sharp$-fields which include
static external fields $\phi^{\mbox{\tiny{ext}}}_{\mbox{\tiny{lab}}}(\sV)$ and $\AV^{\mbox{\tiny{ext}}}_{\mbox{\tiny{lab}}}(\sV)$, and 
(except for an additive constant shift in the energy eigenvalues) it leads to the conventional Schr\"odinger equation for an 
atom acted on by these externally sourced fields by the ``minimal coupling'' rule.}
\bigskip

\noindent\textbf{Appendix to 3.4.3: On the \emph{Zeeman} and \emph{Stark} effects }

{Assuming that electrons and nucleus in a bound state are located in a ball of radius comparable to
(several) Bohr lengths, while the externally generated electric and magnetic fields $\EV^{\mbox{\tiny{ext}}}_{\mbox{\tiny{lab}}}$ 
and $\BV^{\mbox{\tiny{ext}}}_{\mbox{\tiny{lab}}}$ are created by
machines in the laboratory, with very tiny error these external fields will not vary over the distances of separation of electron and proton.  
Thus we may be tempted to suppose that for all practical purposes the correct Schr\"odinger
eigenvalues are obtained by computing with an electron-proton system exposed to a constant electric field 
$\EV^{\mbox{\tiny{ext}}}_{\mbox{\tiny{lab}}}= -\nabla_\sV \phi^{\mbox{\tiny{ext}}}_{\mbox{\tiny{lab}}}(\sV)$ 
and a constant magnetic field $\BV^{\mbox{\tiny{ext}}}_{\mbox{\tiny{lab}}}=\nabla_\sV\times \AV^{\mbox{\tiny{ext}}}_{\mbox{\tiny{lab}}}(\sV)$, 
with $\phi^{\mbox{\tiny{ext}}}_{\mbox{\tiny{lab}}}(\sV)= -\EV^{\mbox{\tiny{ext}}}_{\mbox{\tiny{lab}}}\cdot\sV$ and 
$\AV^{\mbox{\tiny{ext}}}_{\mbox{\tiny{lab}}}(\sV) = \frac12\BV^{\mbox{\tiny{ext}}}_{\mbox{\tiny{lab}}}\times \sV$. 
 This supposition is partly correct, and partly wrong. 

 If one has no electric field, i.e. $\EV^{\mbox{\tiny{ext}}}_{\mbox{\tiny{lab}}}\equiv\nullV$, 
yet a constant applied magnetic field $\BV^{\mbox{\tiny{ext}}}_{\mbox{\tiny{lab}}}\not\equiv\nullV$,
then the just described replacement yields the Schr\"odinger equation for a atom in a constant $\BV^{\mbox{\tiny{ext}}}_{\mbox{\tiny{lab}}}$ field,
which yields the \emph{normal Zeemann effect} of the splitting of the atomic spectral lines.

 If one has no magnetic field, i.e. $\BV^{\mbox{\tiny{ext}}}_{\mbox{\tiny{lab}}}\equiv\nullV$, yet a constant applied electric field 
$\EV^{\mbox{\tiny{ext}}}_{\mbox{\tiny{lab}}}\not\equiv\nullV$, then the just described replacement yields the Schr\"odinger equation for an atom in a 
constant $\EV^{\mbox{\tiny{ext}}}_{\mbox{\tiny{lab}}}$ field, which has \emph{no}
eigenvalues \cite{ReedSimonBOOKiv} --- this however is a consequence of oversimplifying the problem. 
 If on the other hand one uses Schr\"odinger's 1st order perturbation theory to estimate the energy shift in an applied
electric field, starting from an unperturbed eigenstate, this captures the \emph{Stark effect} of the splitting of the 
spectral lines}, which was studied in great detail in \cite{ErwinWMc}.\vspace{-10pt}

\subsection{Emission of a flash of electromagnetic radiation}
 In the following we argue that per our dynamical equations an atom, which initially is 
in an excited $n>1$ eigenstate with angular momentum $\ell>0$ and that is met subsequently by a 
gentle pulse of external electromagnetic vacuum radiation, will begin to 
emit an electromagnetic $\sharp$-field wave with frequencies centered on the $\omega_{n,n'}$,
with $\hbar\omega_{n,n'}$ precisely the difference between the eigen energies of the
usual Schr\"odinger hydrogen problem.

 We will set up a traditional perturbative iteration scheme for the response of the atom and of the
radiation $\sharp$-fields.
 To first order approximation the incoming radiation triggers the atom to 
emit an electromagnetic wave with the right frequency (or frequencies).
 While so far we have not been able to show that this leads to a transition, ultimately to the ground state,
we were able to check that such a dynamical scenario is compatible with our full set of dynamical equations.
 Qualitatively this latter statement is also true for the Schr\"odinger--Maxwell system, but as demonstrated
earlier, that system does not produce quantitatively correct atomic energy spectra whereas our model evidently does.

 To convey the essence of our argument it suffices to consider a hydrogen atom ($N=1=Z$). 
 The many-electron atom can be treated analogously.
\medskip

\noindent
\textbf{Remark:}
\emph{We will show that the emission to first order is a ``flash of light'' in this model,
essentially a thin spherical shell of radiation propagating outward in all directions. 
 This does not qualify as ``the photon as a particle,'' which we announced in the introduction would appear in the model. 
 To arrive at the conclusion that the model accounts for the photon, another change of perspective is required, 
notationally trivial but conceptually radical!
 This will be discussed in a later section. }
 
\subsubsection{Hydrogen radiation} 

 Since the emission of electromagnetic radiation is a dynamical process it is advisable not to complicate the problem
by also allowing time-dependent external sources. 
 Therefore, in the following the  external sources $\rho^{\mbox{\tiny{ext}}}_{\mbox{\tiny{lab}}}(\sV)$ and
$\jV^{\mbox{\tiny{ext}}}_{\mbox{\tiny{lab}}}(\sV)$ are assumed to be smooth, compactly supported, and static; 
recall that the proton's charge density, while also counted as external to the system of electrons, is not included in 
$\rho^{\mbox{\tiny{ext}}}_{\mbox{\tiny{lab}}}$ but is treated as an external source in its own right.
 To facilitate our discussion we first treat the special case in which $\rho^{\mbox{\tiny{ext}}}_{\mbox{\tiny{lab}}}(\sV)\equiv 0$ 
and $\jV^{\mbox{\tiny{ext}}}_{\mbox{\tiny{lab}}}(\sV)\equiv \nullV$, equivalently in the absence of
laboratory-generated static $\EV^{\mbox{\tiny{ext}}}_{\mbox{\tiny{lab}}}$ and $\BV^{\mbox{\tiny{ext}}}_{\mbox{\tiny{lab}}}$.
 A straightforward generalization allows external laboratory sources to be taken into account.
\medskip

\noindent\textbf{3.5.1.i: Explicitly exhibiting the static and the radiation $\sharp$-fields.}
\smallskip

\smallskip

 Having only a single electron and the fixed proton, the external field is the Coulomb field of the proton.
 We thus write 
$\EV^\sharp(t,\sV;\qV) =  \EV^\sharp_{\mbox{\tiny{{el}}}}(t,\sV;\qV) - \nabla_\sV \phi^{(a)}_{\mbox{\tiny{{pr}}}}(|\sV|)$,
where
$\phi^{(a)}_{\mbox{\tiny{{pr}}}}(|\sV|) = e \int_{\Rset^3}\frac{1}{|\sV -\sV'|}\delta^{(a)}_{\nullV}(\sV')\drm^3s'$; thus,
$\phi^{(a)}_{\mbox{\tiny{{pr}}}}(r) = -e/r$ if $r\geq a$, and 
$\phi^{(a)}_{\mbox{\tiny{{pr}}}}(r) = \phi^{(a)}_{\mbox{\tiny{{pr}}}}(0) + \frac12er^2/a^3$  if $0\leq r\leq a$, 
with $\phi^{(a)}_{\mbox{\tiny{{pr}}}}(0)$ determined by continuity at $r=a$. 
 We also have $\BV^\sharp(t,\sV;\qV) = \BV^\sharp_{\mbox{\tiny{{el}}}}(t,\sV;\qV)$.
 The electric $\sharp$-field of the electron,  $\EV^\sharp_{\mbox{\tiny{{el}}}}(t,\sV;\qV)$, 
will be split further into a sum, of its electrostatic Coulomb field with generic position $\qV$, and of its radiation field 
generated by the generic current density vector, 
i.e. $\EV^\sharp_{\mbox{\tiny{{el}}}}(t,\sV;\qV) = 
- \nabla_\sV \phi^{(a)}_{\mbox{\tiny{{el}}}}(|\sV-\qV|) + \EV^\sharp_{\mbox{\tiny{rad}}}(t,\sV;\qV)$, 
where
$\phi^{(a)}_{\mbox{\tiny{{el}}}}(|\sV-\qV|):=- e \int_{\Rset^3}\frac{1}{|\sV -\sV'|}\delta^{(a)}_{\qV}(\sV')\drm^3s'$; 
as for the proton's Coulomb potential, so also for the electron
$\phi^{(a)}_{\mbox{\tiny{{el}}}}(r) = -e/r$ if $r\geq a$, and 
$\phi^{(a)}_{\mbox{\tiny{{el}}}}(r) = \phi^{(a)}_{\mbox{\tiny{{el}}}}(0) + \frac12er^2/a^3$  if $0\leq r\leq a$, 
with $\phi^{(a)}_{\mbox{\tiny{{el}}}}(0)$ determined by continuity at $r=a$. 
 The magnetic $\sharp$-field of the electron is just the magnetic radiation field, i.e.
 $\BV^\sharp_{\mbox{\tiny{{el}}}}(t,\sV;\qV)= \BV^\sharp_{\mbox{\tiny{rad}}}(t,\sV;\qV)$.
 As indicated by the suffix ${}_{\textrm{rad}}$, these contributions to 
$\BV^\sharp_{\mbox{\tiny{{el}}}}(t,\sV;\qV)$ and $\EV^\sharp_{\mbox{\tiny{{el}}}}(t,\sV;\qV)$ 
will account for the phenomenon of electromagnetic radiation; however, included are also ``standing electrical waves''
associated with pulsating spherically symmetric $|\Psi(t,\qV)|^2$. 
 Inserting this decomposition into (\ref{eq:MdotEsharpn})--(\ref{eq:MdivBsharpn}) (with $n=1$ replaced by the suffix $_{\textrm{{el}}}$), 
straightforward vector calculus then yields that the radiation fields satisfy the system of equations
\begin{alignat}{1}\hspace{-.7truecm}
  \partial_t{\EV^\sharp_{\mbox{\tiny{rad}}}} 
+ \bigl(\vV\!\cdot\!\nabla_{\qV}\bigr)\EV^\sharp_{\mbox{\tiny{rad}}}
- c\nabla_\sV\times \BV^\sharp_{\mbox{\tiny{rad}}} 
 &= \label{eq:MdotEsharpRAD}   
-  \nabla_\sV\times \Big(\nabla_\sV\times
\Big(\cI \vV\phi^{(a)}_{\mbox{\tiny{{el}}}}(|\sV-\qV|)\Big)\Big)
,\\
 \nabla_\sV\cdot \EV^\sharp_{\mbox{\tiny{rad}}} &= \label{eq:MdivEsharpRAD}  0,\\
 \partial_t{\BV^\sharp_{\mbox{\tiny{rad}}}} +\bigl(\vV\!\cdot\!\nabla_{\qV}\bigr)\BV^\sharp_{\mbox{\tiny{rad}}}
  + c \nabla_\sV\times\EV^\sharp_{\mbox{\tiny{rad}}}  &= \label{eq:MdotBsharpRAD} \nullV  \, ,  \\ 
    \nabla_\sV\cdot \BV^\sharp_{\mbox{\tiny{rad}}}  &= \label{eq:MdivBsharpRAD} 0\, ,
\end{alignat}
where, for brevity, we again have suppressed the arguments from $\EV^\sharp_{\mbox{\tiny{rad}}}(t,\sV;\qV)$ 
and $\BV^\sharp_{\mbox{\tiny{rad}}}(t,\sV;\qV)$, and from the velocity field $\vV(t,\qV)$, given by 
$\varrho(t,\qV)\vV(t,\qV):=\JV(t,\qV)$, i.e.
\begin{alignat}{1}\label{eq:dBBv}
\vV(t,\qV) = 
\Im \left(\Psi^*(t,\qV)\tfrac{1}{\mEL} \bigl(\hbar \nabla_{\qV} - i\PV^\sharp_{\mbox{\tiny{{el}}}}(t,\qV)\bigr) \Psi(t,\qV)\right)\Big/|\Psi|^2(t,\qV) .
\end{alignat}
 Here, $\Psi$ satisfies the following Schr\"odinger equation,
\begin{alignat}{1}\label{ERWINeqnHYDROcoupleSHARP}
i\hbar\partial_t\Psi(t,\qV)
 = \Bigl(\tfrac{1}{2\mEL}\left(-i \hbar \nabla_\qV - \PV^\sharp_{\mbox{\tiny{{el}}}}(t,\qV) \right)^2 + E^\sharp(t,\qV)\Bigr) \Psi(t,\qV),
\end{alignat} 
with the ${}^\sharp$-field energy $E^\sharp(t,\qV)$ given by (for $|\qV|>2a$)
\begin{alignat}{1}
\frac{1}{8\pi}\int_{\Rset^3}\left(\big|\EV^\sharp\big|^2 + \big|\BV^\sharp\big|^2 \right)\!(t,\sV;\qV)\,\drm^3s =
E^\sharp_{\mbox{\tiny{rad}}}(t,\qV) - \frac{e^2}{|\qV|} + 2\frac35\frac{e^2 }{a},
 \label{eq:FIELDenergyHYDROGENradiatingNOextF} 
\end{alignat}
with
\begin{alignat}{1}
 \label{eq:FIELDenergyHYDROGENradiatingNOextFb} 
 E^\sharp_{\mbox{\tiny{rad}}}(t,\qV) = 
 \frac{1}{8\pi}\int_{\Rset^3}\left(\big|\EV^\sharp_{\mbox{\tiny{rad}}}\big|^2 + \big|\BV^\sharp_{\mbox{\tiny{rad}}}\big|^2\right)\!(t,\sV;\qV)
\,\drm^3s;
\end{alignat}
when $a$ is small enough, we can use r.h.s.(\ref{eq:FIELDenergyHYDROGENradiatingNOextF}) in the Schr\"odinger equation 
\eqref{ERWINeqnHYDROcoupleSHARP} for all $\qV\in\Rset^3$, not just for $|\qV|>2a$, with negligible errors.
 Furthermore, $\PV^\sharp_{\mbox{\tiny{{el}}}}(t,\qV)$ is given by 
(recall that $\BV^\sharp_{\mbox{\tiny{{el}}}}(t,\sV;\qV)= \BV^\sharp_{\mbox{\tiny{rad}}}(t,\sV;\qV)$)
\begin{alignat}{1} \label{eq:FIELDmomentumELECTRONhydrogenPRELIM}
\hspace{-1truecm}
\PV^\sharp_{\mbox{\tiny{{el}}}}(t,\qV) 
& = 
 \tfrac{1}{4\pi c}\int_{\Rset^3} \big(\EV^\sharp_{\mbox{\tiny{{el}}}} \times\BV^\sharp_{\mbox{\tiny{{rad}}}}\big)(t,\sV;\qV)\drm^3s 
\\ \notag
& =
\tfrac{1}{4\pi c}\int_{\Rset^3}\!\!\Big( \EV^\sharp_{\mbox{\tiny{rad}}}\times\BV^\sharp_{\mbox{\tiny{rad}}}
-
\nabla \big( \phi^{(a)}_{\mbox{\tiny{{el}}}}(|\,\cdot\,|)            
       \times\BV^\sharp_{\mbox{\tiny{{rad}}}}\big)(t,\sV;\qV)\drm^3s.
\end{alignat}
        Thus we have explicitly exhibited the static and the radiation degrees of freedom in both the $\sharp$-field equations
and in Schr\"odinger's equation.

 For some computational purposes it is convenient to rewrite the expression involving the Coulomb potentials into a more familiar
looking format, invoking the identity (and purging some subscripts) 
$\nabla\times(\phi\BV^\sharp) = (\nabla\phi)\times\BV^\sharp +\phi \nabla\times\BV^\sharp$, 
aided by Stokes' theorem that converts the integral of the curl into a vanishing integral over the 
``boundary at infinity,'' which leaves an integral over $\phi \nabla\times\BV^\sharp$.
 Next, invoking the $\sV$-solenoidal character of $\BV^\sharp$, we write $\BV^\sharp = \nabla_\sV\times \AV^\sharp$, and 
note that we can demand the Coulomb gauge condition $\nabla_\sV\cdot \AV^\sharp= 0$, which can always be satisfied by
adding the gradient of the solution to a Poisson equation to $\AV^\sharp$ that does not change $\BV^\sharp$. 
 Finally recalling the identity $\nabla\times(\nabla\times\aV) = \nabla(\nabla\cdot\aV) - \Delta\aV$, when $\aV=\AV^\sharp$ 
with vanishing $\sV$-divergence, we find (with the help of Green's theorem) that 
$\int (\nabla\phi)\times\BV^\sharp d^3s = - \int\phi \nabla\times(\nabla\times\AV^\sharp) d^3s
=  \int (\Delta\phi ) \AV^\sharp d^3s$. 
 And so,
\begin{alignat}{1} \label{eq:FIELDmomentumELECTRONhydrogen}
\hspace{-.7truecm}
\PV^\sharp_{\mbox{\tiny{{el}}}}(t,\qV) 
=\tfrac{1}{4\pi c}\!\int_{\Rset^3}\!\!\Big( \EV^\sharp_{\mbox{\tiny{rad}}}\times\BV^\sharp_{\mbox{\tiny{rad}}}\Big)(t,\sV;\qV)\drm^3s 
-  \tfrac{\eEL}{c} 
\Langle\AV^\sharp_{\mbox{\tiny{rad}}}\Rangle_{\qV}      
\!(t,\qV),
\end{alignat}
where 
\begin{equation}
\Langle \AV^\sharp_{\mbox{\tiny{rad}}}\Rangle_{\qV}(t,\qV) := \int_{\Rset^3} \AV^\sharp_{\mbox{\tiny{rad}}}(t,\sV;\qV)\delta^{(a)}_{\qV}(\sV)\drm^3s.
\end{equation}

 Note that l.h.s.(\ref{eq:FIELDmomentumELECTRONhydrogen}) is gauge-invariant, so 
r.h.s.(\ref{eq:FIELDmomentumELECTRONhydrogen}) must be --- except that we employed the Coulomb gauge condition to arrive at 
r.h.s.(\ref{eq:FIELDmomentumELECTRONhydrogen}) so that the remaining gauge transformations need to leave the Coulomb gauge condition intact.

 Turning our attention back to the Schr\"odinger equation \eqref{ERWINeqnHYDROcoupleSHARP}, we note that
the constant at r.h.s.(\ref{eq:FIELDenergyHYDROGENradiatingNOextF}) can be absorbed into the energy eigenvalues by a gauge 
transformation $\Psi\mapsto\widetilde\Psi \equiv e^{i (6/5)(e^2/a)t/\hbar}\Psi$, not changing $\AV^\sharp_{\mbox{\tiny{rad}}}(t,\sV;\qV)$
and $\phi^\sharp_{\mbox{\tiny{}}}(\sV;\qV)$, and which does not change the physical output of the model ---
i.e., $\Psi$ and $\widetilde\Psi$ produce the same $\varrho$ and $\JV$, and also $\EV^\sharp_{\mbox{\tiny{rad}}}(t,\sV;\qV)$ and 
$\BV^\sharp_{\mbox{\tiny{rad}}}(t,\sV;\qV)$ are unaltered.
  Thus, $\widetilde\Psi$ satisfies
\begin{alignat}{1}\label{ERWINeqnHYDROradSHARP}
i\hbar
\partial_t\widetilde\Psi(t,\qV)
 = \big(H_{\mbox{\tiny{{hyd}}}}(\qV) + H_{\mbox{\tiny{int}}}(t,\qV) + H_{\mbox{\tiny{rad}}}(t,\qV)\big)\widetilde\Psi(t,\qV), 
\end{alignat} 
where 
\begin{alignat}{1}\label{Hhyd}
H_{\mbox{\tiny{{hyd}}}}(\qV) := -\frac{\hbar^2 }{2\mEL}\Delta_\qV - e^2\frac{1}{|\qV|}, \qquad\mbox{for}\qquad |\qV|\geq 2a,
\end{alignat} 
(regularized when $|\qV|< 2a$) is Schr\"odinger's Hamiltonian of hydrogen in the Born--Oppenheimer approximation,
where 
\begin{alignat}{1}\label{HhydRAD}
H_{\mbox{\tiny{rad}}}(t,\qV)
: = 
 E^\sharp_{\mbox{\tiny{rad}}}(t,\qV),
\end{alignat} 
and where $H_{\mbox{\tiny{int}}}(t,\qV)$ is defined by (\ref{ERWINeqnHYDROcoupleSHARP})--(\ref{HhydRAD}) as
\begin{alignat}{1}\label{HhydINT}
H_{\mbox{\tiny{int}}}(t,\qV)
: = 
\tfrac{1}{2\mEL}\big|\PV^\sharp_{\mbox{\tiny{{el}}}}\big|^2
+ i\tfrac{\hbar}{2\mEL}\nabla_{\qV}\cdot\PV^\sharp_{\mbox{\tiny{{el}}}}
+ i\tfrac{\hbar}{\mEL}\PV^\sharp_{\mbox{\tiny{{el}}}} \cdot\nabla_{\qV} ,
\end{alignat} 
with $\PV^\sharp_{\mbox{\tiny{{el}}}}$ depending on $(t,\qV)$, as defined above.
\newpage

\noindent\textbf{3.5.1.ii: Solving the radiation $\sharp$-field equations, assuming $\vV(t,\qV)$ is given.}
\smallskip

 Fourier transform $\hat{f}(\kV) := \frac{1}{(2\pi)^{3/2}}\int e^{-i\kV\cdot\sV}f(\sV)d^3s$ 
with conjugate variables $\sV\to\kV$ yields the evolution equations
\begin{alignat}{1}\hspace{-.7truecm}
  \partial_t{\widehat{\EV}^\sharp_{\mbox{\tiny{rad}}}} 
+ \bigl(\vV\!\cdot\!\nabla_{\qV}\bigr)\widehat{\EV}^\sharp_{\mbox{\tiny{rad}}}
- i c\kV\times \widehat{\BV}^\sharp_{\mbox{\tiny{rad}}} 
 &= \label{eq:MdotEsharpRADF}   
-  e 4\pi \tfrac{1\;}{|\kV|^2}\kV\times \Big(\kV\times
{\cI \vV} \widehat{\delta}^{(a)}_{\qV}(\kV)\Big)
,\\
 \partial_t{\widehat{\BV}^\sharp_{\mbox{\tiny{rad}}}} +\bigl(\vV\!\cdot\!\nabla_{\qV}\bigr)\widehat{\BV}^\sharp_{\mbox{\tiny{rad}}}
  + ic \kV\times\widehat{\EV}^\sharp_{\mbox{\tiny{rad}}}  &= \label{eq:MdotBsharpRADF} \nullV  \, , 
\end{alignat}
and the constraint equations
\begin{alignat}{1}\hspace{-.7truecm}
 \kV\cdot \widehat{\EV}^\sharp_{\mbox{\tiny{rad}}} &= \label{eq:MdivEsharpRADF}  0,\\
    \kV\cdot \widehat{\BV}^\sharp_{\mbox{\tiny{rad}}}  &= \label{eq:MdivBsharpRADF} 0\, .
\end{alignat}
 Here, we have suppressed the argument $(t,\kV;\qV)$ from the $\sharp$-fields, and 
the argument $(t,\qV)$ from $\vV$, and where
\begin{alignat}{1}
 \widehat{\delta}^{(a)}_{\qV}(\kV) = \tfrac{3}{4\pi} \tfrac{1}{(a|\kV|)^{3/2}} J^{}_{3/2}(a|\kV|)e^{-i\kV\cdot\qV},
\end{alignat}
with $J_\nu$ a Bessel function of the first kind \cite{AS}.

 It is easily seen that equations \eqref{eq:MdivEsharpRADF} and \eqref{eq:MdivBsharpRADF} are merely constraints on the
initial data; they propagate in time when satisfied initially.
 Thus, let $\dV\in\Rset$ be a non-zero vector, and define the projection onto the direction $\|\dV$ by 
${\boldsymbol{\cP}_{\dV}^\|}:=\tfrac{\dV\otimes\dV}{\dV\cdot\dV}$ 
and its orthogonal complement by ${\boldsymbol{\cP}_{\dV}^\perp}:=\ID-\tfrac{\dV\otimes\dV}{\dV\cdot\dV}$,
where $\ID$ is the identity operator.
 Then our intial data have to be chosen such that
$\widehat{\EV}^\sharp_{\mbox{\tiny{rad}}}(0,\kV,\qV)= \widehat{\EV}^\sharp_{\mbox{\tiny{rad}}}(0,\kV,\qV){}^\perp_{\kV}$,  
and 
$\widehat{\BV}^\sharp_{\mbox{\tiny{rad}}}(0,\kV,\qV)= \widehat{\BV}^\sharp_{\mbox{\tiny{rad}}}(0,\kV,\qV){}^\perp_{\kV}$;
i.e. perpendicular to $\kV$, as in the usual Maxwell--Lorentz theory. 

 Incidentally, recalling that for any vector $\aV$ we have $\kV\times\big( \kV\times \aV\big) = - |\kV|^2 \aV^\perp_{\kV}$,
we see that r.h.s.\eqref{eq:MdotEsharpRADF} can be written more succinctly, viz.
\begin{alignat}{1}
\label{MdotEsharpRADFrhsSIMPLER}
-  \tfrac{1\;}{|\kV|^2}\kV\times \Big(\kV\times
{\cI \vV} \widehat{\delta}^{(a)}_{\qV}(\kV)\Big) = 
  {\cI \vV}^\perp_{\kV} \widehat{\delta}^{(a)}_{\qV}(\kV).
\end{alignat}

 The remaining two equations, \eqref{eq:MdotEsharpRADF} (with \eqref{MdotEsharpRADFrhsSIMPLER})
and \eqref{eq:MdotBsharpRADF}, can each be interpreted as an
inhomogeneous linear first-order transport equation for the Fourier transform of
one of the $\sharp$-fields, given $\vV$ and the Fourier transform of the other $\sharp$-field.
 As such, a general solution can be written as the sum of a special solution to the inhomogeneous equation plus the general
solution to the associated homogeneous equation. 
 For our purposes here we do not need to be that general, though.
 We note that a solution to the source-free 
Maxwell--Lorentz field equations is also a $\qV$-independent solution to 
the associated homogeneous equation of our radiation $\sharp$-field equations. 
 Thus we will only be concerned with solutions to our radiation $\sharp$-field equations which can be written 
as the sum of a source-free Maxwell--Lorentz radiation field (indicated by a superscript ${}^{,f}$ after the 
$\sharp$ symbol, for ``free'') and a special solution to the inhomogeneous radiation $\sharp$-field initial 
value problem with vanishing initial data (indicated by a superscrip ${}^{,s}$ after the $\sharp$ symbol, for ``source'').

 The latter problem we can integrate directly, at least in principle, with the method of characteristics; see e.g. \cite{Kamke}. 
 In this vein, let $u(t,\qV)$ stand for any Cartesian component of any of the Fourier transformed $\sharp$-fields. 
 Then each of the two equations \eqref{eq:MdotEsharpRADF} and \eqref{eq:MdotBsharpRADF} is of the form 
\begin{equation}\label{transportU}
\partial_t u(t,\qV) + \vV(t,\qV)\!\cdot\!\nabla_{\qV} u(t,\qV) = R(t,\qV),
\end{equation}
where for the purpose of this discussion $\vV(t,\qV)$ and $R(t,\qV)$ can be assumed given.
 Suppose $u(t,\qV)$ is a smooth solution to \eqref{transportU}.
 Then we can picture equation \eqref{transportU} geometrically as stating that for each $(t,\qV)\in \Rset_+\times\Rset^3$  
the vector $(1,\vV(t,\qV),R(t,\qV))\in\Rset^5$ is orthogonal to the
vector $(\partial_t u(t,\qV),\nabla_{\qV}u(t,\qV),-1)\in \Rset^5$, in the sense of the usual Euclidean inner product.
 The latter vector is the normal at $(t,\qV)$ to the graph of $u$, which is the
codimension-1 hypersurface $\{(t,\qV, u(t,\qV))\}\subset\Rset^5$. 
 Thus the five-dimensional vector field $(t,\qV)\mapsto (1,\vV(t,\qV),R(t,\qV))$ is everywhere tangent to 
the graph of $u$. 
 Therefore the graph of $u$ can be constructed with the help of the characteristic curves, 
which solve the characteristic equations for \eqref{transportU}, 
\begin{alignat}{1}
 \frac{d}{d\tau} T_\qV(\tau) &= 1,  \label{charT}\\
 \frac{d}{d\tau} \QV_\qV(\tau)& = \vV(\tau,\QV_\qV(\tau)), \label{charQV}\\
 \frac{d}{d\tau} Q_\qV(\tau) &= R(\tau,\QV_\qV(\tau)), \label{charQ}
\end{alignat}
to be solved as \emph{initial-final value problem}, with $T_{\qV}(t)=t$, 
$\QV_{\qV}(t)=\qV$, and $Q_{\qV}(0)= 0$.

 The final value problem for equation \eqref{charT} is trivially solved by $T_\qV(\tau)=\tau$, 
and \emph{assuming a solution to \eqref{charQV} has been found}, also 
the initial value problem to \eqref{charQ} is readily solved by quadrature, 
\begin{alignat}{1}
Q_\qV(\tau) =  \int_0^\tau  R(\theta,\QV_\qV(\theta)) d\theta. \label{charQsol}
\end{alignat}
 The only nontrivial equation is the de Broglie--Bohm-type guiding equation \eqref{charQV}.
 It has been shown in \cite{BerndlETal}, and more generally in \cite{TeuTum}, that for a large
class of solutions to Schr\"odinger's
equation the de Broglie--Bohm velocity field \eqref{eq:dBBv} is regular enough so that \eqref{charQV} with initial
or final values is typically well-posed with a global solution $\tau\mapsto \QV_\qV(\tau)$.
 We expect that a similar result is provable in the context of the present model; this, however, is a rather 
technical problem that will have to be addressed elsewhere.

 Now note, if $u(t,\qV)$ is a smooth solution of \eqref{transportU} that vanishes initially, then for each fixed $\qV$
the function $U_\qV(t):= u(t,\QV_\qV(t)) -Q_\qV(t)$ satisfies $\frac{d}{dt}U_\qV(t)= 0$.                 
 Thus for all $t$ the function $U_\qV(t) = U_\qV(0)$, but $u(0,\qV)=0$ and $Q_\qV(0)=0$, so $U_\qV(t)\equiv 0$, and thus
the solution $u(t,\qV)$ to \eqref{transportU} for vanishing initial data is given by  $u(t,\qV) = Q_\qV(t)$, or explicitly
(recycling $\tau$ now)
\begin{equation}\label{transportUsol}
 u(t,\qV) =\int_0^t  R(\tau,\QV_\qV(\tau)) d\tau.
\end{equation}

 We pause to re-emphasize this important finding: 
\medskip

\noindent
\emph{The de Broglie--Bohm-type final value problem \eqref{charQV}, with $\QV_\qV(t)=\qV$,
is an integral part of the method of characteristics to solve the inhomogeneous transport equations
\eqref{eq:MdotEsharpRADF} and \eqref{eq:MdotBsharpRADF} for the Fourier-transformed radiation $\sharp$-fields.}
\medskip

 Armed with \eqref{transportUsol} we conclude that \eqref{eq:MdotEsharpRADF} (with \eqref{MdotEsharpRADFrhsSIMPLER})
and \eqref{eq:MdotBsharpRADF}, given vanishing initial data, are formally solved by 
\begin{alignat}{1}
&\hspace{-1truecm}
\widehat{\EV}^{\sharp,s}_{\mbox{\tiny{rad}}}(t,\kV;\qV)
 = \label{eq:MdotEsharpRADFsol}   
i c \kV\times\!\! \int_0^t\!\! \widehat{\BV}^{\sharp,s}_{\mbox{\tiny{rad}}}(\tau,\kV,\QV_\qV(\tau)) d\tau 
+  4\pi e \!
 \int_0^t \!\! \cI\vV^\perp_{\kV}(\tau,\QV_\qV(\tau)) \widehat{\delta}^{(a)}_{\QV_\qV(\tau)}(\kV)d\tau ,\\
&\hspace{-1truecm}
\widehat{\BV}^{\sharp,s}_{\mbox{\tiny{rad}}}(t,\kV;\qV)
= - i c\kV\times \int_0^t \widehat{\EV}^{\sharp,s}_{\mbox{\tiny{rad}}} (\tau,\kV,\QV_\qV(\tau)) d\tau.  \label{eq:MdotBsharpRADFsol}  
\end{alignat}
 Now substituting r.h.s.\eqref{eq:MdotBsharpRADFsol} for 
$\widehat{\BV}^{\sharp,s}_{\mbox{\tiny{rad}}}(t,\kV;\qV)$ at r.h.s.\eqref{eq:MdotEsharpRADFsol}, we find
\begin{alignat}{1}\hspace{-.7truecm}
\widehat{\EV}^{\sharp,s}_{\mbox{\tiny{rad}}}(t,\kV;\qV)
 &= \label{eq:MdotEsharpRADFsolsol}   
- c^2 |\kV|^2 \int_0^t \int_0^\tau \widehat{\EV}^{\sharp,s}_{\mbox{\tiny{rad}}} (\theta,\kV,\QV_{\QV_\qV(\tau)}(\theta)) d\theta  
d\tau \\ \notag
& \hspace{5truecm}
 +  4\pi e \int_0^t\!\!  \cI \vV^\perp_{\kV} (\tau,\QV_\qV(\tau)) \widehat{\delta}^{(a)}_{\QV_\qV(\tau)}(\kV)d\tau,
\end{alignat}
and similarly we find
\begin{alignat}{1}\hspace{-.7truecm}
\widehat{\BV}^{\sharp,s}_{\mbox{\tiny{rad}}}(t,\kV;\qV)
 &= \label{eq:MdotBsharpRADFsolsol}   
- c^2 |\kV|^2 \! \int_0^t \int_0^\tau \widehat{\BV}^{\sharp,s}_{\mbox{\tiny{rad}}} (\theta,\kV,\QV_{\QV_\qV(\tau)}(\theta)) d\theta  
d\tau \\ \notag
& \quad - i  4\pi e c\kV\times\!\!
  \int_0^t \int_0^\tau \cI \vV^\perp_{\kV} (\theta,\QV_{\QV_\qV(\tau)}(\theta)) \widehat{\delta}^{(a)}_{\QV_{\QV_\qV(\tau)}(\theta)})d\theta d\tau.
\end{alignat}
 In going from \eqref{eq:MdotEsharpRADFsol} and \eqref{eq:MdotBsharpRADFsol} to
\eqref{eq:MdotEsharpRADFsolsol} and \eqref{eq:MdotBsharpRADFsolsol} we used that by
the constraint equations we have $\kV\times \Big(\kV\times \widehat{\BV}^{\sharp,s}_{\mbox{\tiny{rad}}}\Big)
= - |\kV|^2\widehat{\BV}^{\sharp,s}_{\mbox{\tiny{rad}}}$ and similarly $\kV\times\Big( \kV\times \widehat{\EV}^{\sharp,s}_{\mbox{\tiny{rad}}}\Big)
= - |\kV|^2\widehat{\EV}^{\sharp,s}_{\mbox{\tiny{rad}}}$; 
furthermore, we used that  $\kV\times\aV\perp\kV$, so that for any vector $\aV$
we have $\kV\times\big(\kV\times\big( \kV\times \aV\big)\big) = - |\kV|^2 \kV\times \aV$. 

 Equations \eqref{eq:MdotEsharpRADFsolsol} and \eqref{eq:MdotBsharpRADFsolsol} are fixed point equations for the 
Fourier-transformed radiation $\sharp$-fields; they are decoupled under the
assumption that the de Broglie--Bohm velocity field is given, and so then are the trajectories $t\mapsto \QV_\qV(t)$ by corollary. 
 These equations appear incredibly complicated, with iterated positions ${\QV_{\QV_\qV(\tau)}(\theta)}$ inside
iterated integrals. 
 Yet we note that each trajectory is given by a group, so for $\theta<\tau$ we have ${\QV_{\QV_\qV(\tau)}(\theta)} = \QV_\qV(\theta)$;
recall also that $\QV_\qV(t) = \qV$.

 Thus we see that equations  \eqref{eq:MdotEsharpRADFsolsol} and \eqref{eq:MdotBsharpRADFsolsol} 
are identical to the two fixed point problems
\begin{alignat}{1}\hspace{-.7truecm}
\widehat{\EV}^{\sharp,s}_{\mbox{\tiny{rad}}}(t,\kV,\QV_\qV(t)) 
 &= \label{eq:MdotEsharpRADFsolsolsol}   
- c^2 |\kV|^2 \int_0^t \int_0^\tau \widehat{\EV}^{\sharp,s}_{\mbox{\tiny{rad}}} (\theta,\kV,\QV_\qV(\theta)) d\theta  
d\tau \\ \notag
& \quad +  4\pi e 
 \int_0^t\!  \cI \vV^\perp_{\kV} (\tau,\QV_\qV(\tau)) \widehat{\delta}^{(a)}_{\QV_\qV(\tau)}(\kV)d\tau ,\\
\widehat{\BV}^{\sharp,s}_{\mbox{\tiny{rad}}}(t,\kV,\QV_\qV(t)) 
 &= \label{eq:MdotBsharpRADFsolsolsol}   
 - c^2 |\kV|^2 \int_0^t \int_0^\tau \widehat{\BV}^{\sharp,s}_{\mbox{\tiny{rad}}} (\theta,\kV,\QV_\qV(\theta)) d\theta  
d\tau \\ \notag
& \quad - i  4\pi e c\kV\times
  \int_0^t \int_0^\tau \cI \vV^\perp_{\kV} (\theta,\QV_\qV(\theta)) \widehat{\delta}^{(a)}_{\QV_\qV(\theta)}(\kV)d\theta d\tau ,
\end{alignat}
and these are twice integrated forced classical harmonic oscillator equations.
 Indeed, each Cartesian component of \eqref{eq:MdotEsharpRADFsolsolsol} and \eqref{eq:MdotBsharpRADFsolsolsol} is of the form
\begin{equation}\label{solsolsolEQ}
 f(t) + \omega^2 \int_0^t \int_0^\tau f(\theta) d\theta  d\tau = g(t)
\end{equation}
with $\omega = c|\kV|\in \Rset_+$, and with $g(0)=0$; in the equations for the magnetic components, also $g'(0)=0$. 
 It follows that also $f(0)=0$ (which entered the derivation of these equations, of course), and  $f'(0)=g'(0)$;
so for the magnetic equation it follows that $f'(0)=0$ as well, while for the electric equation in general $f'(0)$ need not vanish. 
 The associated homogeneous problem has only the vanishing solution, so the solution to the inhomogeneous problem \eqref{solsolsolEQ} is 
\begin{equation}\label{solsolsolEQinh}
 f(t) = \int_0^t g'(\tau) \cos(|\kV|c(t- \tau))d\tau + 
\int_0^t g''(\tau) \tfrac{\sin(|\kV|c(t- \tau))}{|\kV|c} d\tau .
\end{equation}

 Moreover, using again that $\QV_\qV(t) = \qV$, and also recalling that $\cI \vV (\tau,\QV_\qV(\tau)) =\frac{d}{d\tau}\QV_\qV(\tau)$ 
(see \eqref{charQV}), we note that only the component of $\tfrac{d}{d\tau}\QV_\qV(\tau)$ 
perpendicular to $\kV$ enters at r.h.s.\eqref{eq:MdotEsharpRADFsolsolsol} and r.h.s.\eqref{eq:MdotBsharpRADFsolsolsol}.
 We thus set 
${\big(\tfrac{d}{d\tau}\QV_\qV(\tau)\big)^{}}_{\kV}^\|(\tau) := {\boldsymbol{\cP}_{\kV}^\|} \cdot \frac{d}{d\tau}{\QV}_\qV(\tau)$
and 
${\big(\tfrac{d}{d\tau}\QV_\qV(\tau)\big)^{}}_{\kV}^\perp(\tau) := {\boldsymbol{\cP}_{\kV}^\perp} \cdot \frac{d}{d\tau}{\QV}_\qV(\tau)$.
 With these abbreviations,
for the Fourier-transformed electric radiation $\sharp$-field we obtain
\begin{alignat}{1}\label{eq:MdotEsharpRADFfinalSOLa}   
&\hspace{-0.7truecm}
\widehat{\EV}^{\sharp,s}_{\mbox{\tiny{rad}}}(t,\kV;\qV) 
= 
 e 4\pi \! \int_0^t\! \cos\!\big(|\kV|c(t- \tau)\big)\cI
{\big(\tfrac{d}{d\tau}\QV_\qV(\tau)\big)^{}}_{\kV}^\perp(\tau)  
\widehat{\delta}^{(a)}_{\QV_\qV(\tau)}(\kV) d\tau \hspace{-5pt}
\\ \notag
&\qquad\qquad\ 
+ e 4\pi 
 \int_0^t \tfrac{\sin\!\big(|\kV|c(t- \tau)\big)}{c|\kV|}\tfrac{d}{d\tau} \Big[
\cI{\big(\tfrac{d}{d\tau}\QV_\qV(\tau)\big)^{}}_{\kV}^\perp(\tau)  
 \widehat{\delta}^{(a)}_{\QV_\qV(\tau)}(\kV)\Big] d\tau , 
\end{alignat}
and for the Fourier-transformed magnetic radiation $\sharp$-field we obtain
\begin{alignat}{1}\label{eq:MdotBsharpRADFfinalSOLa}   
&\hspace{-0.7truecm}
\widehat{\BV}^{\sharp,s}_{\mbox{\tiny{rad}}}(t,\kV;\qV) 
= 
- i 4\pi e c\kV\times\!\!
  \int_0^t\! \cos\!\big(|\kV|c(t- \tau)\big) %
\!\int_0^\tau \!\!\cI
{\big(\tfrac{d}{d\theta}\QV_\qV(\theta)\big)^{}}_{\kV}^\perp(\theta)  
 \widehat{\delta}^{(a)}_{\QV_\qV(\theta)}(\kV)d\theta d\tau 
 \\ \notag
&\qquad \qquad\ - i 4\pi e c\kV\times
  \int_0^t \tfrac{\sin\!\big(|\kV|c(t- \tau)\big)}{|\kV|c} 
\cI{\big(\tfrac{d}{d\tau}\QV_\qV(\tau)\big)^{}}_{\kV}^\perp(\tau)  
\widehat{\delta}^{(a)}_{\QV_\qV(\tau)}(\kV) d\tau . 
\end{alignat}

 Finally, integrating by parts in the second integral at r.h.s.\eqref{eq:MdotEsharpRADFfinalSOLa} and
in the first integral at r.h.s.\eqref{eq:MdotBsharpRADFfinalSOLa}, we obtain
\begin{alignat}{1}\label{eq:MdotEsharpRADFfinalSOLb}   
 &
\hspace{-0.7truecm}
\widehat{\EV}^{\sharp,s}_{\mbox{\tiny{rad}}}(t,\kV;\qV) 
=  e 8\pi  \int_0^t\! \cos\!\big(|\kV|c(t- \tau)\big)
\cI{\big(\tfrac{d}{d\tau}\QV_\qV(\tau)\big)^{}}_{\kV}^\perp(\tau)  
\widehat{\delta}^{(a)}_{\QV_\qV(\tau)}(\kV) d\tau \hspace{-10pt}
  \\ \notag
  &\qquad\qquad\ 
  +  e  4\pi 
\tfrac{\sin\!\big(|\kV|ct\big)}{|\kV|c} \cI{\big(\tfrac{d}{d\tau}\QV_\qV(\tau)\big)^{}}_{\kV}^\perp(\tau)\Big|_{\tau=0}
  \widehat{\delta}^{(a)}_{\QV_\qV(0)}(\kV)
\end{alignat}
and 
\begin{alignat}{1}\label{eq:MdotBsharpRADFfinalSOLb}   
\hspace{-0.7truecm}
\widehat{\BV}^{\sharp,s}_{\mbox{\tiny{rad}}}(t,\kV;\qV) 
 = - i e 8\pi c  \kV\times
  \int_0^t 
\tfrac{\sin\!\big(|\kV|c(t- \tau)\big)}{|\kV|c}
\cI{\big(\tfrac{d}{d\tau}\QV_\qV(\tau)\big)^{}}_{\kV}^\perp(\tau)  
\widehat{\delta}^{(a)}_{\QV_\qV(\tau)}(\kV) d\tau .
\end{alignat}

 We note that for the purpose of computing the radiation Hamiltonian and the Poynting part of
the interaction Hamiltonian, the Fourier representation is just fine, due to the Plancherel and Parseval theorems.

 However, for the interaction Hamiltonian and the velocity field we also need the vector potential.
 We have
$\widehat{\BV}^{\sharp,s}_{\mbox{\tiny{rad}}}(t,\kV;\qV) = i\kV\times \widehat{\AV}^{\sharp,s}_{\mbox{\tiny{rad}}}(t,\kV;\qV)$ and
$\widehat{\EV}^{\sharp,s}_{\mbox{\tiny{rad}}}(t,\kV;\qV) =
 -\frac1c \frac{\partial}{\partial t} \widehat{\AV}^{\sharp,s}_{\mbox{\tiny{rad}}}(t,\kV;\qV)
-
\frac1c \bigl(\vV(t,\qV)\!\cdot\!\nabla_{\qV}\bigr)\AV^{\sharp,s}_{\mbox{\tiny{rad}}}(t,\kV;\qV)$, 
where
\begin{alignat}{1}\label{eq:AsharpRADFfinal}   
\hspace{-0.7truecm}
\widehat{\AV}^{\sharp,s}_{\mbox{\tiny{rad}}}(t,\kV;\qV) 
 = -  e 8\pi 
  \int_0^t \tfrac{\sin\!\big(|\kV|c(t- \tau)\big)}{|\kV|} 
\cI{\big(\tfrac{d}{d\tau}\QV_\qV(\tau)\big)^{}}_{\kV}^\perp(\tau)  
\widehat{\delta}^{(a)}_{\QV_\qV(\tau)}(\kV) d\tau .
\end{alignat}
 This concludes our solution of the Fourier-transformed radiation $\sharp$-field equations, the $\vV$ field considered given.
 The radiation $\sharp$-vector potential itself can now be
obtained by inverse Fourier transform, viz. ${f}(\sV) := \frac{1}{(2\pi)^{3/2}}\int e^{i\kV\cdot\sV}\hat{f}(\kV)d^3k$.

 We can simplify the task by noting that due to the linearity of the $\sharp$-field equations, we can either solve them
with regularized point charge sources in place, or we can solve them with point charge sources and regularize those 
solutions. 
 Thus it suffices to compute the inverse Fourier transform of  \eqref{eq:AsharpRADFfinal} for when $a\to 0$, 
as long as we remember to average the result over the ball of radius $a$ centered at $\qV$.

 Since
\begin{equation}
\lim_{a\to 0} \widehat{\delta}^{(a)}_{\QV_\qV(\tau)}(\kV) = (2\pi)^{-3/2} e^{-i\kV\,\cdot\,\QV_\qV(\tau)},
\end{equation}
for $a\to 0$ the Fourier-transformed sourced radiation $\sharp$-field vector potential becomes
\begin{alignat}{1}\label{eq:AsharpRADFfinalNULLa}   
\hspace{-0.7truecm}
\widehat{\AV}^{\sharp,s}_{\mbox{\tiny{rad}}}(t,\kV;\qV) 
 = -  e \tfrac{8\pi}{(2\pi)^{3/2}}
  \int_0^t \tfrac{\sin\!\big(|\kV|c(t- \tau)\big)}{|\kV|} 
\cI{\big(\tfrac{d}{d\tau}\QV_\qV(\tau)\big)^{}}_{\kV}^\perp(\tau)  
e^{-i\kV\cdot \QV_\qV(\tau)}  d\tau .
\end{alignat}
 First we carry out the angular integrations over $\Sset^2$ in $\kV$ space (i.e. $|\kV|$ is fixed).
 We introduce spherical coordinates in $\kV$ space, with 
$\nV_{\tau,\sV}^{} =\frac{\QV_\qV(\tau) -\sV}{|\QV_\qV(\tau) -\sV|}$ as (north) polar vector for $|\QV_\qV(\tau) -\sV|\neq 0$,
 $\psi$ the polar angle counting from the north pole (i.e. equal to $\pi/2-$latitude), 
and with $\varphi$ the azimuth about ${\nV_{\tau,\sV}^{}}$ (w.r.t. an arbitrary reference azimuth);
the cases of degeneracy can be handled by taking limits.
 We find 
\begin{alignat}{1}
\displaystyle 
&\hspace{-30pt}
\tfrac{1}{4\pi}\int_{\Sset^2}\!\!
{\big(\tfrac{d}{d\tau}\QV_\qV(\tau)\big)^{}}_{\kV}^\perp(\tau)  
 e^{- i \kV\cdot(\QV_{\qV}(\tau) -\sV)} \sin\psi \drm\psi\drm\varphi
=  \label{eq:StwoINTofFourierE} 
{\big(\tfrac{d}{d\tau}\QV_\qV(\tau)\big)^{}}_{\nV_{\tau,\sV}}^\perp\!(\tau) \,\tfrac{\sin\left(|\kV|||\QV_q(\tau) -\sV|\right)}{|\kV||\QV_q(\tau) -\sV|}
  \\ \nonumber
 & +\displaystyle
\left({\big(\tfrac{d}{d\tau}\QV_\qV(\tau)\big)^{}}
_{\nV_{\tau,\sV}}^\perp\!(\tau)-2{\big(\tfrac{d}{d\tau}\QV_\qV(\tau)\big)^{}}_{\nV_{\tau,\sV}}^\|\!(\tau)\right)\!
\left(\tfrac{{\cos\left(|\kV||\QV_q(\tau) -\sV|\right)}}{|\kV|^2|\QV_q(\tau) -\sV|^2}
-
\tfrac{\sin\left(|\kV||\QV_q(\tau) -\sV|\right)}{|\kV|^3|\QV_q(\tau) -\sV|^3}\right)\!.
\end{alignat}
 Note that ${\big(\tfrac{d}{d\tau}\QV_\qV(\tau)\big)^{}}_{\nV_{\tau,\sV}}^\|\!(\tau)$ is invariant under $\nV_{\tau,\sV}\to - \nV_{\tau,\sV}$.

 Next we observe that in the ensuing $|\kV|$ integrations a factor $1/|\kV|^2$ 
cancels vs. the factor $|\kV|^2$ coming from $\drm^3k = |\kV|^2\sin\psi \drm\psi\drm\varphi\drm{|\kV|}$, leading to the integral
\begin{alignat}{1}
\hspace{-30pt}
\tfrac{2}{\pi}
\displaystyle \int_0^\infty\!\!\!\left(
\cos\left(|\kV|R\right)
-
\tfrac{\sin\left(|\kV|R\right)}{|\kV|R}
\right)\!
\tfrac{\sin\bigl(|\kV|c(t-\tau) \bigr)}{|\kV|}\drm{|\kV|}
=  - \tfrac{c(t-\tau)}{R}\ONE^{}_{\{R > c(t-\tau)\}} - \tfrac\pi4 \delta^{}_{R,c(t-\tau)},
\label{eq:Fzwei} 
\end{alignat}
where $\delta^{}_{a,b}$ is the Kronecker $\delta$, not the Dirac $\delta$ (see integrals 3.741\#2 and \#3 on p.414 in \cite{GradRyzh}), 
and the well-known completeness relation
\begin{alignat}{1}
\hspace{-20pt}
\tfrac{2}{\pi}
\displaystyle \int_0^\infty\!\!
 \sin\left(|\kV|R\right)
\sin\bigl(|\kV|c(t-\tau) \bigr) d|\kV|
=   \delta\!\left(c(t-\tau)-R\right) -  \delta\! \left(c(t-\tau)+R\right),
\label{eq:Feins} 
\end{alignat}
with the abbreviation $R= |\QV_q(\tau) -\sV|$.
 Note that (\ref{eq:Fzwei}) is absolutely bounded by $1$.
\newpage

 In the subsequent $\tau$-integration only the retarded $\delta$ function contributes. 
 And so the sourced radiation $\sharp$-field vector potential (in the limit $a\to 0$) reads
\begin{alignat}{1}\label{eq:AsharpRADfinalNULLa}   
\hspace{-0.8truecm}
{\AV}^{\sharp,s}_{\mbox{\tiny{rad}}}(t,\sV;\qV) 
 = &-  2e
  \int_0^{t}
 \tfrac{\cI{\big(\tfrac{d}{d\tau}\QV_\qV(\tau)\big)^{}}_{\nV_{\tau,\sV}}^\perp\!}
{R} \delta\!\left(c(t-\tau)- R\right) 
 d\tau \\ \notag
 & +  2e   \int_0^{t}
\cI\left({\big(\tfrac{d}{d\tau}\QV_\qV(\tau)\big)^{}}
_{\nV_{\tau,\sV}}^\perp\!-2{\big(\tfrac{d}{d\tau}\QV_\qV(\tau)\big)^{}}_{\nV_{\tau,\sV}}^\|\!\right)\!
 \tfrac{c(t-\tau)}{R}\ONE^{}_{\{{R} > c(t-\tau)\}}
 d\tau.
\end{alignat}
 In terms of the ``retarded (instant of) time'' $t^{\mathrm{ret}}(t,\sV;\qV)$,
defined implicitly as solution of $c(t-t^{\mathrm{ret}}) = |\sV-\QV_{\qV}(t^{\mathrm{ret}})|$,
where we also set $\QV_q(t^{\mathrm{ret}})= \QV_q(0)$ if $t^{\mathrm{ret}}<0$, \eqref{eq:AsharpRADfinalNULLa} is
\begin{alignat}{1}\label{eq:AsharpRADfinalNULLb}   
\hspace{-0.8truecm}
{\AV}^{\sharp,s}_{\mbox{\tiny{rad}}}(t,\sV;\qV) 
 = &-  2\tfrac{e}{c}
 \tfrac{\cI{\big(\tfrac{d}{d\tau}\QV_\qV(\tau)\big)^{}}_{\nV_{\tau,\sV}}^\perp\!}
{|\QV_q(\tau) -\sV|} \Big|^{}_{\tau = t^{\mathrm{ret}}(t,\sV;\qV)}
 \\ \notag
 & 
 +  2\tfrac{e}{c}   \int_0^{t}\cI
\left({\big(\tfrac{d}{d\tau}\QV_\qV(\tau)\big)^{}}_{\nV_{\tau,\sV}}^\perp\!
-2{\big(\tfrac{d}{d\tau}\QV_\qV(\tau)\big)^{}}_{\nV_{\tau,\sV}}^\|\!\right)\!
 \tfrac{c(t-\tau)}{|\QV_q(\tau) -\sV|}\ONE^{}_{\{\tau > t^{\mathrm{ret}}(t,\sV;\qV)\}} d\tau .
\end{alignat}
 To obtain the $\sharp$-field vector potential for the $\sharp$-fields with ball sources, 
we need to convolute \eqref{eq:AsharpRADfinalNULLa}, \eqref{eq:AsharpRADfinalNULLb} w.r.t. its $\qV$ variable
with the normalized characteristic function of the ball of radius $a$ centered at $\qV$.


\smallskip

\medskip
\noindent
\textbf{Remark}:
\emph{The first line at r.h.s. \eqref{eq:AsharpRADfinalNULLb} is a counterpart of a Lien\'ard--Wiechert-type potential,
describing the field at time $t$ and space location $\sV$, given the location $\qV$ of the electron at time $t$, 
by looking for the intersection of the electron trajectory with the backward light cone of vertex $\sV$ at time $t$. 
 The second line at r.h.s. \eqref{eq:AsharpRADfinalNULLb} has no analog in the Maxwell--Lorentz field theory. 
 It describes a contribution from \emph{outside that light cone}.}

\medskip
\noindent
\textbf{Remark}:
\emph{Evaluation requires the input of the de-Broglie--Bohm motions, and then the quadratures need to be carried out.
 Unfortunately, due to the short shrift that the de-Broglie--Bohm (dBB) theory has
received in the orthodox quantum-mechanical literature, studies of the dBB motions have been a backwater of research in quantum mechanics.
 The first calculations seem to have been done by Phillipidis, Dewdney, and Hiley} \cite{PhillipidisDewdneyHiley}, \emph{for the very idealized
two-dimensional caricature of the important double-slit experiment; see also} \cite{BoHi}.
 \emph{Remarkably, the very recent reports} \cite{weakM} \emph{on photon trajectories using so-called ``weak-measurements'' are in striking resemblance 
of the idealized dBB trajectories computed in} \cite{PhillipidisDewdneyHiley}; \emph{for a scholarly assessment, see} \cite{ShellySEP}.
 \emph{Moreover, as a mathematical tool for computing solutions of Schr\"odinger's equation, dBB motions have been
put to work in the computational chemistry literature, see} \cite{Wyatt}, \cite{OriolsMompart}.
 \emph{The dBB motions we need as input should be encouragement enough for computationally skilled readers to 
begin computing them!}
\smallskip

 \newpage

\noindent\textbf{3.5.1.iii: The joint initial value problem for $\sharp$-field and wave function}
\smallskip

 We would like to show that an initial state consisting of an excited stationary state with angular momentum eigenvalue $\ell >0$,
complemented with an ``incoming radiation ${}^\sharp$-field,'' will lead to 
an evolution which involves the emission of electromagnetic radiation with the correct Rydberg frequencies,
and the transition of the atomic $\widetilde\Psi$ to a less excited level, ultimately to the ground state wave function, terminating the emission.
 It suffices to focus on a representative case, the Lyman-$\alpha$ line. 

 We stipulate:

\begin{quote}
{The initial state is a \emph{real} eigenfunction $\widetilde\Psi(0,\qV) = \widetilde\Psi_{2,1,m,+}(0,\qV)$
complemented by the electrostatic $\sharp$-field of electron and proton, and by an incoming vacuum radiation field
(a solution to the source-free Maxwell--Lorentz field equations) of finite electromagnetic energy and 
momentum (space integral of its Poynting vector field).}
\end{quote}
\smallskip

\noindent
\textbf{Remark}:
\emph{One should think of such an incoming radiation field as a so-called Gaussian beam, traveling along the $z$-direction,
the $z$ Fourier transform centered on the wave number $\omega_{2,1}/c$, and with a spread of (say) 
$\triangle(k_z) = \sigma_{k_z} / \sqrt{2} =  10^{-3}\omega_{2,1}/c$ in Fourier space. 
 This corresponds to a spread of $\triangle(z) = 1/(\sigma_{k_z}\sqrt{2}) = 500 c/\omega_{2,1} \approx 80 \lambda_{2,1}$ 
of the pulse along the $z$ direction. 
 Now, $\lambda_{2,1}\approx 1.2 \times 10^{-7}$m, and the Bohr radius $a_{\mbox{\tiny{Bohr}}}\approx 5.3\times 10^{-11}$m,
so the spread of the beam in $z$ direction is $\approx 1.8 \times 10^5$ Bohr radii, or $\approx 10,000$ hydrogen diameters. 
 In the lateral directions we can imagine the pulse to be approximately Gaussian with a much bigger spread (also slowly spreading),
say $10^{-3}$m, but this will not be used.
 If we assume that initially the Gaussian $z$ pulse is centered 1m away from the proton, then initially and for a short 
(laboratory) time thereafter the external wave will have a negligible influence on the hydrogen atom, and after the pulse arrives it  
acts in very good approximation like a plane wave of Lyman-$\alpha$ frequency during the period when it passes through the hydrogen
atom; moreover, since the spread of the pulse is $10^4$ hydrogen diameters, it does not act like a shock wave but gently, and so its
effects can to some extent be treated with perturbation theory.
 Finally, we notice that in the Coulomb gauge for $\AV^\sharp$ the Coulomb potential $\phi^\sharp$ is time-independent, and so the
vector potential of the incoming  Gaussian radiation beam pulse satisfies the classical wave equation.
 We remark that generally the total $\AV^\sharp$ itself will satisfy a more complicated wave equation, though.}
\smallskip

 We next support the initial narrative of this subsection using perturbation theory.

\newpage

\noindent\textbf{3.5.1.iv: Perturbation-theoretical approach to the initial value problem}
\smallskip

 Let $0<\eps\ll 1$ be a small parameter. 
 We set up perturbative series in powers $p \in\{0,1,2,...\}$ of $\eps$ for $\widetilde\Psi$ and $\AV^\sharp_{\mbox{\tiny{rad}}}$,
viz. we write $\widetilde\Psi(t,\qV) = \sum_{p=0}^\infty \eps^p\widetilde\Psi_{p}(t,\qV)$
and $\AV^\sharp_{\mbox{\tiny{rad}}}(t,\sV;\qV) =   \sum_{p=0}^\infty\eps^p\widetilde\AV^{\sharp,p}_{\mbox{\tiny{rad}}}(t,\sV;\qV)$;
here, the superscripts $p$ at the $\sharp$-fields should not be misread as powers.
 We assume that at order $\eps^0$ there is no radiation $\sharp$-field, thus $\widetilde\AV^{\sharp,0}_{\mbox{\tiny{rad}}}=0$.
 Inserting this series into the Schr\"odinger equation \eqref{Hhyd} for hydrogen, after
purging $H_{\mbox{\tiny{int}}}(t,\qV) + H_{\mbox{\tiny{rad}}}(t,\qV)$ from \eqref{Hhyd} at order $\eps^0$, we 
can neglect all terms in the perturbative series but $\widetilde\Psi_0(t,\qV)$.
 This yields the Schr\"odinger equation \eqref{eq:ERWINeqnMatterWaveBOhydrogen} for hydrogen in the absence of any radiation $\sharp$-fields.
 The excited eigen state $\widetilde\Psi_{2,1,m,+}(t,\qV)$ is an exact solution.
 So at order $\eps^0$ we set $\widetilde\Psi_0(t,\qV) := \widetilde\Psi_{2,1,m,+}(t,\qV)$ and will consider both $m=0$ and $m=1$;
we remark that the eigen states $\widetilde\Psi_{2,1,1,\pm}(t,\qV)$ yield identical results, so that it suffices
to discuss $\widetilde\Psi_{2,1,1,+}(t,\qV)$ if $m=1$.

 For the above setup to be consistent in the sense of perturbation theory, to order $\eps^0$ the interaction and radiation Hamiltonians
have to be negligible.
 We now assume that the incoming external pulse contributes an $O(\eps)$ term to the radiation $\sharp$-field
$\AV^{\sharp}_{\mbox{\tiny{rad}}}(t,\sV;\qV)$ 
(and analogously for the radiation $\sharp$-fields derived from $\AV^{\sharp}_{\mbox{\tiny{rad}}}(t,\sV;\qV)$).
 However, this is only the (source-)free part $\AV^{\sharp,f}_{\mbox{\tiny{rad}}}(t,\sV)$ of the $O(\eps)$ radiation $\sharp$-field. 
 Since it already produces an $O(\eps)$ velocity field $\vV(t,\qV)=\frac{e}{\mEL c}\cI^{-1}\AV^{\sharp,f}_{\mbox{\tiny{rad}}}(t,\qV)$, 
a pertinent $O(\eps)$ contribution from
$\AV^{\sharp,s}_{\mbox{\tiny{rad}}}(t,\sV;\qV)$ has to be added to $\AV^{\sharp,f}_{\mbox{\tiny{rad}}}(t,\sV)$ in order
to obtain the full $O(\eps)$ radiation $\sharp$-field. 
 In fact, the $\sharp$-field $\AV^{\sharp,s}_{\mbox{\tiny{rad}}}(t,\sV;\qV)$ contributes its own $O(\eps)$ 
share $\eps\frac{e}{\mEL c}\cI^{-1}\widetilde\AV^{\sharp,s,1}_{\mbox{\tiny{rad}}}(t,\qV,\qV)$
to the velocity field and solves a self-consistent problem; see below.
 The beam-produced velocity field is negligible away from the intense pulse region, and since the hydrogen wave functions 
considered by us are exponentially decaying away from the proton with a scale of a few Bohr radii, this velocity field
is initially negligible for all $\qV$ inside the atom, and then also no sourced $\sharp$-field is produced in the atomic
region, yet it will be appreciably large for the brief duration of the radiation beam pulse's passage through the atom. 

 With the starting wave function at $O(1)$, and the starting radiation $\sharp$-fields of $O(\eps)$,
we can insert the perturbative series Ansatz for $\widetilde\Psi(t,\qV)$ and for 
$\AV^{\sharp}_{\mbox{\tiny{rad}}}(t,\sV;\qV)$ into the dynamical equations and sort them by powers of $\eps$,
which yields a consistent hierarchy of coupled linear equations.

 We next discuss the $O(\eps)$ responses to the incoming radiation beam pulse, i.e. the sourced $\sharp$-field 
$\widetilde\AV^{\sharp,s,1}_{\mbox{\tiny{rad}}}(t,\sV;\qV)$, and the perturbation $\widetilde\Psi_{1}(t,\qV)$ of the wave function of the atom. 
 We need to begin with the  $\sharp$-field.
 It should be recalled that only the $\qV$ within a few dozen Bohr radii from the nucleus are of interest. 
\smallskip

 \newpage

\noindent\textbf{3.5.1.v: Response of the $\sharp$-field at first order in perturbation theory.}
\smallskip

 To first order in perturbation theory the $\sharp$-field responds directly to the incoming Maxwell--Lorentz 
radiation beam pulse; no intermediary action by the wave function enters. 
 This is decisively different from Schr\"odinger's matter-wave approach, viz. the Schr\"odinger--Maxwell equations. 
 
 More to the point, and since we only need to consider $\qV$ within a few dozen Bohr radii from the nucleus, 
the incoming radiation can be assumed to be a Gaussian plane-wave pulse\footnote{Often in the literature, 
  by a plane wave one means a monochromatic plane wave; to avoid confusion we speak of a plane-wave pulse.}
$\widetilde\AV^{\sharp,\mbox{\tiny{f}}}_{\mbox{\tiny{rad}}}
= \eV_x \frac{e\mEL c}{\hbar}\exp\big(-\frac{1}{2\sigma_z^2}(z-z_0-ct)^2\big)\cos\big(\frac1c\omega_{2,1}(z-z_0-ct)\big)$,
traveling in the $z$-direction, satisfying the Coulomb gauge, and producing divergence-free electric and magnetic 
radiation pulse fields.
 This maps into the velocity field
\begin{equation}
\widetilde\vV^{\mbox{\tiny{f}}}(t,\qV) = 
\eV_{x} \tfrac{e^2}{\hbar}\exp\big(-\tfrac{1}{2\sigma_z^2}(\qV\cdot\eV_z-z_0-ct)^2\big)\cos\big(\tfrac1c\omega_{2,1}(\qV\cdot\eV_z-z_0-ct)\big).
\end{equation}
 Now define 
\begin{equation}\label{Vdefine}
\widetilde\vV^{(1)}(t,\qV)
:= \widetilde\vV^{\mbox{\tiny{f}}}(t,\qV) + \tfrac{e}{\mEL c}\cI^{-1}\Langle\widetilde{\AV}^{\sharp,s,1}_{\mbox{\tiny{rad}}}\Rangle_{\qV}(t,\qV) .
\end{equation}

\smallskip
\noindent
\textbf{Remark}:
\emph{Since we have taken the limit $a\to 0$ in the source term of the radiation $\sharp$-field equations, 
the sourced $\sharp$-field has a singularity when $\sV=\qV$. 
 Thus we cannot also let $a\to 0$ at r.h.s.\eqref{Vdefine}, because this would lead to the ill-defined 
 $\widetilde{\AV}^{\sharp,s,1}_{\mbox{\tiny{rad}}}(t,\qV;\qV)$.}
\smallskip

 For $\widetilde{\AV}^{\sharp,s,1}_{\mbox{\tiny{rad}}}(t,\sV;\qV)$ we now obtain the following linear fixed point problem,
\begin{alignat}{1}\label{eq:AsharpRADfixpt}   
\hspace{-0.8truecm}
\widetilde{\AV}^{\sharp,s,1}_{\mbox{\tiny{rad}}}(t,\sV;\qV) 
 = &-  2\tfrac{e}{c}
 \tfrac{{\big(\widetilde\vV^{(1)}(\tau,\QV_\qV(\tau)\big)^{}}_{\nV_{\tau,\sV}}^\perp\!}
{|\QV_q(\tau) -\sV|} \Big|^{}_{\tau = t^{\mathrm{ret}}(t,\sV;\qV)}
 \\ \notag
 &\hspace{-2truecm}
 +  2\tfrac{e}{c}   \int_0^{t}\!\!
\left({{\big(\widetilde\vV^{(1)}(\tau,\QV_\qV(\tau)\big)^{}}_{\nV_{\tau,\sV}}^\perp\!}
-2{{\big(\widetilde\vV^{(1)}(\tau,\QV_\qV(\tau)\big)^{}}_{\nV_{\tau,\sV}}^\|\!}
\right)\!
 \tfrac{c(t-\tau)}{|\QV_q(\tau) -\sV|}\ONE^{}_{\{\tau < t^{\mathrm{ret}}(t,\sV;\qV)\}}
 d\tau .
\end{alignat}
 Although linear, this is a formidable integral-delay equation. 
 However, $|\widetilde\vV^{(1)}|/c = O(\alphaS)$, and so
we can expect that a solution can be found by Picard-type iteration, similar to a Born series
in Born's treatment of the quantum-mechanical scattering problem. 
 This leads to an iterative series expansion of $\widetilde{\AV}^{\sharp,s,1}_{\mbox{\tiny{rad}}}(t,\sV;\qV)$.
 In this vein we note that to first approximation in this iteration one neglects the contribution
from $\Langle\widetilde{\AV}^{\sharp,s,1}_{\mbox{\tiny{rad}}}\Rangle_{\qV}(t,\qV)$ at r.h.s.\eqref{Vdefine}
and obtains the first contribution to
$\widetilde{\AV}^{\sharp,s,1}_{\mbox{\tiny{rad}}}(t,\sV;\qV)$ 
by explicit quadrature of r.h.s.\eqref{eq:AsharpRADfixpt}, after solving the final value problem
\begin{equation}\label{charQinBORNapprox}
\tfrac{d}{d\tau} \QV_\qV(\tau) = \eps\cI^{-1} \widetilde\AV^{\sharp,\mbox{\tiny{f}}}_{\mbox{\tiny{rad}}}(\tau, \QV_\qV(\tau) )
\end{equation}
with $\QV_\qV(t) =\qV$.
 This problem reduces further to a simple one-dimensional linear ODE problem of first order, due to the fact that the incoming
plane-wave pulse points in the $\eV_x$ direction while its space-dependence involves only $\qV\cdot \eV_z$, which is fixed 
during a characteristic motion.
 Thus it can be integrated by direct quadrature, 
\begin{eqnarray}\!\!\!\! \notag
 \QV_\qV(\tau) 
\!\!\!&=& \!\!\!
\qV - \eV_{x} \eps  \tfrac{e^2}{\hbar}\!\!\int_\tau^t \!\!
\exp\big(-\tfrac{1}{2\sigma_z^2}(\qV\cdot\eV_z-z_0-c\theta)^2\big)\cos\big(\tfrac1c\omega_{2,1}(\qV\cdot\eV_z-z_0-c\theta)\big)d\theta \\
\label{QinBORNapprox}
\!\!\! &=&\!\!\!
\qV - \eV_{x} \eps  \alphaS \!\!\int_{c\tau - \qV\cdot\eV_z+z_0}^{ct -\qV\cdot\eV_z+z_0}\!\!
\exp\big(-\tfrac{1}{2\sigma_z^2}\xi^2\big)\cos\big(\tfrac1c\omega_{2,1}\xi)\big)d\xi ,
\end{eqnarray}
which can be expressed in terms of known functions, involving the error function, but the representation \eqref{QinBORNapprox} is explicit enough
for our purposes.

 Inserting \eqref{QinBORNapprox} into \eqref{eq:AsharpRADfixpt} with $\eps\widetilde\vV^{(1)}\!\big(\tau,\QV_\qV(\tau)\big)$ 
given by r.h.s.\eqref{charQinBORNapprox} is still an involved formula, due to the retardations. 
 However, since the relevant $\qV$ are restricted to the atomic vicinity of the nucleus, and 
since it is clear that the velocity field within a few dozen Bohr radii of the nucleus 
will be of a brief transient character, and since during this transient period $T$ it oscillates with the Lyman-$\alpha$ frequency,
our formula \eqref{eq:AsharpRADfixpt} in this Born-type approximation reveals that 
$\widetilde{\AV}^{\sharp,s,1}_{\mbox{\tiny{rad}}}(t,\sV;\qV)$ will point in the $\eV_x$ direction, and far away from
the atom one essentially has a spherical shell of radiation of thickness $cT$, centered on the nucleus, moving outward 
at the speed of light, plus some dispersive part due to the contributions from outside the light cone. 
 Moreover, and most importantly, the sourced radiation $\sharp$-field in this Born-type approximation 
oscillates itself with Lyman-$\alpha$ frequency.


\smallskip

\noindent\textbf{3.5.1.vi: Response of the atom at first order in perturbation theory.}
\smallskip

 With radiation ${}^\sharp$-fields of $O(\eps)$, the interaction Hamiltonian 
$H_{\mbox{\tiny{int}}}(t,\qV)$ is $O(\eps)$ relative to $H_{\mbox{\tiny{{hyd}}}}(\qV)$   
because of the linear contribution from the vector potential; it is then negligible at $O(\eps^0)$ in the Schr\"odinger equation,
vindicating our original assumption.
 The radiation Hamiltonian and the integrated Poynting vector in the interaction Hamiltonian are only
$O(\eps^2)$ operators, hence negligible at both $O(\eps^0)$ and $O(\eps)$ in our Schr\"odinger equation for hydrogen.


 To $O(\eps)$ included we have
$\PV^\sharp_{\mbox{\tiny{{el}}}}(t,\qV) = 
- \eps \tfrac{\eEL}{c}  
\Langle \widetilde\AV^{\sharp,\mbox{\tiny{1}}}_{\mbox{\tiny{rad}}}\Rangle_{\qV} 
\Rangle_{\nullV} 
(t,\qV)$.

 Furthermore, the magnetic vector potential of the incoming radiation field is independent of $\qV$, and 
since we stipulated the Coulomb gauge condition in the $\sV$ variable, we find
$\nabla_\qV\cdot\Langle \AV^{\sharp,\mbox{\tiny{f}}}_{\mbox{\tiny{rad}}}\Rangle_{\qV}(t,\qV)= 0 $ because
\begin{alignat}{1}
\label{divCALC}
\nabla_\qV\cdot \int_{\Rset^3} \AV^{\sharp,\mbox{\tiny{f}}}_{\mbox{\tiny{rad}}}(t,\sV)\delta^{(a)}_{\qV}(\sV)\drm^3s
& = 
 \int_{\Rset^3} \AV^{\sharp,\mbox{\tiny{f}}}_{\mbox{\tiny{rad}}}(t,\sV)\cdot \nabla_\qV \delta^{(a)}_{\qV}(\sV)\drm^3s \\ 
& = -  \int_{\Rset^3} \AV^{\sharp,\mbox{\tiny{f}}}_{\mbox{\tiny{rad}}}(t,\sV)\cdot \nabla_\sV \delta^{(a)}_{\qV}(\sV)\drm^3s \\ 
& =  \int_{\Rset^3} \big(\nabla_\sV\cdot \AV^{\sharp,\mbox{\tiny{f}}}_{\mbox{\tiny{rad}}}(t,\sV)\big) \delta^{(a)}_{\qV}(\sV)\drm^3s = 0.
\end{alignat}
 So we have shown that $\nabla_{\qV}\cdot\PV^{\sharp,\mbox{\tiny{f}}}_{\mbox{\tiny{{el}}}}(t,\qV) = 0$.
 Similarly, for the averaged sourced $\sharp$-field vector potential we know that its $\sV$-divergence vanishes, and this entails
\begin{alignat}{1}
\label{divCALCsourceA}
0 & =  \int_{\Rset^3} \big(\nabla_\sV\cdot \AV^{\sharp,\mbox{\tiny{s}}}_{\mbox{\tiny{rad}}}(t,\sV;\qV)\big) \delta^{(a)}_{\qV}(\sV)\drm^3s \\
\label{divCALCsourceB}
& = - \int_{\Rset^3} \AV^{\sharp,\mbox{\tiny{s}}}_{\mbox{\tiny{rad}}}(t,\sV;\qV)\cdot \nabla_\sV \delta^{(a)}_{\qV}(\sV)\drm^3s \\ 
\label{divCALCsourceC}
& =   \int_{\Rset^3} \AV^{\sharp,\mbox{\tiny{s}}}_{\mbox{\tiny{rad}}}(t,\sV;\qV)\cdot \nabla_\qV \delta^{(a)}_{\qV}(\sV)\drm^3s \\ 
& = -  \int_{\Rset^3} \big(\nabla_\qV\cdot \AV^{\sharp,\mbox{\tiny{s}}}_{\mbox{\tiny{rad}}}(t,\sV;\qV)\big) \delta^{(a)}_{\qV}(\sV)\drm^3s \\
& = - \nabla_\qV\cdot \int_{\Rset^3} \AV^{\sharp,\mbox{\tiny{s}}}_{\mbox{\tiny{rad}}}(t,\sV;\qV)\delta^{(a)}_{\qV}(\sV)\drm^3s 
+ \int_{\Rset^3} \AV^{\sharp,\mbox{\tiny{s}}}_{\mbox{\tiny{rad}}}(t,\sV;\qV)\cdot \nabla_\qV \delta^{(a)}_{\qV}(\sV)\drm^3s \\ 
& = - \nabla_\qV\cdot \int_{\Rset^3} \AV^{\sharp,\mbox{\tiny{s}}}_{\mbox{\tiny{rad}}}(t,\sV;\qV)\delta^{(a)}_{\qV}(\sV)\drm^3s,
\end{alignat}
the last equality because of the sequence of equalities \eqref{divCALCsourceA}--\eqref{divCALCsourceC}.
 This shows that its $\qV$-divergence vanishes, too. 

 The interaction Hamiltonian (\ref{HhydINT}) at $O(\eps)$ simplifies to $\eps H^{\mbox{\tiny{(1)}}}_{\mbox{\tiny{int}}}(t,\qV)$, with
\begin{alignat}{1}\label{HhydINTreduced}
H^{\mbox{\tiny{(1)}}}_{\mbox{\tiny{int}}}(t,\qV)
 = 
- i\tfrac{\hbar \eEL}{\mEL c} 
 \Langle\widetilde\AV^{\sharp,\mbox{\tiny{f}}}_{\mbox{\tiny{rad}}} + \widetilde\AV^{\sharp,\mbox{\tiny{s,1}}}_{\mbox{\tiny{rad}}}\Rangle_{\qV}(t,\qV)
\cdot\nabla_{\qV} .
\end{alignat} 
 To establish  the consistency of the perturbative setup we need to assume that $H^{\mbox{\tiny{(1)}}}_{\mbox{\tiny{int}}}(t,\qV)$
itself is of the same order as $H_{\mbox{\tiny{hyd}}}(t,\qV)$, which can always be arranged by choosing the amplitude of
the  incoming radiation field appropriately.
\newpage

\noindent
\textbf{Remark}:
\emph{Since the incoming radiation $\sharp$-fields can be assumed to be smooth, and since the radius, $a$, of the mollification
of the point electron is as tiny as we please, we have}
$\Langle\widetilde\AV^{\sharp,\mbox{\tiny{f}}}_{\mbox{\tiny{rad}}}\Rangle_{\qV}(t,\qV) 
 \approx \widetilde\AV^{\sharp,\mbox{\tiny{f}}}_{\mbox{\tiny{rad}}}(t,\qV)$
\emph{with arbitrary precision. 
 Therefore, for all practical purposes we can replace}
$\Langle\widetilde\AV^{\sharp,\mbox{\tiny{f}}}_{\mbox{\tiny{rad}}}\Rangle_{\qV}(t,\qV)$ with 
$\widetilde\AV^{\sharp,\mbox{\tiny{f}}}_{\mbox{\tiny{rad}}}(t,\qV)$ \emph{in \eqref{HhydINTreduced}.}
\smallskip

 Letting therefore $a\to 0$ in the incoming-radiation term of the perturbative expansion of the Schr\"odinger equation \eqref{eq:ERWINeqnNEW} for
hydrogen exposed to an incoming radiation field, 
we find (after dropping the argument $(t,\qV)$ and $\widetilde{\quad}$
from the $\widetilde\Psi$s and $\widetilde\AV^{\sharp,1}_{\mbox{\tiny{rad}}}$)
\begin{equation}
\Big(\! i \hbar \partial_t + \tfrac{\hbar^2 }{2\mEL}\Delta_\qV + \tfrac{e^2}{|\qV|}\Big)\Psi_1
=   
- i\tfrac{\hbar \eEL}{\mEL c} 
\left(
\AV^{\sharp,\mbox{\tiny{f}}}_{\mbox{\tiny{rad}}} 
 +  \Langle\AV^{\sharp,\mbox{\tiny{s,1}}}_{\mbox{\tiny{rad}}}\Rangle_{\qV} 
 \right) \cdot\nabla_{\qV}  \Psi_0.
 \label{eq:ERWINeqnNEWfirstORDER}
\end{equation}
 We have arrived at a traditional first-order formulation of time-dependent perturbation theory of the Schr\"odinger
equation of hydrogen in an external electromagnetic radiation field, albeit in the Schr\"odinger picture, while
textbooks usually present the unitarily equivalent formulation in the interaction picture. 
 Since this problem has been sorted out in great detail, it suffices to summarize the established facts;
and since the two ``pictures'' are unitarily equivalent and we here consider only first-order theory,
we may as well stay in the Schr\"odinger picture and describe the results.

 We now restrict the further discussion to the situation in which the time-evolved wave function $\Psi(t,\qV)$
remains in the negative energy subspace of the usual hydrogen Hamiltonian 
$H_{\mbox{\tiny{hyd}}} = -\frac{\hbar^2}{2\mEL}\Delta_\qV - \frac{e^2}{|\qV|}$.

\medskip
\noindent
\textbf{Remark}:
\emph{We should be prepared to find that such a scenario is an oversimplification, and that a positive energy admixture may be unavoidable,
as suggested by Dirac's 1927 calculations known as ``Fermi's Golden Rule.''
 In this sense, we are not assuming that it is always possible to describe the electron wave function dynamics in the 
negative energy subspace of $H_{\mbox{\tiny{hyd}}}$, but that we are \emph{restricting} the discussion to those situations when it is.
 It should be a valid assumption if the atom is initially in its ground state and the incoming wave has a frequency less than the 
ionization energy $/\hbar$.}
\medskip
 
 Technically speaking, if $\Pi_{\ul{E},\ol{E}}$ denotes the projection on the subspace of Hilbert space $L^2(\Rset^3)$
such that $\langle \Psi, H_{\mbox{\tiny{{hyd}}}}\Psi\rangle \in [\ul{E},\ol{E}]$ whenever
$\Psi = \Pi_{\ul{E},\ol{E}}\Psi$, 
then what we assume is that to first order in perturbation theory included, $\Psi = \Pi_{E_1,0}\Psi$,
 where $E_1$ is the ground state energy for $H_{\mbox{\tiny{{hyd}}}}$.
 By the completeness of the hydrogen wave eigenfunctions in this subspace we then can expand the solution, to first order included, thus: 
$\Psi(t,\qV) = \Psi_0(t,\qV) + \Psi_1(t,\qV)$, with
\begin{equation}
\Psi_1(t,\qV) = 
\sum_{n\in\Nset} e^{-i E_n t/\hbar}\sum_{\ell=0}^{n-1}\sum_{m=0}^\ell\sum_{\varsigma\in\pm}
 c^{}_{n,\ell,m,\varsigma}(t)\psi^{}_{n,\ell,m,\varsigma}(\qV),
\label{eq:PSIboundGENERALagain}
\end{equation}
where $c^{}_{n,\ell,m,\varsigma}(0)=0$.
\newpage

 Inserting this expansion into \eqref{eq:ERWINeqnNEWfirstORDER}, then multiplying by the real
$\psi^{}_{n',\ell',m',\varsigma'}(\qV)$ and integrating over $\Rset^3$ in $\qV$, 
and abbreviating
$\AV^{\sharp,\mbox{\tiny{f}}}_{\mbox{\tiny{rad}}} 
 +  \Langle\AV^{\sharp,\mbox{\tiny{s,1}}}_{\mbox{\tiny{rad}}}\Rangle_{\qV} 
 =: \AV^{\sharp,\mbox{\tiny{fs1}}}_{\mbox{\tiny{rad}}}$,
we obtain
\begin{equation}
e^{-i E_{n'} t/\hbar} i\hbar \frac{d}{dt} c^{}_{n',\ell',m',\varsigma'}(t)
= 
- i\hbar \tfrac{\eEL}{\mEL c} e^{-i E_{2} t/\hbar}
\big\langle\psi^{}_{n',\ell',m',\varsigma'}\big|\AV^{\sharp,\mbox{\tiny{fs1}}}_{\mbox{\tiny{rad}}}\cdot\nabla_{\qV}\big|\psi^{}_{2,1,m,+}\big\rangle,
\label{cCOEFFodes}
\end{equation}   
where the angular brackets are the usual Dirac notation.
 Since $H_{\mbox{\tiny{int}}}(t,\qV)$ is given we can integrate \eqref{cCOEFFodes} with vanishing initial data, 
obtaining 
\begin{equation}
c^{}_{n',\ell',m',\varsigma'}(t)
= 
- \tfrac{\eEL}{\mEL c} \int_0^t e^{-i( E_{2}-E_{n})\tau/\hbar}\big\langle\psi^{}_{n',\ell',m',\varsigma'}
\big|\AV^{\sharp,\mbox{\tiny{fs1}}}_{\mbox{\tiny{rad}}}\cdot\nabla_{\qV}\big|\psi^{}_{2,1,m,+}\big\rangle {d\tau} .
\label{cCOEFFs}
\end{equation} 
 We next discuss the contributions from the incoming (free) and the outgoing (sourced) $\sharp$-fields separately.
 Thus we write 
$c^{}_{n',\ell',m',\varsigma'}(t) = c^{\mbox{\tiny{f}}}_{n',\ell',m',\varsigma'}(t) + c^{\mbox{\tiny{s}}}_{n',\ell',m',\varsigma'}(t)$,
in self-explanatory notation.

\emph{The source-free contribution}.

 For the incoming Gaussian beam pulse, the $z$-dependent factor of
$\AV^{\sharp,\mbox{\tiny{f}}}_{\mbox{\tiny{rad}}}$
is to very good approximation given by $\exp\big(-\frac{1}{2\sigma_z^2}(z-z_0-ct)^2\big)\cos\big(\frac1c\omega_{2,1}(z-z_0-ct)\big)$,
where $z_0$ is the center of the pulse at time $t=0$, and $\omega_{2,1} = (E_2 - E_1)/\hbar$. 
 So it is clear that many coefficients $c^{\mbox{\tiny{f}}}_{n',\ell',m',\varsigma'}(t)$ will rapidly (on our everyday time scale)
transit from $0$ to some generally nonzero complex final value $c^{\mbox{\tiny{f}}}_{n',\ell',m',\varsigma'}(\infty)$ as $t\to\infty$;
some may vanish identically, though, due to symmetries.

 Now, writing the cosine as real part of the exponential of $i\big(\frac1c\omega_{2,1}(z-z_0-ct)\big)$,
 we see that the interaction Hamiltonian is a sum of terms with factors $\exp(\pm i \omega_{2,1}t)$ and
$\exp(\pm 2 i \omega_{2,1}t)$, and a non-oscillating (w.r.t. $t$) term. 
 Under the integral the term $\propto \exp( i \omega_{2,1}\tau)$ will cancel the factor $\exp(- i\omega_{2,n'}\tau)$ 
at r.h.s.\eqref{cCOEFFs} iff $n'=1$, and in this case that integral will have no time-oscillatory terms left.
 All other coefficients $c^{\mbox{\tiny{f}}}_{n',\ell',m',\varsigma'}(t)$ are given by integrals over oscillatory functions 
with one of the Rydberg frequencies of hydrogen (in particular, the $n'=2$ terms have the Lyman-$\alpha$ frequency),
or frequencies which are higher than the ionization frequency (which should make the smallest contributions).
 Thus it can be expected that $|c^{\mbox{\tiny{f}}}_{1,0,0,+}(\infty)|$ is either identically 0 by symmetry, or the largest magnitude of 
the coefficients, while all other coefficients remain smaller in magnitude than these.

 In the following we report our results for the $c^{\mbox{\tiny{f}}}_{n',\ell',m',\varsigma'}(t)$ as given by
\eqref{cCOEFFs} with $H^{(1)}_{\mbox{\tiny{int}}}(\tau,\,.\,)$ computed from the stipulated Gaussian beam pulse. 
 For convenience we restrict the study to coefficients with small enough $n'$ values so that the pertinent hydrogen eigenfunctions
are exponentially decaying in $r$ with scale $n'a_{\mbox{\tiny{B}}}$ that is smaller than the lateral spread of the beam pulse.
 In 
$\big\langle\psi^{}_{n',\ell',m',\varsigma'}\big|\AV^{\sharp,\mbox{\tiny{f}}}_{\mbox{\tiny{rad}}}\cdot\nabla_{\qV}\big|\psi^{}_{2,1,m,+}\big\rangle$
we then can approximately set $\AV^{\sharp,\mbox{\tiny{f}}}_{\mbox{\tiny{rad}}}
=  \frac{e\mEL c}{\hbar}\eV_x\exp\big(-\frac{1}{2\sigma_z^2}(z-z_0-ct)^2\big)\cos\big(\frac1c\omega_{2,1}(z-z_0-ct)\big)$,
a plane-wave pulse traveling in the $z$-direction, satisfying the Coulomb gauge, and producing divergence-free electric and magnetic 
radiation pulse fields; 
viz., we have
$\AV^{\sharp,\mbox{\tiny{f}}}_{\mbox{\tiny{rad}}}\cdot\nabla_{\qV} = 
 \frac{e\mEL c}{\hbar}\exp\big(-\frac{1}{2\sigma_z^2}(z-z_0-ct)^2\big)\cos\big(\frac1c\omega_{2,1}(z-z_0-ct)\big)\partial_{q_x}$.

 Taking $n'=1,\ell'=0=m',\varsigma=+$, the bra vector is the ground state,
and its wave function is spherically symmetric. 
 Then we obtain the following results for when $m=0$ versus when $m=1$.

\smallskip
\noindent
($m=0$) 
 For the exited initial state, $\partial_{q_x}\psi^{}_{2,1,0,+}\propto \sin\vartheta \cos\varphi$, so the $\varphi$-integration yields
$\big\langle\psi^{}_{1,0,0,+}\big|\AV^{\sharp,\mbox{\tiny{f}}}_{\mbox{\tiny{rad}}}\cdot\nabla_{\qV}\big|\psi^{}_{2,1,0,+}\big\rangle =0$.
 Thus to first-order perturbation theory this initial state does not excite the ground state amplitude when the
atom is traversed by the stipulated Gaussian beam, viz. $c^{\mbox{\tiny{f}}}_{1,0,0,+}(t)=0\forall\, t$.
 If the $O(\eps^2)$ terms in $H_{\mbox{\tiny{int}}}(t,\qV)$ due to the incoming radiation $\sharp$-field would have been retained,
$c^{\mbox{\tiny{f}}}_{1,0,0,+}(t)$ would be generally nonzero, but this is second order in perturbation theory.

\smallskip
\noindent
($m=1$)   When the exited initial state is $\psi^{}_{2,1,1,+}$,
its partial $q_x$-derivative is a sum of a spherically symmetric function $\times (1+ q_x^2/r)$, 
and this implies that the integrand
$\big\langle\psi^{}_{1,0,0,+}\big|\AV^{\sharp,\mbox{\tiny{f}}}_{\mbox{\tiny{rad}}}\cdot\nabla_{\qV}\big|\psi^{}_{2,1,1,+}\big\rangle$
yields a generally nonzero result. 
 The time-dependence can be conveniently evaluated with the help of a computer, but 
$\lim_{t\to\infty}c_{1,0,0,+}^{}(t)$ can be computed explicitly to excellent approximation if one realizes that the 
lower limit of the time integral can be extended from $0$ to $-\infty$ with a superexponentially small error. 
 In that case, by an application of the Tonelli-Fubini theorem, one can carry out the time integration over $\tau$ first,
which becomes just the Fourier transform of the plane-wave pulse with conjugate variables $\tau\mapsto \omega$. 
 We use that
\begin{equation}\label{FourierGAUSSa}
\int_{\Rset} e^{-i\omega_{2,1}\tau}e^{-\frac{1}{2\sigma_z^2}(z-z_0-c\tau)^2}e^{ - i(\frac1c\omega_{2,1}(z-z_0-c\tau)}d\tau
= \tfrac{\sqrt{2\pi} \sigma_z}{c} e^{ - i \frac1c\omega_{2,1}(z-z_0)} 
\end{equation}
and that a Gaussian maps to a Gaussian, so that 
\begin{equation}\label{FourierGAUSSb}
\int_{\Rset} e^{-i\omega_{2,1}\tau}e^{-\frac{1}{2\sigma_z^2}(z-z_0-c\tau)^2}e^{ i(\frac1c\omega_{2,1}(z-z_0-c\tau)}d\tau
=  
\tfrac{\sqrt{2\pi} \sigma_z}{c}  e^{- 2\frac{\sigma_z^2 }{c^2}\omega_{2,1}^2} e^{ - i \frac1c\omega_{2,1}(z-z_0)} 
\end{equation}
and the Gaussian at r.h.s.\eqref{FourierGAUSSb}, for our choice of data, is $\approx\exp(- 10^6)$, hence zero for all practical purposes
relative to the contribution from r.h.s.\eqref{FourierGAUSSa}.

 Similarly it follows that all other amplitude coefficients are practically as good as zero, as long as the plane-wave pulse approximation
is valid; at larger $n'$ the amplitude coefficients cannot be computed with the plane-wave pulse approximation, yet after taking the
lateral cutoff into account the remaining time integration argument applies with minor adjustments and again produces essentially
vanishing magnitudes.

\emph{The sourced contribution (Born-type approximation)}.

 The coefficients $c^{\mbox{\tiny{s}}}_{n',\ell',m',\varsigma'}(t)$ are more difficult to evaluate, and our results are rather preliminary
at this point. 
 A few insights are available.

 First of all, even though the velocity field generated by the incoming Gaussian plane-wave pulse always points along $\eV_x$,
the sourced $\sharp$-field vector potential, in Born-type approximation, points in (almost) all directions.
 This essentially spherical emission consists of a retarded Lien\'ard--Wiechert type part that leaves the realm of the atomic 
region as quickly as the plane-wave pulse passes through the atom, and
a contribution from outside the light cone that lingers a bit longer; yet, since in Born-type approximation the velocity field in the
atomic region is only active in transient while the incoming Gaussian beam pulse passes through the atom, also the contribution 
from outside the light cone is very brief in time.

 Next, the velocity field contribution to the sourced $\sharp$-field vector potential in Born-type approximation 
oscillates with Lyman-$\alpha$ frequency, but the contribution from $\QV_q(t^{\mathrm{ret}}(t,\sV;\qV))$ in
the numerator of r.h.s.\eqref{eq:AsharpRADfixpt} may cause a spectrum of frequencies. 
 
 The upshot is: We may expect that the emission of the sourced $\sharp$-field radiation in first order of perturbation theory,
and in first Born-type approximation, makes a non-zero contribution
$c^{\mbox{\tiny{s}}}_{1,0,0,+}(t)$ to the ground state amplitude, in addition to $c^{\mbox{\tiny{f}}}_{1,0,0,+}(t)$.
 Moreover, one should be prepared to find that this emission process will also contribute to other amplitudes; in particular,
it may contribute to the de-activation of the excited initial state, while contributions to other eigenstate amplitudes should
be vanishingly small for similar numerical reasons as offered above already.
 To sort this out rigorously will require a detailed and lengthy evaluation of the retarded contributions and of those from outside the
light cone. 
 This, unfortunately,  
must be relegated to the to-do list as a high-priority item. 

\smallskip
\noindent
\textbf{Conclusion}: \emph{In the present model, first order perturbation theory predicts that 
after the passing through of an electromagnetic radiation beam pulse with Lyman-$\alpha$ frequency,
a hydrogen atom that initially is in its first excited state with $n=2$, will have responded as follows:}

$\bullet$ \emph{If $\ell=1$ and $m=1$, the atom will essentially be in a superposition of this state}

\ \ \ \emph{and the ground state};

$\bullet\;$ \emph{If $\ell=1$ and $m=0$, or if $\ell=0$, the atom will remain in that state}.

\noindent
\emph{In any of these situations, there will be a negligible admixtures of other eigenstates.}
\medskip

\noindent
\textbf{Remark}:
\emph{Since $\AV^{\sharp,\mbox{\tiny{f}}}_{\mbox{\tiny{rad}}}$ is independent of ${\qV}$, 
we have 
$$
\big\langle\psi^{}_{1,0,0,+}\big|\AV^{\sharp,\mbox{\tiny{f}}}_{\mbox{\tiny{rad}}}\cdot\nabla_{\qV}\big|\psi^{}_{2,1,1,+}\big\rangle
=
- \big\langle\psi^{}_{2,1,1,+}\big|\AV^{\sharp,\mbox{\tiny{f}}}_{\mbox{\tiny{rad}}}\cdot\nabla_{\qV}\big|\psi^{}_{1,0,0,+}\big\rangle,
$$
and so we can conclude that the incoming Lyman-$\alpha$ frequency wave, after passing through the atom that initially is in its
ground state, will leave the atom in a superposition of the ground state and the first excited level state $(2,1,1)$.}

\newpage

\medskip
\noindent
\textbf{Remark}:
\emph{Since  to first order in perturbation theory the incoming Gaussian radiation beam pulse generates 
a flash of outgoing $\sharp$-field radiation already 
at the level of the $\sharp$-field equations \emph{without} the mediation of the wave function of the atom, we now have
the apparently paradoxical situation that to first order in perturbation theory the atom, when initially in the first excited state
with $m=0$, will remain in that state and yet a flash of radiation is emitted. 
 However, the first excited $m=0$ state is spherically symmetric, and so upon taking the quantum-mechanical expected value
this outgoing radiation $\sharp$-field flash will average to zero, compatible with the empirical results. 
 When the atom is initially in a not spherically symmetric first excited $m=1$ state, then the quantum-mechanical average 
to first order in perturbation theory will not vanish.}
\medskip

We will compute the $t$-dependence of $|c^{\mbox{\tiny{f}}}_{1,0,0,+}(t)|^2$ numerically for some illustrative parameter values, 
when the initial state of the atom is $\Psi^{}_0 = \psi^{}_{2,1,1,+}$; see Fig. TBA.

\bigskip

\vfill\vfill
 This concludes our discussion of the first-order response of the hydrogen atom to the incoming radiation field pulse
which, as we have shown, reduces to the same type of problem as in orthodox quantum mechanics, already studied by Schr\"odinger in 1926,
except that he used monochromatic plane waves and not a Gaussian plane-wave pulse. 
 But there is such a large literature on Gaussian beam pulses, e.g. \cite{EncyclopediaOPTICS} and references therein.

  Next we address the more subtle issue of what happens as a result of the feedback of the
emitted radiation $\sharp$-fields into the Hamiltonian.

\newpage

\smallskip
\noindent\textbf{3.5.1.vii: On the transition to the ground state.}
\smallskip

 In this section we assume the initial state of the atom was the first excited state $(n=2,\ell=1,m=1,\varsigma=+)$, and
that the stipulated Gaussian radiation beam pulse with Lyman-$\alpha$ frequency has moved through the atom already.
 Since to first order in perturbation theory our model exhibits all the contributions that also the usual Schr\"odinger-type
perturbation theory features, plus some additional contributions, we may take guidance from Fermi's Golden Rule and conclude
that the atom will have ended up in a superposition of this state and the ground state, plus perhaps some admixture of a 
continuum state --- which we ignored in our perturbative computations, and which we continue to ignore for simplicity also
in this subsection.
 The question now is whether the transition to the ground state indeed happens in our theory.
 To find out one has to compute in higher-order of perturbation theory as a consequence of the interaction Hamiltonian caused by
 the emission, or without perturbation arguments altogether. 
 In the following we give a non-perturbative argument for why we expect that this transition to the ground state indeed
happens in this theory.


 Our Hamiltonian is the generator of the unitary dynamics for ${\Psi}$,
 but its expected value
$\left\langle H_{\mbox{\tiny{{hyd}}}} + H_{\mbox{\tiny{{int}}}} \right\rangle + \big\langle E^\sharp_{\mbox{\tiny{rad}}}(t,\qV)\big\rangle$
does not seem to have the significance of furnishing a conserved energy for the system as a whole.
 If this was a conserved quantity, or if one had a conserved quantity similar to \eqref{eq:SMenergyFctl}, say 
$\left\langle H_{\mbox{\tiny{{hyd}}}} + H_{\mbox{\tiny{{int}}}} \right\rangle 
+  \frac{1}{8\pi}\int_{\Rset^3}\!\bigl(\abs{\EV_{\mathrm{el}}^{\mbox{\tiny{rad}}}(t,\sV)}^2 + 
\abs{\BV_{\mathrm{el}}^{\mbox{\tiny{rad}}}(t,\sV)}^2 \bigr)d^3\!s$,
then one could feel inspired by the reasoning in Schr\"odinger--Maxwell theory (recall its discussion in section \ref{sec:Erwin})
and try to show that the field energy necessarily increases, which then must come at the expense of the 
first two terms, and if one can also show that the expected interaction Hamiltonian goes to zero as time increases, 
then one has the basis for showing that the atom has made a transition to a lower energy state, and when starting
in the first excited $\ell=1$ state, it would have to be the ground state. 

 On the positive side, we can offer three potentially useful observations.


 First of all,  $\frac{1}{8\pi}\int_{\Rset^3}\!\big(\abs{\EV_{\mathrm{el}}^{\mbox{\tiny{rad}}}(t,\sV)}^2 + 
\abs{\BV_{\mathrm{el}}^{\mbox{\tiny{rad}}}(t,\sV)}^2 \big)d^3\!s$ is a lower bound to $\big\langle E^\sharp_{\mbox{\tiny{rad}}}(t,\qV)\big\rangle$,
by Jensen's inequality. 
 Initially these two agree, for the incoming vacuum radiation beam is $\qV$-independent.
 Thus initially the energy of the expected electromagnetic radiation fields is just the field energy of the incoming radiation,
and the emission will increase it.
 So also $\langle  H_{\mbox{\tiny{rad}}} \rangle$ increases from its initial value, even more so because the emitted $\sharp$-fields do
depend on $\qV$.

 Second, in our model the Hamiltonian $H = H_{\mbox{\tiny{{hyd}}}} + H_{\mbox{\tiny{{int}}}}+ H_{\mbox{\tiny{{rad}}}}$, so
\begin{equation}\label{ddtHhyd}
\tfrac{\drm \ }{\drm t} 
\left\langle H_{\mbox{\tiny{{hyd}}}} \right\rangle 
= \left\langle \tfrac{1}{i\hbar}[H_{\mbox{\tiny{{hyd}}}}, (H_{\mbox{\tiny{{int}}}} + H_{\mbox{\tiny{rad}}})]\right\rangle,
\end{equation}
and r.h.s.(\ref{ddtHhyd}) is not manifestly zero, clearly.
 Of course, this observation does not imply that $\langle H_{\mbox{\tiny{hyd}}} \rangle$ will decrease when the atom radiates,
but the atom could not transit from an excited state to the ground state if 
$\langle H_{\mbox{\tiny{hyd}}} \rangle$ was conserved, as in Schr\"odinger's equation (\ref{eq:ERWINeqnMatterWaveBOhydrogen}) 
where it has a conserved expected value
$\langle H_{\mbox{\tiny{hyd}}} \rangle$ as per (\ref{eq:ERWINeqnMatterWaveBOhydrogen}).

 What we can conclude, however, is that \emph{if} $\langle H_{\mbox{\tiny{hyd}}} \rangle$ \emph{decreases as a consequence of 
the emission of electromagnetic radiation then}, since $H_{\mbox{\tiny{{hyd}}}}$ is bounded below, 
\emph{the evolution of $\Psi(t,\qV)$, starting in our initial state, will inevitably approach the ground state asymptotically
in time}. 
 Indeed, note that by our hypothesis that $\langle H_{\mbox{\tiny{hyd}}} \rangle$ decreases as a consequence of the emission
of radiation,  $\langle H_{\mbox{\tiny{hyd}}} \rangle$ cannot settle down to a value between the ground state and the
first excited state, for then $\Psi$ would have to be in a superposition of eigenstates, which inevitably would lead to the
emission of radiation (as shown above) and to the further decrease of $\langle H_{\mbox{\tiny{hyd}}} \rangle$.

 Third, for it to be possible that $\langle H_{\mbox{\tiny{hyd}}} \rangle$ approaches its ground state value asymptotically,
(\ref{ddtHhyd}) would have to approach zero asymptotically.  
 We will now see that this scenario is compatible with the dynamical equations.
 
 As shown earlier, if the atom is in its ground state $\Psi$, then the associated velocity field $\vV$ vanishes. 
 The radiation $\sharp$-field still contributes to the velocity field, but despite still lingering $\QV_q(\tau)$ motions
the main dynamics of the $\sharp$-field is an outward moving flash of radiation --- at least this is suggested by first
order perturbation theory. 
 Thus the $\sharp$-field equations should become $\qV$-independent and 
the emission of radiation inevitably would fade away, compatible with $\langle H_{\mbox{\tiny{hyd}}} \rangle$ 
ending its decrease. 
 Although the already emitted $\sharp$-field radiation is time- and space-dependent, 
since this electromagnetic radiation leaves the Bohr-radius-sized region of the atom  at the speed of light, 
it very soon after the essential ending of the emission process will become effectively $\qV$-independent.
 The large amplitude region of the $\sharp$-field radiation will be concentrated around a spherical shell of radius $ct$
away from the origin, and the integrals $E^\sharp_{\mbox{\tiny{rad}}}$ and $\PV^\sharp_{\mbox{\tiny{{el}}}}$ should become essentially
independent of $\qV$ for $\qV$ in the atomic vicinity of the proton (origin), and exponentially (in $|\qV|$) suppressed 
in the expected value functional. 
 Moreover, $E^\sharp_{\mbox{\tiny{rad}}}$ and $\PV^\sharp_{\mbox{\tiny{{el}}}}$ should approach
time-independency because of the energy-momentum conservation for free Maxwell radiation fields (which is what $\qV$-independent
$\sharp$-fields are).
 As such, the expected commutator $\left\langle \tfrac{1}{i\hbar}[H_{\mbox{\tiny{{hyd}}}}, 
(H_{\mbox{\tiny{{int}}}} + H_{\mbox{\tiny{rad}}})]\right\rangle\to 0$, and
$\langle H_{\mbox{\tiny{hyd}}} \rangle$ approaches a constant.

 With the just described scenario the radiation Hamiltonian becomes effectively a constant 
number that is being added to $H_{\mbox{\tiny{hyd}}}$.
 But this Hamiltonian has the same wave eigenfunctions as the initial Hamiltonian. 
 Thus, the assumption of the atom settling down to the ground state wave function of the traditional hydrogen 
Hamiltonian, accompanied by the dynamical emission-of-``a-flash-of''-radiation scenario, at least is well compatible with our
dynamical quantum-mechanical equations.
\newpage

\noindent\textbf{3.5.1.viii: Incorporating electron spin and static $\EV^{\mbox{\tiny{ext}}}_{\mbox{\tiny{lab}}}$ and $\BV^{\mbox{\tiny{ext}}}_{\mbox{\tiny{lab}}}$}
\smallskip

 Everything we discussed in the previous subsection generalizes to the case of hydrogen when the $\sharp$-fields include
laboratory-generated static $\EV^{\mbox{\tiny{ext}}}_{\mbox{\tiny{lab}}}$ and $\BV^{\mbox{\tiny{ext}}}_{\mbox{\tiny{lab}}}$.
 Of course, since we have neglected electron spin so far,  the \emph{anomalous Zeeman effect} of hydrogen would not show up. 

 We can easily generalize all this to an electron with spin by switching from Schr\"odinger's to Pauli's equation.
 Thus, $\Psi(t,{\qV})$ becomes a two-component Pauli spinor, and the Schr\"odinger-type equation 
(\ref{ERWINeqnHYDROcoupleSHARP}) is replaced by a Pauli-type equation
\begin{equation}
\left(i \hbar \partial_t - E^\sharp (t,{\qV})\right)\Psi(t,{\qV}) 
= \tfrac{1}{2\mEL}\left(\boldsymbol\sigma\cdot\left(-i \hbar \nabla_{\qV} -
 \PV^\sharp_{\mathrm{el}}(t,{\qV}) \right)\right)^2 \Psi(t,{\qV}).
 \label{eq:PAULIeqn}
\end{equation}
 For a spinor, the density $\varrho = \Psi^\dagger\Psi = |\Psi_+|^2 + |\Psi_-|^2$, where the suffix ${}_\pm$ indicates the 
upper and lower components of the Pauli spinor, and the probability current density 
 \begin{equation}
\JV (t,\qV) 
= 
\Im \left(\Psi^\dagger\tfrac{1}{\mEL} \bigl(\hbar \nabla_{\qV} - i\PV^\sharp_{\mbox{\tiny{{el}}}}\bigr) \Psi\right)(t,\qV)
+ 
\tfrac{1}{2\mEL} \hbar \nabla_{\qV} \times \bigl(\Psi^\dagger\boldsymbol\sigma\Psi\bigr)(t,\qV);
 \label{eq:PAULIspinorJ}
\end{equation}
the curl term is optional, yet suggested by Dirac's equation (see \cite{BoHi}). 
 The velocity field is again defined by $\JV = \varrho \vV$.
 One now gets the correct nonrelativistic hydrogen spectra, including the anomalous Zeeman effect and the Stark effect. 
 Relativistic corrections, such as spin-orbit coupling, are of course not included.

\section{Systems with multiple electrons and nuclei}\label{manyatoms}

 Everything we discussed in the previous section on hydrogen radiation
 generalizes to the case of a many-electron atom coupled with the $\sharp$-fields.
 Since the spectrum of a many-electron atom won't come out right without electron spin, even in the absence of an applied laboratory-generated 
magnetic field, we have to replace the $N$-electron Schr\"odinger-type equation (\ref{eq:ERWINeqnNEW})
with the $N$-electron Pauli-type equation
\begin{equation}
\boxed{\left(i \hbar \partial_t - E^\sharp (t,\qVv)\right)\Psi(t,\qVv) 
= {\textstyle\sum\limits_{n=1}^N}
\tfrac{1}{2\mEL}\left(\boldsymbol{\sigma}_n\cdot\left(-i \hbar \nabla_{\qV_n} - \PV^\sharp_n(t,\qVv) \right)\right)^2 \Psi(t,\qVv)},
 \label{eq:PAULIeqnNEW}
\end{equation}
with $\Psi(t,\qVv)$ an $N$-body Pauli spinor wave function that is antisymmetric under the permutation group $S_N$. 
 
 For the bound-state spectrum in the presence of laboratory-generated static external fields, the de facto 
Hamiltonian extracted from this equation is a sum of  the many-electron Hamiltonian at l.h.s.(\ref{eq:ErwinEQstationaryRELext})
plus Pauli terms $- \frac{e \hbar}{\mEL c}\boldsymbol\sigma_j\cdot \BV^{\mbox{\tiny{ext}}}_{\mbox{\tiny{lab}}}(\qV_j)$.
 By `de facto' we mean that an irrelevant additive constant has been subtracted from $H$.

 In the general dynamical situation, also an interaction Hamiltonian $H_{\mbox{\tiny{int}}}$ that is a sum of similar expressions as
before (one for each electron) emerges from (\ref{eq:PAULIeqnNEW}), as well as the by now familiar radiation Hamiltonian.

 Next we can generalize this generalization even further, to the many-nuclei problem of molecules and solids, i.e. 
as long as we employ the Born--Oppenheimer approximation.
 In that case the generalization from a many-electron atom to a system of many electrons and many nuclei is entirely straightforward. 
 Indeed, the many-electron Pauli equation (\ref{eq:PAULIeqnNEW}) governs the evolution of the 
$N$-electron wave function with spin unchanged in its appearance. 
 What changes is that the $\PV_n^f$ and $E^\sharp$ are now computed from solutions to the $\sharp$-field equations in which the source term
of the constraint equation (\ref{eq:aMdivEsharp}) includes $K$ nuclei rather than only one, i.e. (\ref{eq:aMdivEsharp}) changes to
\begin{alignat}{1}\hspace{-.3truecm}
\nabla_\sV\cdot\EV^\sharp &=\label{eq:aMdivEsharpMANYnuclei} 
4\pi e\Bigl({\textstyle\sum\limits_{k=1}^K}Z_k \delta^{(a)}_{\qV_k^+}(\sV) - {\textstyle\sum\limits_{n=1}^N} \delta^{(a)}_{\qV_n}(\sV) \Bigr),
\end{alignat}
where the positions of the nuclei are distinguished from those of the electrons by the superscript ${}^+$.
 The energy of the pertinent electrostatic $\sharp$-field solution, provided no two charged balls of radius $a$ overlap, is 
\begin{alignat}{1}\hspace{-.3truecm}
\frac{1}{8\pi} \int_{\Rset^3} \big|\EV^\sharp(\sV;\qVv |\qVv^+)\big|^2 \drm^3s
 =  \label{eq:HAMfromFIELDenergyMANYz}
& 
\\ \notag
E_{\mbox{\tiny{self}}} + {\textstyle\sum\sum\limits_{\hskip-.7truecm  1 \leq  j < k \leq K} } \frac{z_jZ_ke^2}{|\qV_j^+-\qV_k^+|} 
& - {\textstyle{\sum\limits_{k=1}^K\sum\limits_{n=1}^N}} \frac{Z_k e^2}{|\qV_n- \qV_k^+|} 
+ {\textstyle{\sum\sum\limits_{\hskip-.7truecm  1 \leq  j < k \leq N}}} \frac{e^2}{|\qV_j-\qV_k|},
\end{alignat}
where $E_{\mbox{\tiny{self}}} =  \frac35\frac{e^2 }{a} \Big( N + {\textstyle{\sum\limits_{k=1}^K}}Z_k^2 \Big)$ is a constant; 
for smaller distances the Coulomb interactions are regularized.
 In the absence of laboratory-generated static external fields this is the correct Schr\"odinger potential of a many-nuclei many-electron
system in Born--Oppenheimer approximation \cite{MaxOppi}. 
 Laboratory-generated static external fields can be included also; we skip the calculation of the pertinent effective Hamiltonian.
 
 We close this brief section by noting that so far the locations of the $\qV^+$ variables have been treated as given parameters, yet
of course most arbitrary choices will not result in a physically accurate mathematical problem. 
 Born and Oppenheimer \cite{MaxOppi} proposed to treat the quantum-mechanical expected value of \eqref{eq:HAMfromFIELDenergyMANYz}
as a stand-in for the Hamiltonian of a classical Newtonian $K$-body problem. 
 The ground state problem, for instance, would then require first minimization of the quantum-mechanical Hamiltonian, given the 
$\qV_+$ variables, followed by the minimization of it's expected value w.r.t. those $\qV_+$ variables.
 There is a huge literature on this; see also \cite{MaxKerson}.

\newpage
\section{Photons}

 Concerning the dynamics of an excited atom (and for simplicity, say, hydrogen) coupled to the $\sharp$-fields, even the most favorable outcome 
in the model so far describes a scenario in which the atom transits into the ground state while emitting a flash of electromagnetic 
$\sharp$-field radiation which has a quantum-mechanical expected value that corresponds to the overall empirical emission of a 
large number of independently radiating hydrogen atoms when registered far away from the region that these hydrogen atoms occupy. 
 This expected value of radiation is an essentially spherical shell of radius $ct$ of Maxwell fields. 
 As we have shown, the flash of electromagnetic $\sharp$-field radiation itself will, for each generic position $\qV$ of the electron, 
consist of a similar spherical shell, centered on $\qV$ (significant only for $\qV$ in the Bohr-radius size vicinity of the nucleus),
plus a lingering contribution from ``outside the light cone.''
 Clearly this is not what seems to happen in experiments: an atom which transits from an excited to its ground state seems to do so
under the emission of photons, which get registered in localized photon detectors. 
 The radiation $\sharp$-field, being spread out, cannot in itself represent such a localized event. 

 However, the following, notationally trivial but conceptually radical change of perspective brings the photon into the model.

 Thus, we note that the $\sharp$-fields, depending in addition to their variables $t$ and $\sV$ also 
on the configuration space variable $\qV$ of the hydrogen's electron, or more generally on the generic configuration 
space vector $\qVv$ of $N$ electrons in case a many electron system is considered,
are more reminiscent of quantum-mechanical many-body wave functions than of a classical field. 
 It is therefore very suggestive to contemplate that the variable $\sV$ in the $\sharp$-fields and their $\sharp$-field equations
does not represent a generic point in physical space but instead represents the generic position of a photon.
 In the next subsection we pursue this lead.

\subsection{Systems with a single photon}

 To emphasize this radical change of perspective, that the emitted electromagnetic $\sharp$-field wave now is re-interpreted
as a kind of quantum-mechanical wave function, we replace $\sV$ in the $\sharp$-fields and their equations
by $\qV_{\mathrm{ph}}$, and we set $\qV\mapsto\qV_{\mathrm{el}}$ (respectively, $\qVv\mapsto\qVv_{\mathrm{el}}$)
to distinguish the two types of position variables clearly.

 Next, comparing the $\sharp$-field wave equations and the Schr\"odinger equation, one is struck by the fact that
the feedback from $\sharp$-fields into the Schr\"odinger or Pauli equation is through bilinear (and square of bilinear) functionals of
the $\sharp$-fields, but the  Schr\"odinger or Pauli $\Psi$ enters the $\sharp$-field equations via $\vV$, computed
from $\Psi$ (the ratio of two bilinear expressions in $\Psi$).
 Yet if we recall that $\JVv=\varrho \vVv$ with each three-dimensional component $\JV$ given in (\ref{eq:PAULIspinorJ}),
and (following Heinrich Weber \cite{Weber}; see endnote 114 in \cite{KTZphoton}) set
$\EV^\sharp(t,\qV_{\mathrm{ph}};\qVv_{\mathrm{el}})+i\BV^\sharp(t,\qV_{\mathrm{ph}};\qVv_{\mathrm{el}})=:
  \eEL\hbar\boldsymbol{\Psi}(t,\qV_{\mathrm{ph}};\qVv_{\mathrm{el}})$, 
we obtain
\begin{alignat}{1}\hspace{-1truecm}
%
\Big[ 
i\hbar\partial_t+c\hbar\bigl(\nabla_{\qV_{\mathrm{ph}}}\times\bigr)\!
 +i\hbar \vVv(t,{\qVv_{\mathrm{el}}})\bcdot\nabla_{\qVv_{\mathrm{el}}}\!\Big]\boldsymbol{\Psi}(t,\qV_{\mathrm{ph}};\qVv_{\mathrm{el}}) 
 = \label{eq:MdotPSIsharp}
  4\pi \vVv(t,{\qVv_{\mathrm{el}}})\bcdot \delta^{(a)}_{{\qVv_{\mathrm{el}}}}(\qV_{\mathrm{ph}}),&
 \\
\hbar \nabla_{\qV_{\mathrm{ph}}}\bcdot \boldsymbol{\Psi}(t,\qV_{\mathrm{ph}};\qVv_{\mathrm{el}})
= \label{eq:MdivPSIsharp} 
 4\pi 
\Bigl(\delta^{(a)}_{\nullV}(\qV_{\mathrm{ph}}) - {\textstyle\sum\limits_n}\delta^{(a)}_{\qVv_{\mathrm{el},n}}(\qV_{\mathrm{ph}})
\!\Bigr)\!,&
\end{alignat}
where $\vVv\bcdot\delta^{(a)}_{\qVv_{\mathrm{el}}}$ is shorthand for $\sum_n \vV_n \delta^{(a)}_{\qVv_{\mathrm{el},n}}$.
 If we now multiply (\ref{eq:MdotPSIsharp}) 
by $\varrho$, and recall that $\varrho\vVv=\JVv$,
we exhibit a bilinear feedback from the $\Psi$ equation into the $\boldsymbol{\Psi}$ equations; thus
the coupled system of $\Psi$ and $\boldsymbol{\Psi}$ equations appears more on an equal footing.

 Yet when $\boldsymbol{\Psi}(t,\qV_{\mathrm{ph}};\qVv_{\mathrm{el}})$ lives on the joint configuration space for electron and photon,
it is very tempting to let oneself be inspired by the speculations of de Broglie, Born, and Bohm, and to think 
of $\boldsymbol{\Psi}(t,\qV_{\mathrm{ph}};\qVv_{\mathrm{el}})$ as a guiding field for the photon. 
 Thus we need also the guiding equation for the actual position of the photon in physical space.
 Let $\qV_{\mathrm{ph}}(t)$ be its position at time $t$. 
 Then it's suggestive in this semi-relativistic setting to postulate that the photon moves according to the guiding equation
\begin{alignat}{1}\label{eq:photonGUIDING}
\frac{\drm \qV_{\mathrm{ph}}(t)}{\drm t}= 
c\frac{\Im\left(\boldsymbol{\Psi}^*(t,\qV_{\mathrm{ph}};\qVv_{\mathrm{el}}(t))\times\boldsymbol{\Psi}(t,\qV_{\mathrm{ph}};\qVv_{\mathrm{el}}(t))\right)}
{\quad \boldsymbol{\Psi}^*(t,\qV_{\mathrm{ph}};\qVv_{\mathrm{el}}(t))\cdot\boldsymbol{\Psi}(t,\qV_{\mathrm{ph}};\qVv_{\mathrm{el}}(t))}
\Biggr|_{\qV_{\mathrm{ph}}=\qV_{\mathrm{ph}}(t)}.
\end{alignat}
 Note that the r.h.s. is homogeneous of degree 0, so even while $\int |\boldsymbol\Psi|^2\drm^3q_{\mathrm{ph}}\drm^3q_{\mathrm{el}}$ is
generally not conserved, a changing $\int |\boldsymbol\Psi|^2\drm^3q_{\mathrm{ph}}\drm^3q_{\mathrm{el}}$ does not affect the law of motion.
\smallskip

 Note furthermore that the magnitude of the guiding velocity field is generally less than $c$, but equals $c$ in plane wave solutions.
 So if this guiding law contains a grain of truth then photons would only appear to move roughly at the speed of light in experiments
where one can in good approximation assume there is a plane wave. 
 \smallskip

\noindent
\textbf{Remark}:
\emph{Einstein pondered a guiding field for photons (his ``quanta of light'') on \emph{physical spacetime}, obeying a relativistic field equation.
 In subsection \ref{sec:sharpFIELDS} we noted that the evaluation of the $\sharp$-fields with the actual electron position 
$\qV_{\mathrm{el}}(t)$ in place of the generic $\qV_{\mathrm{el}}$ turns the $\sharp$-fields into solutions of the classical
Maxwell--Lorentz field equations for point charges, viz. (\ref{eq:MdotEactual})--(\ref{eq:MdivBactual}). 
 In this sense it would seem that with the guiding equation (\ref{eq:photonGUIDING}) one
comes as close as one can get to realizing Einstein's surmise that the classical electromagnetic field guides the photons.}
\smallskip

 Our next step is to upgrade to a quantum mechanical model of radiating atoms in which 
a highly excited atoms may emit several photons while cascading down to its ground state. 
\newpage
\subsection{Systems with many photons}

 Photons are thought to not interact with each other directly but only with charged particles.
 In quantum electrodynamics this includes virtual electron-positron pairs, which effectively allow photon-photon
scattering without a real charged particle mediating the interaction; but electron-positron pair creation / annihilation is 
not part of the semi-relativistic so-called standard model of everyday matter, and not part of our purely quantum-mechanical model. 
 Therefore we will implement many photons in such a way that they do not interact with each other but 
only with the real (i.e. not virtual) charged particles of the model.

 There are a number of requirements which a generalization of our model to 
a system of equations for an atom in the presence of many photons needs to satisfy.
 First of all, since photons are spin $1$ bosons, their quantum-mechanical $L$-photon wave function $\PsiVL$ has to be permutation-symmetric. 
 Thus, the generalized $L$-photon $\sharp$-field $\PsiVL(t,\qVv_{\mathrm{ph}};\qVv_{\mathrm{el}})$ 
takes values in the closure of the $L$-fold symmetrized tensor products of
 single-photon $\boldsymbol{\Psi}(t,\qV_{\mathrm{ph}}^{\ell};\qVv_{\mathrm{el}})$ over $\ell =1,...,L$. 
 Second, the stationary states must produce the correct atomic (molecular, etc.) energy spectra.

 It is straightforward to verify that our single-photon $\sharp$-field equations (\ref{eq:MdotPSIsharp}), (\ref{eq:MdivPSIsharp}) 
(in Weber notation) are the single-photon special case of the following equations for the $L$-photon 
wave function $\PsiVL$, \emph{conditioned} on the generic $N$-electron configuration, 
\begin{alignat}{1}
&\hspace{-0.8truecm} 
\Bigl(i\hbar\partial_t+
c\hbar{\textstyle{\sum\limits_\ell}}\nabla_{\qV_{\mathrm{ph}}^\ell}\times_\ell\Bigr)\PsiVL(t,\qVv_{\mathrm{ph}};\qVv_{\mathrm{el}})  
+ i\hbar \left(\vVv(t,{\qVv_{\mathrm{el}}})\bcdot\nabla_{\qVv_{\mathrm{el}}}\right)\PsiVL(t,\qVv_{\mathrm{ph}};\qVv_{\mathrm{el}})
\notag \\
 \label{eq:MdotPSIsharpL}
& \phantom{nixnixnixnixnix} = 4\pi  \tfrac{1}{\sqrt{L}}{\textstyle{\sum\limits_\ell}}
\PsiVa(t,\qVv_{\mathrm{ph}}^{\hat\ell};\qVv_{\mathrm{el}}) \otl
\vVv(t,{\qVv_{\mathrm{el}}})\bcdot\delta^{(a)}_{{\qVv_{\mathrm{el}}}}(\qV_{\mathrm{ph}}^\ell),
\\
&\hspace{-0.7truecm}  \label{eq:MdivPSIsharpL}   
 \hbar \nabla_{\qVv_{\mathrm{ph}}}\bcdot \PsiVL(t,\qVv_{\mathrm{ph}};\qVv_{\mathrm{el}})
= 4\pi  
\tfrac{1}{\sqrt{L}} {\textstyle{\sum\limits_\ell}}
\PsiVa(t,\qVv_{\mathrm{ph}}^{\hat\ell};\qVv_{\mathrm{el}})
\Bigl(\delta^{(a)}_{\nullV}(\qV_{\mathrm{ph}}^\ell) - {\textstyle{\sum\limits_n}}\delta^{(a)}_{\qV_{\mathrm{el}},n}(\qV_{\mathrm{ph}}^\ell)
\Bigr).
\end{alignat}
 Here, $\PsiVa(t,\qVv_{\mathrm{ph}}^{\hat\ell};\qV_{\mathrm{el}})$ is an $L-1$ photon wave function, 
\emph{conditioned} on the generic $N$-electron configuration
$\qVv_{\qVv_{\mathrm{el}}}$,  and $\qVv_{\mathrm{ph}}^{\hat\ell}$ is a $3(L-1)$-dimensional generic configuration space position of $L-1$ photons, 
obtained from $\qVv_{\mathrm{ph}}$ by removing $\qV_{\mathrm{ph}}^\ell$.
 Moreover, $\nabla_{\qVv_{\mathrm{ph}}}\bcdot \PsiVL(t,\qVv_{\mathrm{ph}};\qV_{\mathrm{el}})$
is a sum over $\ell\in\{1,...,L\}$ of the $\ell$-th divergence operator acting on the $\ell$-th factor of $\PsiVL$. 
 Finally, $\PsiVa(t,\qVv_{\mathrm{ph}}^{\hat\ell};\qVv_{\mathrm{el}})\otl \vVv(t,{\qV_{\mathrm{el}}}) 
\bcdot\delta^{(a)}_{{\qVv_{\mathrm{el}}}}(\qV_{\mathrm{ph}}^\ell)$ manifestly  
resembles an $L$-photon wave function obtained from an $L-1$-photon wave function, both conditioned on the $N$-electron configuration, 
by applying a ``single-photon creation operator'' in which $\vVv\bcdot\delta^{(a)}_{{\qVv_{\mathrm{el}}}}(\qV_{\mathrm{ph}}^\ell)$ 
takes the place of the $\ell$-th factor.
 Therefore, tentatively we may consider (\ref{eq:MdotPSIsharpL}), (\ref{eq:MdivPSIsharpL}) as a plausible $L$-photon wave equation,
conditioned on the generic $N$-electron configuration. 
\newpage

\noindent
\textbf{Remark}: 
\emph{Of course, in writing down (\ref{eq:MdotPSIsharpL}), (\ref{eq:MdivPSIsharpL}) we had the benefit of hindsight offered
by the well-developed  creation / annihilation operator formalism.
 Yet also this started in the 1920s, with the papers of Jordan \cite{Jordan}, enhanced jointly with Wigner \cite{JW}, and by Dirac \cite{GoldenRule}.
 Moreover, it is remarkable that the $L=1$ special case is just the $\sharp$-field equations obtained from interpreting
the Maxwell--Lorentz field equations that Schr\"odinger had written down as quantum-mechanical expected values w.r.t. the
Born probability measure for the $N$ electrons, $\varrho = |\Psi|^2$.}
\medskip

 With the creation operator formalism occurring in the single-photon equations for $\PsiV$ as per re-interpretation of the 
original charged source terms in the electromagnetic $\sharp$-field interpretation, and easily generalized to the $L$-photon
equations for $\PsiVL$, we next note that the annihilation operator formalism already occurs in the $\Psi$ equations when 
re-interpreted accordingly.
 Indeed, it suffices to note that the ``$\sharp$-field energy''
\begin{equation}
E^\sharp(t,\qVv_{\mathrm{el}})=\frac{1}{8\pi}\int_{\Rset^3} \big(\PsiV^* \cdot \PsiV\big) (t,\qV_{\mathrm{ph}};\qVv_{\mathrm{el}})\drm^3q_{\mathrm{ph}},
\label{eq:FIELDenergyPHOTON}
\end{equation}
i.e. $E^\sharp$ at l.h.s.(\ref{eq:ERWINeqnNEW}) rewritten in Weber notation,
is an ``annihilation operator'' acting on the one-photon wave function $\PsiV$ conditioned on the $N$-electron configuration, 
with the ``annihilation'' effected by the same $\PsiV$ itself.

 To generalize this to conditioned $L$-photon wave functions $\PsiVL$, it turns out that one has several options if the only
requirement is that for stationary states the same empirically correct Pauli energy spectra should be obtained. 
 Therefore one needs further guidance in ones choice. 
 A suggestive requirement to impose is to retain the bilinear type of feedback. 
 In this vein, let $\Pi_\ell$ denote the projector onto the $\ell$-th factor of $\PsiVL$. 
 Then it is readily verified that
\begin{equation}
E_L^\sharp(t,\qVv_{\mathrm{el}}):=
\frac{1}{8\pi L}
\sum_\ell \int_{\Rset^{3}} \Big(\big(\Pi_\ell{\PsiVL}\big)^* \bcdot \big(\Pi_\ell\PsiVL\big)\Big) (t,\qV^\ell_{\mathrm{ph}};\qVv_{\mathrm{el}})\drm^{3}q_{\mathrm{ph}}
  \label{eq:FIELDenergyLphotonsELLsum}
\end{equation}
produces the correct atomic (etc.) Hamiltonian in the absence of radiation (i.e. static $\sharp$-fields). 
 To see that this leads to the same spectrum as previously discussed, for convenience here only in the absence of
laboratory-generated external fields, we observe that the stationary states $\Psi$ with purely electrostatic $\PsiVL$ 
admit a separation of variables, so $\PsiVL$ is a static Hartree state, consisting of $L$ copies of the Coulomb field.
 For now this is our tentative proposal for the $\sharp$-field feedback term when exactly a single $\PsiVL$ is coupled to $\Psi$. 

 To back up our claim that one has other options, we note that without the bilinarity requirement also 
\begin{equation}
\widetilde{E}_L^\sharp(t,\qVv_{\mathrm{el}}):=
\frac{1}{8\pi}\Big(\int_{\Rset^{3L}} \big({\PsiVL}^* \bcdot \PsiVL\big) (t,\qVv_{\mathrm{ph}};\qVv_{\mathrm{el}})\drm^{3L}q_{\mathrm{ph}}\Big)^{1/L}
  \label{eq:FIELDenergyLphotons}
\end{equation}
yields the same Pauli spectra. 
 Indeed, for the static Hartree state the integral at r.h.s.(\ref{eq:FIELDenergyLphotons}) is the $L^{th}$ power of the integral at 
r.h.s.(\ref{eq:FIELDenergyPHOTON}), and the outer power $1/L$ in (\ref{eq:FIELDenergyLphotons}) corrects this. 
 This observation may just be a curiosity, but it is helpful to realize that in the further development of the model one has to make
choices. 
 
 It is also readily possible now to couple all possible $\PsiVL$ to $\Psi$.
 Thus, let $w^{}_\ell\geq 0$ and $\sum_{\ell\in\Nset} w^{}_\ell = 1$. 
 Then  a weighted sum of the $L$-photon terms, viz.
\begin{equation}
E^\sharp(t,\qVv_{\mathrm{el}}):= \sum_{\ell\in\Nset} w^{}_\ell E_L^\sharp(t,\qVv_{\mathrm{el}}),
  \label{eq:FIELDenergyALLphotons}
\end{equation}
is the most general non-negative ``$\sharp$-field energy'' feedback term that produces the correct Hamiltonian in the absence of radiation
and which retains bilinearity.

 One also needs to extract  corresponding three-dimensional $\sharp$-field momentum vectors $\PV_n^\sharp$ from the $\PsiVL$ that replace
the current single-photon coupling terms in Schr\"odinger's, respectively Pauli's equation. 
 We temporarily postpone this to a later revised and enlarged version of this paper. 

 This concludes our demonstration that there seems to be a natural path to generalizing our radiating-atom model from the single-photon
to a many-photon version. 

\subsection{Creation / annihilation of photons vs. their activation}

 It is remarkable that the source terms in (\ref{eq:MdotPSIsharpL}), (\ref{eq:MdivPSIsharpL}), which in our setup essentially
suggest themselves as logical generalizations of the empirical charge and current density source terms in the Maxwell--Lorentz field
equations, look very much like regularized boson creation operators in ``non-relativistic QFT.'' 
 So this may well suggest that a mandatory next step is to consider the Fock space of all $L$-photon sectors, $L\in\Nset\cup\{0\}$,
with a hierarchy of equations of the type (\ref{eq:MdotPSIsharpL}), (\ref{eq:MdivPSIsharpL}) or similar. 
 This then would also seem to mandate the incorporation of the analogues of annihilation operators into the $\PsiVL$ equations,
in a similarly logically compelling manner. 
 In keeping with the spirit of the whole quantum-mechanical approach, this should be done without invoking the second-quantization formalism.

 There is a different possibility, though, namely that what has the appearance of creation operators are really merely source terms for the 
photon wave function, not for the photons themselves. 
 Also, the annihilation operators already made an appearance in the $\Psi$ equations, where they originally were implemented as 
$\sharp$-field energy (and momentum) operators without originally thinking of these as annihilation operators. 
 Moreover, in our model so far the single-photon $\PsiV$ is never identically zero due to the 
constraint equations, and in this sense there is no ``no photon'' state.
 Instead, what transpires is to not think of creation / annihilation of photons themselves but of their ``activation,'' viz. their
being set in motion by $\PsiV$ or rather $\PsiVL$ getting excited above the electrostatic level.
 There may be $\infty$ many photons all the time, none getting destroyed or created, but perhaps only $L$ of them are in motion, or 
possibly different $L$ participate but with different weights $w_\ell^{}$.
 Clearly this is not sorted out but definitely seems worth pursuing.

\vspace{-5pt}
\section{Summary and Outlook}\vspace{-10pt}

\subsection{Summary}
 In this paper we have developed a tentative semi-relativistic quantum-mechanical model of electrons and photons which interact with
each other and with fixed atomic nuclei.
 The model accurately reproduces all the atomic and molecular (etc.) energy spectra of the so-called standard model of everyday matter,
and it also describes the emission / absorption of photons by atoms. 
 It also seems to capture at least some the details of a single photon emission processes accurately to the extent that can be expected 
from a semi-relativistic theory.
 Whether it captures all details accurately, and whether the many-photon generalization already captures the physics of a radiating atom 
qualitatively correctly and quantitatively accurately (to the extent which can reasonably be demanded from a semi-relativistic model of atoms and photons) 
is a different, and difficult question which can only be answered after further careful analysis of the equations.

 Yet, the semi-relativistic tentative quantum mechanics of electrons and photons developed in this paper 
gets so many things right already, qualitatively and quantitatively, that it seems reasonable to pursue this model further.
 We expect that it will serve as an intermediate stepping stone on the way to a completely satisfactory, \emph{macroscopically relativistic},
QM of electrons, photons, and their anti-particles --- indeed we believe that such a theory is feasible.

 By ``macroscopically relativistic'' we mean the intriguing possibility that relativity theory may only be valid 
as a ``quantum-mechanical expected value,'' as suggested by our model. 
 The underlying theory itself would not be Lorentz covariant, yet its equations be such that their quantum expected value is.
 Since macroscopic matter consists of a huge number of particles, by a law of large numbers the expected values would be essentially sharp
in all macroscopic phenomena.
 Thus relativity theory would appear to be a law of nature only for all practical purposes, similar to thermodynamics, 
yet not reflect a fundamental symmetry of nature.
 This would offer a way out of the apparent conflict between Einstein's relativity theory (no influence outside
of the lightcone) and quantum nonlocality as established by Bell \cite{BellBOOK}.
 \newpage

\subsection{Outlook}
 Such a ``macroscopically relativistic'' quantum mechanics should be formulated with a single joint wave function of all these particles, 
not a coupled system of various partial wave functions. 
 This joint wave function should obey a single linear wave equation.

 Before one gets there, one first may want to generalize our model by replacing
Schr\"odinger's, respectively Pauli's equations by a Dirac equation. 
 Thus, in the example of hydrogen, or a hydrogenic atom / ion, we should replace (\ref{eq:PAULIeqn}) by
\begin{equation}
\left(i \hbar \partial_0 - P_0^f
\right)\Psi 
= 
\boldsymbol\alpha\cdot\left(-i \hbar \nabla_{\qV_{\mathrm{el}}} - \PV^\sharp 
 \right) \Psi 
+\mEL c\beta \Psi 
,\hspace{-.2truecm}
 \label{eq:DIRACeqn}
\end{equation}
where $\boldsymbol\alpha$ and $\beta$ are the conventional Dirac matrices, $\Psi$ now is a Dirac bi-spinor (cf. \cite{Thaller}),
and where we have dropped the argument $(t,{\qV_{\mathrm{el}}})$ from $P_0$, $\PV^\sharp$, and $\Psi$.
 This makes it plain that the square of a bilinear expression in the $\sharp$-fields which entered
Schr\"odinger's, respectively Pauli's equations, was just a consequence of the non-relativistic approximation to a Dirac equation.

 Next, also the $\sharp$-field equations, generalizations of Maxwell's field equations, presumably will have to be replaced by 
counterparts more deserving of the name \emph{wave equation for a photon}; see \cite{KTZphoton} for such an equation describing a single free photon,
which is a Dirac type equation acting on rank-two bi-spinors which can be decomposed into a system of four equations, part of which
comprises two copies of formally Maxwell's field equations. 
 This photon wave equation yields a Hamiltonian with the correct photon energies of $\hbar \omega = \hbar c|\kV|$, where $\omega$ is the 
angular frequency of a plane photon wave function of wave vector $\kV$, something we have not yet extracted from the current $\sharp$-field 
equations, and like Dirac's equation for the electron, it has positive and negative energies. 

 We have yet to address its generalization to the $L$-photon formalism. (TBA)

 But even with such improved quantum-mechanical wave equations that for a single free particle \emph{are} properly Lorentz covariant, 
we do need to couple them, and to avoid the familiar infinite self-energies of point charges, we in this paper simply regularized
them with tiny charged balls of radius $a$. 
 Balls of fixed radius $a$ are not a Lorentz-covariant concept, and thus the quantum-expectation value of our formalism cannot be
completely Lorentz covariant.
 To fix this there are at least two options, both of which worthy of pursuit.

 First, one can pick up on the ideas of Max Born, soon joined by Infeld, that the culprit is Maxwell's electromagnetic vacuum law. 
 The Born--Infeld (BI) proposal is very intriguing, but the intimidating nonlinearity of their vacuum law makes it difficult to work with;
cf. \cite{KieJSPa}, \cite{KieJSPb}.  
 An alternative proposal was made by Fritz Bopp, and soon thereafter by Land\'e--Thomas, and then Podolsky (BLTP), to replace Maxwell's vacuum
law with another linear law that, however, now involves second order partial differential operators (in fact, Klein--Gordon operators). 
 The classical BLTP electrodynamics of $N$ point charges has been proved to be locally well-posed as a joint initial value problem for
fields and charges, and is fully Lorentz covariant; see \cite{KiePRD}, \cite{KTZonBLTP}.
 It would seem to suggest itself as a way to replace the tiny balls of radius $a$ by true point particles, and thus to pave the ground
for a quantum mechanics of charged point electrons, point nuclei, and photons that is macroscopically fully Lorentz covariant.

 Another intriguing possibility has been pioneered in \cite{TeufelTumulka}.
 There the UV infinities are removed thanks to what they call ``interior-boundary'' condition, which does not mean a boundary condition at
some interior boundary of a space (like, e.g., the center of a punctured disk), but refers to a relationship between an (indeed) internal 
boundary of $N>2$ particle configuration space (namely the set of coincidence points of two particles) and the interior region of the $N-2$ 
particle sector (note the hyphen!).
 [I suggest, if I may, to reverse the order to ``boundary-interior'' condition, to avoid confusion.]
 No UV cutoff like balls or such are needed, but this approach now does not preserve the $L^2$ norm of the $N$-particle wave functions 
and works in Fock space, with creation and annihilation operators acting in the usual manner. 

 Finally, it may yet be feasible to implement Lorentz covariance at the microscopic many-body level with the help of a multi-time
formalism; see \cite{KLTZ} for a recent Lorentz-covariant model of an 
interacting electron-photon system in 1+1 dimensions using multi-time formalism, and the references theirein.
 The synchronization would yield not manifestly Lorentz covariant equations, and whether these can be matched with the types
of equations developed in this paper has yet to be seen.
\bigskip

\textbf{ACKNOWLEDGEMENT}: TBA

\newpage

\appendix

\section*{Appendices}
\addcontentsline{toc}{section}{Appendices}
\renewcommand{\thesubsection}{\Alph{subsection}}
\numberwithin{equation}{subsection}

\subsection{Real hydrogen wave functions}

 Equation (\ref{eq:ERWINeqnMatterWaveBOhydrogen}) is precisely Schr\"odinger's equation for hydrogen in Born--Oppenheimer approximation.
 Thus, introducing $\alphaS = \frac{e^2}{\hbar c} \approx \frac{1}{137.036}$ (Sommerfeld's fine structure constant), 
the eigenvalues are
\begin{eqnarray}
E_n = 
-\tfrac12 \alphaS^2 \mEL c^2 \tfrac{1}{n^2}    ,\quad  n \in\Nset.
\label{eq:BOHRspec} 
\end{eqnarray}
 For each $n\in\{1,2,...\}$ there are $n^2$ linear independent real eigenfunctions, indexed by 
$\ell\in\{0,1,...,n-1\}$ and $m\in\{0,...,\ell\}$ and, for each $m>0$ a parity index $\varsigma\in\{\pm\}$,
while $\varsigma=+$ if $m=0$.
   In terms of the usual physics notation, if $Y_\ell^m(\vartheta,\varphi)$ with $\ell\in\{0,1,2,3,...\}$ and 
$m\in\{-\ell,...,0,...,\ell\}$ denotes the three-dimensional spherical harmonics, which satisfy
\begin{eqnarray}
-\Delta_{\Sset^2}^{} Y_\ell^m (\vartheta,\varphi)
= \label{eq:SPHEREharmLAP}
\ell(\ell+1)Y_\ell^m (\vartheta,\varphi),
\end{eqnarray}
and which are of the form $Y_\ell^m(\vartheta,\varphi) = \Theta_\ell^m(\vartheta) e^{im\varphi}$,
where $\Theta_\ell^m(\vartheta)$ is real and satisfies 
\begin{eqnarray}
- \frac{1}{\sin\vartheta}\frac{\drm}{\drm\vartheta}\left(\sin\vartheta\frac{\drm}{\drm\vartheta}
\Theta_\ell^m (\vartheta) \right) + \frac{m^2 }{\sin^2\vartheta} \Theta_\ell^m (\vartheta) 
= \label{eq:SPHEREharmLAPtheta}
\ell(\ell+1)\Theta_\ell^m (\vartheta),
\end{eqnarray}
known as \emph{Legendre's differential equation}, then only the real, resp. imaginary parts of the 
$Y_\ell^m (\vartheta,\varphi)$ are to be used to obtain an eigenstate with vanishing velocity field.
 Thus, the real eigenfunctions are (for $n\in\Nset$, $\ell\in\{0,...,n-1\}$, and $m\in\{0,1,...,\ell\}$)
\begin{eqnarray}
\psi_{n,\ell,m,+}(r,\vartheta,\varphi)  = \label{eq:PSInlmPLUS}
 R_{n,\ell}(r) \Theta_\ell^m (\vartheta)\cos(m\varphi) ,
\end{eqnarray}
\begin{eqnarray}
\psi_{n,\ell,m,-}(r,\vartheta,\varphi)  = \label{eq:PSInlmMINUS}
 R_{n,\ell}(r) \Theta_\ell^m (\vartheta)\sin(m\varphi) .
\end{eqnarray}
 Here, 
\begin{eqnarray}
 R_{n,\ell}(r)
  = \label{eq:RellnCOULOMB}
 \sqrt{\frac{(n-1-\ell)!}{2n(n+\ell)!}}\;
\left(\frac{2\alpha}{n}\right)^{\frac32}\left(\frac{2\alpha}{n}\frac{r}{\lambda_{\mbox{\tiny{C}}}}\right)^\ell 
L_{n-1-\ell}^{2\ell+1}\left(\frac{2\alpha}{n}\frac{r}{\lambda_{\mbox{\tiny{C}}}}\right)  e^{-\alpha r/n\lambda_{\mbox{\tiny{C}}}} ,
\end{eqnarray}
with $\lambda_{\mbox{\tiny{C}}}$ the reduced Compton length of the electron,
is the familiar radial part of the Schr\"odinger hydrogen eigenfunctions, 
where $L_\nu^{\kappa}(\xi)$ with $\nu\in\{0,1,2,...\}$ and $\kappa\in\Rset_+$ is the 
associated Laguerre polynomial obtained from the generating function (cf. \cite{AS})
\begin{eqnarray}
\frac{e^{-\frac{\tau\xi}{1-\tau}}}{(1-\tau)^{\kappa +1}} =\sum_{\nu=0}^\infty \tau^\nu L_\nu^{\kappa}(\xi) .
\end{eqnarray}

\newpage

\subsection{$\sharp$-field energy-momentum identities}


 Taking the inner product of (\ref{eq:MdotBsharp}) with  $\BV^\sharp(t,\sV;\qV)$
and the inner product of (\ref{eq:aMdotEsharp}) with  $\EV^\sharp(t,\sV;\qV)$, then integrating each of the resulting
equations over $\Rset^3$ w.r.t. $\drm^3s$, adding the results, then multiplying by $\varrho(t,\qV)$ and
integrating over $\Rset^3$ w.r.t. $\drm^3q$, and recalling the definition \eqref{eq:FIELDenergyZ} and
the abbreviation $\langle\,\cdot\,\rangle = \int_{\Rset^3}(\,\cdot\,)\varrho d^3q$, we find 
\begin{alignat}{1}
\langle\partial_t E^\sharp\rangle(t) + \left\langle \left(\vV\cdot\nabla_\qV\right)E^\sharp\right\rangle(t)
= \label{eq:aveEfperSHARP} 
 \eEL  \left\langle \Langle \EV^\sharp\Rangle_a^{}\cdot\vV\right\rangle(t) .
\end{alignat} 

\noindent
\textbf{Remark}: 
\emph{The limit $\lim_{a\to 0} \Langle \EV^\sharp\Rangle_a^{}$ is generally not well-defined.}
 
 On the other hand, for the expected field energy (\ref{eq:FIELDenergyZ}) we also compute
\begin{alignat}{1}
\frac{\drm \ }{\drm t} \langle E^\sharp\rangle(t) 
&= \label{eq:aveEf}  
\langle\partial_t E^\sharp\rangle(t) + \int_{\Rset^3}\big(\partial_t\varrho(t,\qV)\big)E^\sharp(t;\qV)\drm^3q \\ 
&=\langle\partial_t E^\sharp\rangle(t) -
\int_{\Rset^3}\big(\nabla_\qV\cdot[\varrho(t,\qV)\vV(t,\qV)]\big)E^\sharp(t;\qV)\drm^3q \\ 
&= \langle\partial_t E^\sharp\rangle(t) +
\int_{\Rset^3}\varrho(t,\qV)\left(\vV(t,\qV)\cdot\nabla_\qV\right)E^\sharp(t;\qV)\drm^3q \\
 &= \langle\partial_t E^\sharp\rangle(t) + \left\langle \left(\vV\cdot\nabla_\qV\right)E^\sharp\right\rangle(t).\label{eq:aveEfend} 
\end{alignat} 

  Comparing (\ref{eq:aveEf})--(\ref{eq:aveEfend}) with (\ref{eq:aveEfperSHARP}) we conclude that
\begin{alignat}{1}
\boxed{\tfrac{\drm \ }{\drm t} \langle E^\sharp\rangle(t) 
= \label{eq:aveEftimeDERIV} 
 \eEL \left\langle \Langle \EV^\sharp\Rangle_a^{}\cdot\vV\right\rangle(t) }.
\end{alignat} 

Similarly, for the generic field momentum (\ref{eq:FIELDmomentumZ}) we compute
\begin{alignat}{1}
\frac{\drm \ }{\drm t} \langle \PV^\sharp\rangle(t) 
 = \label{eq:avePf}  
\langle\partial_t \PV^\sharp\rangle(t) + \left\langle \left(\vV\cdot\nabla_\qV\right)\PV^\sharp\right\rangle(t).
\end{alignat} 
 At rhs(\ref{eq:avePf}) we now substitute rhs(\ref{eq:FIELDmomentumZ}) for $\PV^\sharp$, use the $\sharp$-field
equations, and find
 \begin{alignat}{1}\hspace{-0.7truecm}
\langle\partial_t \PV^\sharp\rangle(t) + \langle\left(\vV\cdot\nabla_\qV\right)\PV^\sharp\rangle(t)
 = \label{eq:avePfAGAIN}  
 \eEL \left(\left\langle \Langle \EV^\sharp\Rangle_a^{}\right\rangle(t) 
+ \left\langle \tfrac1c\vV\times \Langle \BV^\sharp\Rangle_a^{}\right\rangle(t) \right).
\end{alignat} 
 And so,
 \begin{alignat}{1}
\boxed{\tfrac{\drm \ }{\drm t} \langle \PV^\sharp\rangle(t) 
 = \label{eq:avePfAGAINagain}  
 \eEL \left(\left\langle \Langle \EV^\sharp\Rangle_a^{}\right\rangle(t) 
+ \left\langle\tfrac1c\vV\times \Langle \BV^\sharp\Rangle_a^{}\right\rangle\right)(t) }.
\end{alignat} 

 From (\ref{eq:avePfAGAIN}) we also obtain 
 \begin{alignat}{1}
\boxed{\tfrac{\drm \ }{\drm t} \langle \tfrac12|\PV^\sharp|^2\rangle(t) 
 = \label{eq:avePfsquare}  
 \eEL \left(\left\langle \PV^\sharp\cdot \Langle \EV^\sharp\Rangle_a^{}\right\rangle(t) 
+ \left\langle\PV^\sharp\cdot \left(\tfrac1c\vV\times \Langle \BV^\sharp\Rangle_a^{}\right)\right\rangle\right)(t) }.
\end{alignat} 

\newpage

\subsection{\!\!\!On Hartree units for hydrogen in an external $\AV$ field}

 In the usual atomic units introduced by Hartree, Schr\"odinger's equation \eqref{eq:ERWINeqnAphi} with an external radiation field 
$\AV_{\mbox{\tiny{ext}}}$ in place of $\AV_{\mathrm{el}}$ \emph{appears} to furnish a small $O(\alphaS)$ perturbation 
$H_{\mbox{\tiny{int}}}(t,\qV) + H_{\mbox{\tiny{rad}}}(t,\qV)$ relative to both terms in the 
hydrogen Hamiltonian $H_{\mbox{\tiny{{hyd}}}}(\qV)$; here, $\alphaS\approx 1/137.036$ is Sommerfeld's fine structure constant.
 Because of this appearance it could seem that incoming radiation can always be treated as a small perturbation, but this appearance 
is misleading, as we now explain.

In the dimensionless atomic units of Hartree:

\qquad$\bullet$ 
charge is a multiple of $e$, 

\qquad$\bullet$ 
length is a multiple of the Bohr radius $\hbar/\mEL c \alphaS$, 

\qquad$\bullet$ 
time is a multiple of $\hbar/\mEL c^2 \alphaS^2$, 

\qquad$\bullet$ 
energy is a multiple of $\alphaS^2\mEL c^2$,

\qquad$\bullet$ 
$\AV_{\mbox{\tiny{rad}}}$ is a multiple of $\alphaS e \mEL c/\hbar$. 

\noindent
 It follows that in these units 
\begin{equation}
\label{HamHydHARTREEunits}
H_{\mbox{\tiny{{hyd}}}}(\qV) = -\tfrac12\Delta - \tfrac1r
\end{equation}
and (in the Coulomb gauge)
$$
H_{\mbox{\tiny{int}}}(t,\qV)
=
- \alphaS i \AV_{\mbox{\tiny{rad}}}\cdot\nabla_{\qV} + \alphaS^2\tfrac{1}{2}\big|\AV_{\mbox{\tiny{rad}}}\big|^2 .
$$

 Thus it appears that the radiation field furnishes a small $O(\alphaS)$ perturbation 
$H_{\mbox{\tiny{int}}}(t,\qV)$ {relative} to the hydrogen Hamiltonian $H_{\mbox{\tiny{{hyd}}}}(\qV)$, which is $O(1)$
in both terms, in these units.
 However, this is only apparent, because the Hartree units themselves depend on $\alphaS$, and because the
amplitude of $\AV_{\mbox{\tiny{rad}}}$ is a free parameter.
 Since the physical electromagnetic coupling constants of the linear Coulomb and the linear radiation-interaction Hamiltonian
are the same, the negligibility of the radiation-interaction Hamiltonian has to be vindicated by the relative
smallness of the radiation field amplitudes, not by artificially generating different coupling constants through a 
convenient choice of units which obscures the physical assumptions that enter into the perturbation theory. 

 To make this point explicitly clear, consider $\alphaS$ not as a physical constant but as a real parameter that can
be sent to $0$. 
 In the Hartree units, the Hamiltonian 
$H = -\tfrac12\Delta - \tfrac1r - \alphaS i \AV_{\mbox{\tiny{rad}}}\cdot\nabla_{\qV} + \alphaS^2\tfrac{1}{2}\big|\AV_{\mbox{\tiny{rad}}}\big|^2 $
then converges to the hydrogen Hamiltonian \eqref{HamHydHARTREEunits}, but of course this is not the case in any dimensionless units which do not
themselves depend on $\alphaS$  --- in those units the limit $\alphaS\to 0$ would lead to the Hamiltonian $-\frac12\Delta$.

\newpage



\begin{thebibliography}{[99999991]}\footnotesize{

\bibitem[AbSt1972]{AS}
\vskip-5pt
        Abramowitz, M., and Stegun, I.,
        ``Handbook of Mathematical Functions,'' 9th ed.,
        Dover, New York (1972).

\bibitem[AGTZ2011]{ValiaETal}
\vskip-5pt
                  Allori, V.,
                  Goldstein, S.,
                  Tumulka, R.,
                  and Zangh\`{\i}, N.,
 \textit{Many-Worlds and Schr\"o\-dinger's First Quantum Theory},
 British J. Phil. Sci. \textbf{62}, 1--27 (2011).

\bibitem[ApKi2001]{AppKieAOP}
\vskip-5pt
	Appel, W., and Kiessling, M.K.-H.,
        	{\sl Mass and spin renormalization in Lorentz electrodynamics},
    	Annals Phys. (N.Y.) \textbf{289}, 24-83 (2001).

\bibitem[BaVa2009]{Solvay}
\vskip-5pt
 Bacciagaluppi, G.,
 and
 Valentini, A.,
 \emph{Quantum Theory at the Crossroads: Reconsidering the 1927 Solvay Conference},
 Cambridge Univ. Press, Cambridge, UK (2009);
 an updated version is available as arXiv:quant-ph/0609184v2.


\bibitem[BaSe1974]{BazleySeydel}
\vskip-5pt
Bazley, N.,
and
Seydel, R.,
   ``{Existence and bounds for critical energies of the Hartree operator},''
   Chem. Phys. Lett. \textbf{24}, 128--132 (1974).

\bibitem[Bell87/04]{BellBOOK}
\vskip-5pt
	Bell, J.S.,
		{\sl Speakable and unspeakable in quantum mechanics},
	Cambridge University Press, Cambridge, UK (1987);
	$2^{nd}$ ed. (2004).

\bibitem[BFZ1976]{BFZ}
\vskip-5pt
Benci, V.,
Fortunato, D., 
Zirilli, F.,
  ``Exponential decay and regularity properties of the Hartree approximation to the bound state wave functions of the helium atom,'' 
J. Math. Phys. \textbf{17}, 1154--1155 (1976). 

 \bibitem[BBL1981]{BenguriaBrezisLieb}
\vskip-5pt
Benguria, R.,
Brezis, H.,
 and Lieb, E.H.,
``The Thomas--Fermi--von Weizs\"acker theory of atoms and molecules,''
Commun. Math. Phys. \textbf{79}, 167--180 (1981).

 \bibitem[BeLi1983]{BenguriaLieb}
\vskip-5pt
Benguria, R., and Lieb, E.H.,
``Proof of stability of highly negative ions in the absence of the Pauli principle,'' 
Phys. Rev. Lett. \textbf{50}, 1771--1774 (1983).

\bibitem[Betal1995]{BerndlETal}
\vspace{-5pt}
Berndl, K., D\"urr, D., Goldstein, S., Peruzzi, G. and Zangh\`{\i}, N.
``On the global existence of Bohmian mechanics,''
\textit{Commun. Math. Phys.}, {\bf 173}, 647--673 (1995).


\bibitem[Bohm1951]{BohmBOOKa}
\vskip-5pt
  Bohm, D.,
  \textit{Quantum Theory},
  Prentice Hall.

\bibitem[Bohm1952a]{BohmsHIDDENvarPAPERS}
\vskip-5pt
	Bohm, D.,	
		{\sl A suggested interpretation of the quantum
			theory in terms of ``hidden'' variables. Part I},
	Phys. Rev. \textbf{85},  166-179 (1952);
	         {\sl Part II}, 
	ibid.,  180-193 (1952).
\bibitem[Bohm1952b]{BohmsREPLYtoCRITICSa}
\vskip-5pt
        Bohm, D., 
                {\sl Reply to a criticism of a causal re-interpretation of the quantum theory,}
        Phys. Rev. \textbf{87}, 389-390 (1952).
\bibitem[Bohm1953]{BohmsREPLYtoCRITICSb}
\vskip-5pt
        Bohm, D., 
	        {\sl Comments on an article of Takabayashi concerning the
          formulation of quantum mechanics with classical pictures,}
         Prog. Theor. Phys. \textbf{9}, 273-287 (1953).

\bibitem[BoHi1993]{BoHi}
  Bohm, D.,
  and 
  Hiley, B.,
   ``{The Undivided Universe},''
  Routledge, London (1993).

\bibitem[Bohr1913]{BohrHatomA}
\vskip-5pt
 Bohr, N.,
  ``On the constitution of atoms and molecules,'' Philos. Mag. \textbf{26}, 1--25 (1913).



\bibitem[Born1926a]{BornsPSISQUAREpapersA}
\vskip-5pt
	Born, M., 	
	         {\sl Zur Quantenmechanik der Stossvorg\"ange},
	Z. Phys. {\textbf{37}},  863--867 (1926).

\bibitem[Born1926b]{BornsPSISQUAREpapersB}
\vskip-5pt
	Born, M., 	
	         {\sl Quantenmechanik der Stossvorg\"ange},
	Z. Phys. {\textbf{38}},  803--827 (1926).  

\bibitem[Born1926c]{BornsPSISQUAREpapersC}
\vskip-5pt
	Born, M., 	
	         {\sl Das Adiabatenprinzip in der Quantenmechanik},
	Z. Phys. {\textbf{40}},  167--192 (1926).  

\bibitem[Born1926d]{BornsPSISQUAREpapersD}
\vskip-5pt
	Born, M., 	
	         {\sl Zur Wellenmechanik der Stossvorg\"ange},
                 Nachr. Ges. Wiss. G\"ottingen {\textbf{38}},  146--160 (1926).  

\bibitem[BoOp1927]{MaxOppi}
\vskip-5pt
	Born, M., 	
        and
        Oppenheimer, J.R.,
       ``Zur Quantentheorie der Molekeln,'' 
        Annalen Phys. \textbf{389}: 457--484 (1927). 

\bibitem[BoHu1954]{MaxKerson}
\vskip-5pt
 Born, M., 
 and
 Huang, K., \textit{Dynamical Theory of Crystal Lattices,} New York: Oxford Univ. Press (1954). 

\bibitem[Bric2016]{BricmontBOOK}
\vskip-5pt
  Bricmont, J.
  \textit{Making sense of quantum mechanics},
  Springer (2016).


\bibitem[deBr1928]{deBroglieSOLVAY}
\vskip-5pt
        de Broglie, L.V.,
	        {\sl La nouvelle dynamique des quanta,}
        in ``Cinqui\`eme Conseil de Physique Solvay (Bruxelles, 1927),''
        J. Bordet, ed. (Gauthier-Villars et Cie., Paris, 1928);
        English transl.: ``The new dynamics of quanta'', p.374-406 in:
        Bacciagaluppi, G., and Valentini, A.,
        {\sl Quantum Theory at the Crossroads}, (Cambridge Univ. Press, 2009).






\bibitem[CBSM2020]{CBSM}
\vskip-5pt
Cox, H., 
Baskerville, A.L., 
Syrjanen, V.J.J., 
and
Melgaard, M., 
``The bound state stability of the hydride ion in Hartree-Fock theory,''
Adv. Quant. Chem. \textbf{81}, 167--189 (2020).

\bibitem[Darw1928]{Darwin}
\vskip-5pt
       Darwin, C.G.,
       \textit{The wave equations of the electron},
       Roy. Soc. Proc., A, \textbf{118}, 654--680 (1928).

\bibitem[PDH1979]{PhillipidisDewdneyHiley}
\vskip-5pt
Philippidis, C.,
 Dewdney, C.,
and Hiley, B.J.,
 ``Quantum Interference and the Quantum Potential,'' 
 Il Nuovo Cimento \textbf{B 52}, 15--28 (1979).


\bibitem[Dira1927]{GoldenRule}
 Dirac, P. A. M.,
 ``The Quantum Theory of Emission and Absorption of Radiation,'' 
         Proc. Royal Soc. A. \textbf{114}: 243--265 (1927).  

\bibitem[D\"uTe2009]{DuerrTeufelBOOK}
\vskip-5pt
        D\"urr, D.,
	and
	Teufel, S.,
	         {\sl Bohmian Mechanics. The Physics and Mathematics of Quantum Theory},
	Springer Verlag, Berlin (2009).

\bibitem[DGZ1992]{DuerrEtalA}
\vskip-5pt
        D\"urr, D.,     
	Goldstein, S.,  
	and 
	Zangh\`{\i}, N. 
	{\sl Quantum equilibrium and the origin of absolute uncertainty}, 
	J. Stat. Phys. \textbf{67},  843-907 (1992). 

\bibitem[Eins1909]{EinsteinPHOTON}
\vskip-5pt
        Einstein, A.,    
                {\sl \"Uber die Entwicklung unserer Anschauung \"uber das Wesen und die Konstitution der Strahlung},
        Phys. Z. \textbf{10}, 817--826 (1909).

\bibitem[Ency2005]{EncyclopediaOPTICS} 
 \textit{Encyclopedia of Modern Optics}, Elsevier (2005).



\bibitem[FrGo2009]{GeroETal}
\vskip-5pt
Friesecke, G., 
and 
Goddard, B.D.,
``Explicit large nuclear charge limit of electronic ground states for Li, Be, B, C, N, O, F, Ne, 
and basic aspects of the periodic table,''
  SIAM J. Math. Anal. \textbf{41}, 631--664 (2009).  

\bibitem[Fr\"oh2004]{JuergA}
\vskip-5pt
 Fr\"ohlich, J.,
 \textit{The electron is inexhaustible},
  Lecture notes (lectures at ETH Z\"urich, and at Universitas Lipsiensis (Univ. Leipzig)), 
  copy of May 24, 2004.


\bibitem[Gold2017]{ShellySEP}
\vskip-5pt
       Goldstein, S.,
 \textit{Bohmian Mechanics},
 Stanford Encyclopedia of Philosophy (Summer 2017 Ed.), 
Edward N. Zalta (ed.), available at the URL: https://plato.stanford.edu/archives/sum2017/entries/qm-bohm/.

\bibitem[GrRy80]{GradRyzh}
\vskip-.3truecm
	Gradshteyn, I.S., and Ryzhik, I.M.,
        \textit{Table of Integrals, Series, and Products}, 4th. ed.,
        Academic Press, New York (1980).

\bibitem[Guer1995]{GuerraBOOK}
\vskip-5pt
Guerra, F.,
  \textit{Introduction to Nelson Stochastic Mechanics as a Model for Quantum Mechanics},
   Fundamental Theories of Physics book series \textbf{71}, Springer (1995).

\bibitem[Hart1928]{Hartree}
\vskip-5pt
Hartree, D.,
 ``The wave mechanics of an atom with a non-Coulomb central field. Part I. Theory and methods,'' 
 Proc. Camb. Phil. Soc. \textbf{24}, 89--132 (1928).

\bibitem[Hill1963]{Hill}
\vskip-5pt
Hill, R.N.,
 ``Proof that the H$^-$ ion has only one bound state. Details and extension to finite nuclear mass,''
 J. Math. Phys. \textbf{18}, 2316--2330 (1977).

\bibitem[Holl1993]{Holland}
\vskip-5pt
Holland, P.R,
 \emph{The quantum theory of motion (An account of the de Broglie--Bohm causal interpretation of quantum mechanics},  
 Cambridge Univ. Press, Cambridge (1993).

\bibitem[Hyll1963]{Hylleraas}
\vskip-5pt
Hylleraas, E.A.,
 ``Reminiscences from early quantum mechanics of two-electron atoms,''
Rev. Mod. Phys. \textbf{35}, 
 421--430 (1963).


\bibitem[Jack1975]{JacksonBOOKb}
\vskip-5pt
        Jackson, J.D.,   
                {\sl Classical electrodynamics}, 
		$2^{nd}$ ed., J. Wiley \& Sons, New York (1975).

\bibitem[Jord1927]{Jordan}
\vskip-5pt
Jordan, P., 
``\"Uber Wellen und Korpuskeln in der Quantenmechanik,''
  Zeits. Phys. \textbf{45}, 765--775 (1927).


\bibitem[JoWi1928]{JW}
\vskip-5pt
Jordan, P., 
and
Wigner, E.P.,
``\"Uber das Paulische \"Aquivalenzverbot,''
  Zeits. Phys. \textbf{47}, 631--651 (1928).

\bibitem[Kamk1979]{Kamke}
\vskip-5pt
Kamke, E.,
 \textit{Differentialgleichungen: L\"osungsmethoden und L\"osungen. II: Partielle Differentialgleichungen erster Ordnung
f\"ur eine gesuchte Funktion,} 6th Ed., Teubner, Stuttgart (1979).



\bibitem[KaKr2012]{KK}
\vskip-5pt
Kawohl, B., and Kr\"omer, S.,
``Uniqueness and symmetry of minimizers of Hartree type equations with external Coulomb potential,''
Adv. Calc. Var. \textbf{5}, 427--432 (2012).


\bibitem[Kies1999]{KiePLA} 
\vskip-5pt
	Kiessling, M.K.-H.,
		{\sl Classical electron theory and conservation laws},
	Phys. Lett. \textbf{A 258}, 197--204  (1999).

\bibitem[Kies2004a]{KieJSPa} 
\vskip-5pt
	Kiessling, M.K.-H.,
		{\sl Electromagnetic field theory without divergence problems. 1. The Born legacy},
	J. Stat. Phys. \textbf{116}, 1057--1122  (2004).

\bibitem[Kies2004b]{KieJSPb} 
\vskip-5pt
	Kiessling, M.K.-H.,
		{\sl Electromagnetic field theory without divergence problems. 
		        2. A least invasively quantized theory},
	J. Stat. Phys. \textbf{116}, 1123--1159 (2004).

\bibitem[Kies2019]{KiePRD}
\vskip-5pt
 Kiessling, M.K.-H.,
  \textit{Force on a point charge source of the classical electromagnetic fields},
  Phys. Rev. D \textbf{100}, art065012 (19pp) (2019); erratum ibid. \textbf{101}, art. 109901E (2020).

\bibitem[KLTZ2020]{KLTZ}
\vskip-5pt
 Kiessling, M.K.-H.,
 Lienert, M.,
 and
 Tahvildar-Zadeh, A.S.,
  \textit{A 
     Lorentz-covariant model of an interacting electron-photon system in 1+1 spacetime dimensions},
 Lett. Math. Phys. \textbf{110}, 3153--3195 (2020). 
https://doi.org/10.1007/s11005-020-01331-8



\bibitem[KTZ2018]{KTZphoton}
\vskip-5pt
 Kiessling, M.K.-H.,
 and
 Tahvildar-Zadeh, A.S.,
 \textit{On the quantum mechanics of a single photon}, J. Math. Phys. \textbf{59}, art.112302 (2018).

\bibitem[KTZ2020]{KTZonBLTP}
\vskip-5pt
 Kiessling, M.K.-H.,
 and
 Tahvildar-Zadeh, A.S.,
 \textit{Bopp--Land\'e--Thomas--Podolsky electrodynamics as initial value problem},
  in preparation (2021).

\bibitem[Ketal2011]{weakM}
\vskip-5pt
Kocsis, S., 
Braverman, B., 
 Ravets,  S., 
 Stevens, M.J., 
Mirin,  R.P., 
Shalm, L.K., 
 and 
Steinberg, A.M.,  
 ``Observing the Average Trajectories of Single Photons in a Two-Slit Interferometer,'' Science \textbf{332}, 1170--1173 (2011).

\bibitem[Kome2019]{KomechLECT}
\vskip-5pt
Komech, A.I.,
\textit{Lectures on Quantum Mechanics for mathematicians}, e-print at
 arXiv:1907.05768v2 math-ph (2019).



 \bibitem[LiSi1977]{LiebSimon}
\vskip-5pt
Lieb, E.H., and Simon, B.,
``Hartree-Fock theory for Coulomb systems,''
Commun. Math. Phys. \textbf{53}, 185--194 (1977).

 \bibitem[Lie1984]{Lieb}
\vskip-5pt
Lieb, E.H.,
``Atomic and molecular negative ions,''
Phys. Rev. Lett. \textbf{52}, 315--317 (1984).



\bibitem[Lou1986]{Loura}
\vskip-5pt
de Loura, L.,
``A numerical method for the Hartree equation of the helium atom,''
        Calcolo \textbf{23}, 185--207 (1986).

\bibitem[Nels1967]{NelsonBOOKa}
\vskip-5pt
  Nelson, E. 
  \textit{Dynamical Theories of Brownian Motion}, Princeton University Press, Princeton (1967).

\bibitem[Nels1985]{NelsonBOOKb}
\vskip-5pt
        Nelson, E., 
	       {\sl Quantum fluctuations},
	Princeton Series in Physics, Princeton University Press, Princeton (1985).

\bibitem[vNeu1932]{vN}
\vskip-5pt
 v. Neumann, J.,
  \textit{Mathematical foundations of quantum mechanics},
  Princeton Univ. Press (1955).



\bibitem[OrMo2012]{OriolsMompart}
\vskip-5pt
Oriols, X.,
 and 
Mompart, J., 
\textit{Applied Bohmian Mechanics: From Nanoscale Systems to Cosmology}, Pan Standford Publishing, Singapore (2012).

\bibitem[Paul1926]{PauliH}
\vskip-5pt
Pauli, W.E.,
  \textit{\"Uber das Wasserstoffspektrum vom Standpunkt der neuen Quantenmechanik},
 Z. Phys. \textbf{36}, 336--363 (1926). 

\bibitem[Paul1927]{PauliEQ}
\vskip-5pt
Pauli, W.E.,
  \textit{Zur Quantenmechanik des magnetischen Elektrons}, 
 Z. Phys. \textbf{43}, 601--623 (1927). 

\bibitem[Rau1996]{Rau}
\vskip-5pt
Rau, A.R.P.,
 ``The negative ion of hydrogen,''
 J. Astrophys. Astron. \textbf{17}, 113--145 (1996).

\bibitem[ReSi1980]{ReedSimonBOOKi}
\vskip-5pt
	Reed, M., 	
	and 
	Simon, B.,	
		{\sl Functional analysis} 
		(Methods of modern mathematical physics I), revised ed.,
	Acad. Press, Orlando (1980).

\bibitem[ReSi1975]{ReedSimonBOOKii}
\vskip-5pt
	Reed, M., 	
	and 
	Simon, B.,	
		{\sl Fourier analysis} 
		(Methods of modern mathematical physics II),
	Acad. Press, Orlando (1975).

\bibitem[ReSi1979]{ReedSimonBOOKiii}
\vskip-5pt
	Reed, M., 	
	and 
	Simon, B.,	
		{\sl Scattering theory} 
		(Methods of modern mathematical physics III),
	Acad. Press, Orlando (1979).

\bibitem[ReSi1978]{ReedSimonBOOKiv}
\vskip-5pt
	Reed, M., 
	and 
	Simon, B.,
		{\sl Analysis of operators} 
		(Methods of modern mathematical physics IV),
	Acad. Press, Orlando (1978).

\bibitem[Reek1970]{Reeken}
\vskip-5pt
	Reeken, M., 
   ``{General theorem on bifurcation and its application to the Hartree equation of the helium atom},''
        J. Math. Phys. \textbf{11}, 2505--2512 (1970).

\bibitem[Renn2013]{Renn}
\vskip-5pt
	Renn, J., 
   ``{Schr\"odinger and the Genesis of Wave Mechanics},''
        Max Planck Inst. f. Wissenschaftsgeschichte, Preprint \textbf{437}, 29pp. (2013).


\bibitem[Schr1926a]{ErwinWMa}
\vskip-5pt
	Schr\"odinger, E.,
		{\sl Quantisierung als Eigenwertproblem (Erste Mitteilung)},
	Annalen Phys. \textbf{79}, 361--376 (1926).

\bibitem[Schr1926b]{ErwinWMb}
\vskip-5pt
	Schr\"odinger, E.,
		{\sl Quantisierung als Eigenwertproblem (Zweite Mitteilung)},
	Annalen Phys. \textbf{79}, 489--527 (1926).
\bibitem[Schr1926c]{ErwinWMc}
\vskip-5pt
	Schr\"odinger, E.,
		{\sl \"Uber das Verh\"altnis der Heisenberg-Born-Jordanschen Quantenmechanik zu der meinen},
	Annalen Phys. \textbf{79}, 734--756 (1926).

\bibitem[Schr1926d]{ErwinWMd}
\vskip-5pt
	Schr\"odinger, E.,
		{\sl Quantisierung als Eigenwertproblem. (Dritte Mitteilung)},
	Annalen Phys. \textbf{80}, 437--490 (1926).

\bibitem[Schr1926e]{ErwinWMe}
\vskip-5pt
	Schr\"odinger, E.,
		{\sl Quantisierung als Eigenwertproblem. (Vierte  Mitteilung)},
	Annalen Phys. \textbf{81}, 109--140 (139?) (1926).


\bibitem[Schr1982]{WaveMechanics}
\vskip-5pt
   Schr\"odinger, E.,
   \textit{Collected Papers on Wave Mechanics},
   3rd augmented ed.,  AMS Chelsea Pub., AMS (1982)

\bibitem[Somm1916]{fineSTRUKTUR}
\vskip-5pt
  Sommerfeld, A.,
  \textit{Zur Quantentheorie der Spektrallinien},
  Ann. Physik \textbf{51}, 1--94 \&\ 125--167 (1916).

\bibitem[Spoh2004]{SpohnBOOKb}
\vskip-7pt
       Spohn, H.,
       \textit{Dynamics of charged particles and their radiation fields},
       Cambridge UP (2004).

\bibitem[Stru2015]{Ward}
\vskip-7pt
       Struyve, W.,
       \emph{Semi-classical approximations based on Bohmian mechanics},
        Preprint arXiv:1507.04771 (2015). 


\bibitem[Thal1992]{Thaller}
\vskip-5pt
 Thaller, B.,
 ``The Dirac equation,''
  Springer, New York (1992).

\bibitem[TeTu2005]{TeuTum}
\vspace{-5pt}
  Teufel, S.,
  and 
Tumulka, R.,
  ``Simple Proof for Global Existence of Bohmian Trajectories,''
{\em Commun. Math. Phys.}, {\bf 258}:349--365 (2005).


\bibitem[TeTu2020]{TeufelTumulka}
\vskip-5pt
 Teufel, S.,
 and
 Tumulka, R.,
 \textit{Hamiltonians Without Ultraviolet Divergence for Quantum Field Theories},
 Quantum Studies: Math. and Found., in press (2020).

\bibitem[Webe1901]{Weber}
\vskip-5pt
  Weber, H.,
  ``{Die partiellen Differentialgleichungen der mathematischen Physik nach Riemann's Vorlesungen},''
  vol.2, 4th ed., Vieweg, Braunschweig (1901).\vspace{-4pt}

\bibitem[Wein1995]{WeinbergBOOKqft}
\vskip-5pt
  Weinberg, S.
  ``The quantum theory of fields, Vol. I,''
  Cambridge Univ. Press, (1995).

\bibitem[Wyat2005]{Wyatt}
\vskip-5pt
Wyatt, R.E.,
\textit{Quantum Dynamics with trajectories}, Springer, New York (2005).

\bibitem[Yagh2006]{Yaghjian}
\vskip-5pt
Yaghjian, A.,
  \textit{Relativistic Dynamics of a Charged Sphere: Updating the Lorentz--Abraham Model}, Springer, New York (2006).

}
\end{thebibliography}
\end{document}